\documentclass[a4paper,usenatbib]{mnras}

\usepackage{newtxtext,newtxmath}

\usepackage[T1]{fontenc}
\usepackage{ae,aecompl}

\usepackage{url}
\usepackage{lscape}
\usepackage{longtable}

\usepackage{graphicx}	
\usepackage{amsmath}	
\usepackage{amssymb}	

\title[Chronos and KAIROS: On the Selection of $z\sim1$ Post-Starburst Group and Cluster Galaxies]{Chronos and KAIROS: MOSFIRE Observations of Post-Starburst Galaxies in $z\sim$1 Clusters and Groups} 
\author[B.~C.\ Lemaux et al.]{B.C. Lemaux$^{1}$\thanks{E-mail:
bclemaux@ucdavis.edu},  A.~R.\ Tomczak$^{1}$, L.~M.\ Lubin$^{1}$, P-F.\ Wu$^{2}$, R.~R.\ Gal$^{3}$, N.\ Rumbaugh$^{1,6}$, \newauthor D.~D.\ Kocevski$^{4}$, \& G.~K.\ Squires$^{5}$\\
$^{1}$ Department of Physics, University of California, Davis, One Shields Ave., Davis, CA 95616, USA \\
$^{2}$ Max-Planck Institut f\"{u}r Astronomie, K\"{o}nigstuhl 17, D-69117, Heidelberg, Germany\\
$^{3}$ University of Hawai'i, Institute for Astronomy, 2680 Woodlawn Drive, Honolulu, HI 96822, USA \\
$^{4}$ Department of Physics and Astronomy, Colby College, Waterville, ME 04961, USA \\ 
$^{5}$ Spitzer Science Center, California Institute of Technology, M/S 220-6, 1200 E. California Blvd., Pasadena, CA 91125, USA \\
$^{6}$ National Center for Supercomputing Applications, University of Illinois, 1205 West Clark St., Urbana, IL 61801, USA}

\date{Accepted XXX. Received YYY; in original form ZZZ}

\pubyear{2017}

\begin{document}
\label{firstpage}
\pagerange{\pageref{firstpage}--\pageref{lastpage}}
\maketitle

\begin{abstract}
We present an exploration of $\sim500$ spectroscopically-confirmed galaxies in and around two large scale structures (LSSs) at $z\sim1$ drawn
from the ORELSE survey, an ongoing, wide-field photometric and spectroscopic campaign targeting a large ensemble of LSSs at $0.6<z<1.3$. A sub-sample of these galaxies 
($\sim150$) was targeted for the initial phase of a near-infrared MOSFIRE spectroscopic campaign investigating the
differences in selections of galaxies which had recently ended a burst of star formation and/or had rapidly quenched (i.e., post-starburst/K+A galaxies). Selection
with MOSFIRE utilizing the H$\alpha$ and [NII] emission features resulted in a post-starburst sample more than double that selected by traditional $z\sim1$ 
(observed-frame optical) methods even after the removal of the relatively
large fraction of dusty starburst galaxies selected through traditional methods. While the traditional post-starburst fraction increased with increasing global density, 
the MOSFIRE-selected post-starburst fraction was found to be constant across field, group, and cluster environments. However, this fraction computed relative to the number of star-forming galaxies was observed to elevate in the cluster environment. Post-starbursts selected with MOSFIRE exhibited 
moderately strong [OII] emission originating from activity other than star formation. Such galaxies, termed K+A with ImposteR [OII]-derived Star formation (KAIROS) galaxies, 
were found to be younger than and likely undergoing feedback absent or diminished in their optically-selected counterparts. 
A comparison between the environments of the two types of post-starbursts suggested a picture in which the evolution of a post-starburst galaxy  
is considerably different in cluster environments than in the more rarefied environments of a group or the field.
\end{abstract}




\section{Introduction}

It has been evident since both the earliest observations of nearby clusters of galaxies and the initial modeling of these clusters that the environment in which 
a galaxy resides dictates, to some extent, its 
fate \citep{gunn72,oemler74, bo1978, dressler80}. Studies of enormous samples of field, group, and cluster galaxies residing in the local universe made possible by the Sloan Digital Sky 
Survey (SDSS, \citealt{york00}) have reaffirmed these early observations; galaxy populations which reside in higher density environments almost universally have 
larger fraction of red, quiescent early-type galaxies than those populations inhabiting more rarefied environments (e.g., \citealt{goto03a, goto03b, gomez03, hansen09}). While 
this correlation appears, generally, to hold up to $z\sim1$ \citep{casey10,strazzullo10, cooke16} at least for the most massive overdensities, strong differential evolution is 
observed in the quiescent fraction between overdense and more typical regions of the universe (e.g., \citealt{mcoopz07, olga16}). 
At earlier epochs, with some exceptions, the fraction of redder or passive galaxies appears to decrease appreciably in overdense environments reaching 
levels indistinguishable from or beneath that of the field population (e.g., \citealt{taowang16,lihwai16}). Correspondingly, an increase in the average star formation rate (SFR) of 
galaxies residing in overdense environments is observed, with the relationship between SFR and density flattening (e.g., \citealt{greg13,ziparo14}) or, in some cases, reversing 
(e.g., \citealt{tran10, santos14,santos15, dey16}) at $z\ga1$ relative to the anti-correlated behavior observed in the local universe. In tandem, strong differential evolution 
between the field and overdense environments is observed
in the fraction of galaxies undergoing strong transient activity, such as those hosting a particularly prodigious star-formation event \citep{dirtydale11b, webb13} or 
a powerful active galactic nuclei (AGN, \citealt{martini13, alberts16}). Given the large number of processes that serve to either induce or quench star formation 
activity in group and cluster environments, which are either not present or considerably less effective in more typical environments, it is easy 
to adopt the naive view that environment is \emph{the} fundamental quantity in governing the evolution of a galaxy both at early and late times in the history of the universe. 

However, this simplistic view is challenged in a variety of respects. Large spectroscopic and photometric samples of galaxies from $0<z<4$ predominantly located in 
field environments have unequivocally shown that separate processes which appear largely independent of or only circumstantially connected with environment can act 
effectively to transform blue, star-forming, late-type galaxies to red, quiescent, early types \citep{bundy06, faber07,pozzetti10, ilbert10,ilbert13, iary13}. Such processes 
appear to be intimately connected, whether circumstantially or causally, to the stellar mass content of galaxies. It is observed in both the local \citep{peng10,deng11} and 
the higher-redshift universe \citep{muz12,darvish16} that the average star formation
rate per unit stellar mass (specific star formation rate, SSFR) of star-forming galaxies is constant at fixed stellar mass as a function of environment (though this trend may depend on 
the method used to estimate environment, see \citealt{noble16}). Conversely, the average SSFR is found to be a strong function of stellar mass in both the local 
and distant universe. In a study of several thousands of galaxies from $z=0-1$, \citet{peng10} inferred based on a simple empirical model that (stellar) massive galaxies 
($\log(\mathcal{M}_{\ast}/M_{\odot})$>10.2) are much more likely to be quenched via stellar-mass-related processes than those related to environment. Lending credence 
to this picture is the observed similarity in the stellar mass function of \emph{both} quiescent and star-forming galaxies in the field relative to those galaxies
residing in more dense environments (e.g., \citealt{vanderburg13, vulcani13}). These observations seem to require that, if 
environmentally-driven quenching is to occur, 
it must, once begun, operate over a relatively rapid timescale in order to preserve these trends. Such a scenario is supported by several recent studies 
which have, through a variety of different methods, attempted to constrain the star formation histories (SFHs) of local and $z\sim1$ group and cluster galaxies 
\citep{wetzel13, muz14, mok14, balogh16} finding that, after a relatively long delay, truncation of star formation must be rapid ($\sim0.1-0.8$ Gyr) to simultaneously fit  
data, models, and simulations (though see, e.g., \citealt{taranu14} for an alternative view). Conversely, such a feature is not required in the SFHs of field galaxies. 
Thus, it appears that the most promising avenue of inquiry to observe and constrain environmentally-driven quenching is in galaxies which have undergone recent, 
dramatic changes in their star formation properties.  

While several candidate populations exist, e.g., red spirals \citep{moran07, masters10, bundy10}, galaxies selected with colours intermediate to star-forming and quiescent populations 
\citep{balogh11, kschawin14}, rejuvenated lenticular or spheroid galaxies \citep{treu05,sheilaK09}, one population in particular has been given particular attention 
over the past several decades. Early studies of intermediate redshift clusters revealed a modestly large population of galaxies with spectral features 
indicative of a lack of ongoing star formation and a large number of recently formed stars added to an older underlying 
stellar population \citep{dressler1983, couch1987, dressler1992}. Such a spectrum is only possible for a galaxy which has recently undergone a 
star-formation event vastly exceeding its past-averaged star formation activity (hereafter ``starburst") or rapid quenching (or both), necessarily 
meaning such galaxies have necessarily undergone a violent transformation in the recent past. These galaxies, eventually termed ``K+As" after the 
two primary stellar types observed in their spectra\footnote{Though these galaxies are sometimes termed ``post-starburst" it should be clear from the 
definition that a starburst is not required to generate the K+A phase. Regardless, we irresponsibly use the two terms interchangeably throughout the paper.}, 
became the subject of intensive searches across all environments in both the local and distant universe. 
Initial searches found a considerable fractional excess of K+A galaxies inhabiting massive clusters at moderate redshift relative to the coeval 
field (e.g., \citealt{belloni95, dressler99, tran03}), which led to speculation that cluster-related processes were essential to induce the strength
of burst or the rapidity of the quenching needed to induce a K+A phase. Later studies showed that inhabiting a cluster environment 
was not necessarily a requisite condition, as K+As were also found in more rarefied environments such as groups 
and, in some cases, the field \citep{zabludoff96}. While some more recent studies have found that K+A galaxies are, by fraction, more 
likely to inhabit the cluster environment at $z\sim1$ \citep{pog09, muz12, pwu14}, several studies have found a relatively large fraction of K+As in the field at these redshifts 
\citep{yan09,wild09, vergani10}, again precluding the possibility that cluster-specific processes are solely responsible for generating this evolutionary phase. 

Howevever, the extreme rarity of the K+A population and the lack of campaigns targeting overdense environments at $z\sim1$ with coverage analogous to wide-scale coeval field surveys at $z\sim1$  
(e.g., \citealt{lilly07,dong13,new13}) make such trends highly subject to sample variance as well as the depth and breadth of the spectroscopic coverage. 
Perhaps the most pernicious difficulties in interpreting these trends comes in the form of the choice of metrics used to define 
environment (local vs. global), various controls or lack thereof on the sample (volume-limited vs. flux-limited, luminosity-limited vs. stellar mass limited), and 
the method by which K+A galaxies are selected. Approached carefully, the former two issues are perhaps easier to mitigate. While quenching mechanisms
appear to have a complex relationship with local density, halo mass, dynamical state of the parent halo, stellar mass, and various photometric and 
spectroscopic limits, such limitations can be broadly controlled for with relative ease by making completeness corrections to the sample, incorporating appropriate 
sample variance uncertainties, or by making proper internal comparisons. The issue of differing K+A selection, however, is not correctable by these approaches as 
differing selections of galaxies classed ``K+A" can fundamentally change the galaxy population being probed and, by consequence, the conditions that the
selected population is experiencing. While promising progress has been made on selecting K+A populations photometrically \citep{wild14, maltby16}, such methods
are still maturing, and, thus, we limit our discussion here to K+As selected spectroscopically. There are several issues with the canonical selection
of K+A galaxies at higher redshift ($z\sim1$), and they are primarily related to the requirement that K+A galaxies have no ongoing star formation. While 
exceptions exist, the vast majority of $z\sim1$ studies of K+A galaxies require the absence of the [OII] $\lambda$3727\AA\ line in order to classify 
a galaxy as K+A. Such a requirement carries with it a variety of issues related to purity and completeness. The criterion or criteria used to set
the limit for the non-detection of [OII] line is highly dependent on the signal to noise (S/N) and resolution of the observed spectra. A minimal change 
at fixed S/N and resolution can lead to the selection of dramatically different populations (e.g., \citealt{pwu14}). Equally importantly, the [OII] 
line is highly subject to differential extinction and can be emitted, copiously so, by processes other star formation leaving open the possibility that 
even those $z\sim1$ K+A samples selected using an [OII] cut appropriate for their data will be comprised of a large number of dusty starburst galaxies (false 
positives) and will exclude a large number of galaxies with post-starburst features that are emitting [OII] for a reason other than star formation (false negatives). 

In this study we investigate a large population ($\sim500$) of galaxies in and around two large scale structures at $z\sim0.8$ targeted with 
observed-frame optical spectroscopy, of which $\sim$150 were followed up with near-infrared (NIR) spectroscopy. These samples are additionally complemented 
by deep 10+ band optical/NIR and X-ray imaging. With these observations, we investigate the effects of completeness and purity in traditionally-selected 
K+A populations in both the field and in overdense environments and the consequences 
for the inferences on quenching mechanisms in such environments. The paper is organized as follows. Section \S\ref{tarnobs} discusses the structures targeted in this study 
and lists the properties of our optical/NIR imaging and spectroscopy. In \S\ref{methods} we discuss the various analysis used to approach the analysis 
of K+A and other galaxy types. In \S\ref{PSB} we discuss the results of our investigation including those on the purity and completeness of traditional K+A selection
and the differences between galaxies selected using traditional means and those selected with our observations. Finally, in \S\ref{concl} we summarize all
of our results. Throughout this paper all magnitudes, including those in the IR, are presented in the AB system \citep{okengunn83,fukugita96}
All equivalent width measurements are presented in the rest frame and we adopt the convention of negative equivalent widths corresponding
to a feature observed in emission. All distances are quoted in proper units. We adopt a concordance $\Lambda$CDM 
cosmology with $H_{0}$ = 70 km s$^{-1}$ Mpc$^{-1}$, $\Omega_{\Lambda}$ = 0.73, and $\Omega_{M}$ = 0.27. 

\section{Targets and Observations}
\label{tarnobs}

The subject of this study is the galaxy population in and surrounding two large scale structures (LSSs) at $z\sim0.8$, SG0023 and RXJ1716, 
drawn from the Observations of Redshift Evolution in Large Scale Environments (ORELSE; \citealt{lub09}) survey. The ORELSE survey is a massive ongoing 
photometric and spectroscopic campaign dedicated to mapping out and characterizing the galaxy population in and around $\sim$20 large 
scale structures at $0.6\le z \le1.3$. These two LSSs were chosen from the full ORELSE sample to be the maiden fields for MOSFIRE observations 
due to (a) their similarity in redshift, (b) their similar extensive coverage in optical/NIR imaging and optical spectroscopy, (c) their 
encompassing the full range of properties of LSSs targeted by the ORELSE survey at these redshifts, and (d) falling at a redshift where 
the H$\alpha$ $\lambda$6563\AA\ and the [NII] $\lambda$6583\AA\ features are comfortably situated away from bright OH lines in the J-band sky. The 
optically-selected SG0023 supergroup at $z\sim0.83$ is comprised of at least five distinct groups ($\sigma_{v}<550$ 
km s$^{-1}$) characterized by relatively low dynamical masses ($\log(\mathcal{M}_{vir}/\mathcal{M}_{\odot}) = 12.7-13.9$), a lack of discernible diffuse X-ray 
emission originating from a hot medium \citep{rum13}, and a galaxy population primarily composed of star-forming and starbursting galaxies \citep{lub09}.
In stark contrast, the massive ($\log(\mathcal{M}_{vir}/\mathcal{M}_{\odot})=15.2$), X-ray selected RXJ1716 cluster at $z\sim0.81$ is characterized 
by a strong, regular diffuse intracluster medium (ICM) emission ($L_{X,\ bol}=9.3\pm0.4\times10^{44}$ ergs s$^{-1}$; \citealt{vikhlinin02}; Rumbaugh et al.\		
2016), though with low level, but significant, structure near its outskirts, a core of massive ($\log(\mathcal{M}_{\ast}/\mathcal{M}_{\odot}) > 11$), 
quiescent members, and a large overall fraction of quiescent member galaxies\footnote{When referring to a particular cluster or group, the definition of 
a ``member galaxy" is given in Table \ref{tab:LSSs}. For LSSs, a member galaxy is defined more loosely as simply a galaxy in the redshift range of that LSS 
within the spatial constraints of our DEIMOS coverage.} ($\sim50$\%) to the stellar mass completeness limit of the 
spectroscopic survey ($\log(\mathcal{M}_{\ast}/\mathcal{M}_{\odot})\ge 10$). However, despite what appears to be an evolved, isolated structure, 
the LSS appears to house appreciable spatial and velocity sub-structure and estimates of its halo mass from lensing \citep{clowe98} and X-ray \citep{ettori04} 
($3-4\pm1\times10^{14}$ $\mathcal{M}_{\odot}$) are both considerably below the estimate made from the dynamics of its member population (see Table \ref{tab:LSSs}) indicating at 
least a moderate departure from virialization. The general properties of the member groups and cluster of the two LSSs are given in Table \ref{tab:LSSs}. In this section 
we briefly discuss the imaging and spectroscopic data taken of these LSSs and their surrounding fields.

\begin{table*}
        \centering
        \caption{Cluster and Group Properties} 
        \label{tab:LSSs}
        \begin{tabular}{lllllll} 
                \hline
                Structure & $\alpha_{J2000}^{a}$ & $\delta_{J2000}^{a}$ & $\langle z \rangle$&  $N_{mem, \, spec}^{b}$ & $\sigma_{v}^{c}$ & $M_{vir}$ \\
                          & [$\deg$] & [$\deg$] & &  & [km s$^{-1}$] & [$\log{\mathcal{M}_{\odot}}$]\\
                \hline
                RXJ1716 & 259.2016 & 67.1392 & 0.8116 & 144 & 1150$\pm$162 & 15.2$\pm$0.2\\
                SG0023A & 6.0256 & 4.3590 & 0.8396 & 29 & 507$\pm$126 & 13.8$\pm$0.6\\
                SG0023B1 & 5.9757 & 4.3884 & 0.8290 & 11 & 106$\pm$51 & 12.7$\pm$0.3\\
                SG0023B2 & 5.9697 & 4.3820 & 0.8453 & 17 & 231$\pm$54 & 13.3$\pm$0.3\\
                SG0023C & 5.9247 & 4.3807 & 0.8466 & 70 & 544$\pm$59 & 13.7$\pm$0.3\\
                SG0023M & 5.9674 & 4.3199 & 0.8472 & 17 & 487$\pm$85 & 13.9$\pm$0.3\\
                \hline
        \end{tabular}
	\begin{flushleft}
$a$: $I^{+}$/$i^{\prime}$-luminosity-weighted centre of member galaxies calculated using the method described in \citet{begona14}.
$b$: Defined as galaxies with $|\Delta_{v}|\le3\sigma_{v}$ from $\langle z \rangle$ and $R_{proj}<2R_{vir}$ from the optical spatial centre, except for SG0023B1/B2 where
coherent structure in differential velocity was observed out to $R>>R_{vir}$, for which we adopted $R_{proj}\le0.5$ $h_{70}^{-1}$ Mpc. 
$c$: The measured line-of-sight (LOS) galaxy velocity dispersion measured using the method of \citet{lem12}.
\end{flushleft}

\end{table*}

\subsection{Imaging and Photometry}
\label{phot}

The wealth of imaging data available as well as their depth for both SG0023 and RXJ1716 is given in Table \ref{tab:imaging}. 
Here we briefly summarize the observations and reduction of these data. A full description of the reduction process of these data will be given in
Tomczak et al.\ (\emph{submitted}). Optical imaging of the two LSSs was taken from our own
observations with the Large Format Camera (LFC; \citealt{simcoe00}) on the Palomar 5-m telescope and our own and archival imaging with 
Suprime-Cam \citep{miyazaki02} mounted on the Subaru 8-m telescope. Reduction of the LFC data was done in the Image Reduction and Analysis Facility 
(\texttt{IRAF}, \citealt{tody93}) and follows the methods outlined in \citep{gal08}. Reduction of the Suprime-Cam data was performed 
with the \texttt{SDFRED2} pipeline \citep{ouchi04} supplemented by several Traitement \'{E}l\'{e}mentaire R\'{e}duction et Analyse des PIXels
(\texttt{TERAPIX\footnote{http://terapix.iap.fr}}) routines. Photometric calibration in all cases was performed from observations of 
\citet{landolt92} standard star fields taken on the same night of each observation. 

Near-infrared imaging in the $J$ and $K$ bands was taken with the Wide-Field Camera (WFCAM; \citealt{casali07}) 
mounted on the United Kingdom Infrared Telescope (UKIRT) and the Wide-field InfraRed Camera (WIRCam; \citealt{puget04}) mounted on the 
Canada-France-Hawai'i Telescope (CFHT) for SG0023 and RXJ1716, respectively. The UKIRT data were processed using the standard UKIRT 
processing pipeline courtesy of the Cambridge Astronomy Survey Unit\footnote{http://casu.ast.cam.ac.uk/surveys-projects/wfcam/technical} and
the CFHT data through the I'iwi pre-processing routines and \texttt{TERAPIX}. The photometric calibration of the mosaics output by
both pipelines was done selecting bright ($m<15$), non-saturated objects with existing Two Micron All Sky Survey (2MASS; \citealt{skrutskie06})
photometry. Further infrared imaging was obtained with the \emph{Spitzer} \citep{wer04} InfraRed Array Camera (IRAC; \citealt{fazio04}) in all four channels
for RXJ1716 and the two non-cryogenic channels ($[3.6]/[4.5]$) for SG0023. The basic calibrated data (cBCD) images provided by the \emph{Spitzer} 
Heritage Archive were reduced using the MOsaicker and Point source EXtractor (MOPEX; \citealt{makovoz06}) package in conjunction with several 
custom Interactive Data Language (\texttt{IDL}) scripts written by J. Surace. Further details will be given in Tomczak et al.\ (\emph{submitted}). All Spitzer imaging is 
provided flux-calibrated in units of MJy/sr.
 
For the ground-based imaging, photometry was obtained by running Source Extractor (\texttt{SExtractor}; \citealt{BertinArn96}) in dual-image mode using an 
inverse-variance-weighted $R^{+}I^{+}$ and $R_{C}I^{+}Z^{+}$ stack as the detection image for SG0023 and RXJ1716, respectively. Prior to running \texttt{SExtractor} 
in each field, all images are registered to a common 
grid and convolved to the worst point spread function (PSF) for that field ($1\arcsec$ and $0.9\arcsec$ for SG0023 and RXJ1716, respectively) estimated
from stacked point sources in each image using the Richardson \& Lucy algorithm in \texttt{scikit-image} to generate the convolution kernel. Fixed aperture 
photometry (1.3$\times$ the FWHM of the homogenized PSF) was then performed on these PSF-matched images ensuring that an identical fraction of the light of each object is measured 
in all broadband images. An aperture correction was made to the measured magnitudes by scaling the ratio of aperture and AUTO flux 
densities as measured in the detection image, a similar practice to that commonly adopted in other large surveys (e.g., \citealt{laigle16, thibaud16}). 
Magnitude uncertainties were calculated from adding, in quadrature, 
\texttt{SExtractor} uncertainties to our own estimates of background noise drawn from the 1$\sigma$ root mean square (RMS) scatter of measurements 
in hundreds of blank sky regions for each band. Photometry for the \emph{Spitzer}/IRAC images
was treated separately due to the appreciably larger and differently shaped PSF ($\sim2\arcsec$) in these bands relative to the ground-based images. 
The package \texttt{T-PHOT} \citep{merlin15} was used to translate the segmentation map in the detection image for each field to its equivalent 
in the \emph{Spitzer}/IRAC images using a given kernel and to mitigate blending through optimal scaling of the resultant convolved segmentation map
estimated from a fit to the observed data. Flux densities are then extracted from the scaled best fit model of each object. A summary of all imaging
data and associated depths for both fields is given in Table \ref{tab:imaging}. 

\begin{table}
        \centering
        \caption{Imaging Data}
        \label{tab:imaging}
        \begin{tabular}{lll} 
                \hline
                Band & Telescope/Instrument & Depth$^{a}$ \\
                \hline
                RXJ1716 &  &  \\ 
                \hline
		\hline
		$B$ & Subaru/Suprime-Cam & 25.5 \\
		$V$ & Subaru/Suprime-Cam & 26.0 \\
		$R_{C}$ & Subaru/Suprime-Cam & 25.8\\
                $I^{+}$ & Subaru/Suprime-Cam & 25.1\\
                $Z^{+}$ & Subaru/Suprime-Cam & 24.3 \\
                $J$ & CFHT/WIRCam & 21.9 \\
                $K_s$ & CFHT/WIRCam & 22.3 \\
                $[3.6]$ & \emph{Spitzer}/IRAC & 22.5 \\
                $[4.5]$ & \emph{Spitzer}/IRAC & 22.1 \\
		$[5.8]$ & \emph{Spitzer}/IRAC & 21.6 \\
                $[8.0]$ & \emph{Spitzer}/IRAC & 20.6 \\	
		\hline
		SG0023 &  &  \\ 
                \hline
		\hline
		$B$ & Subaru/Suprime-Cam & 26.4 \\
                $V$ & Subaru/Suprime-Cam & 25.9 \\
		$R^{+}$ & Subaru/Suprime-Cam & 25.2\\
		$r^{\prime}$ & Palomar/LFC & 25.1 \\
		$I^{+}$ & Subaru/Suprime-Cam & 24.6\\
		$i^{\prime}$ & Palomar/LFC & 24.5 \\
		$z^{\prime}$ & Palomar/LFC & 23.1 \\
		$J$ & UKIRT/WFCAM & 22.0 \\
		$K$ & UKIRT/WFCAM & 22.0 \\
		$[3.6]$ & \emph{Spitzer}/IRAC & 22.2 \\
		$[4.5]$ & \emph{Spitzer}/IRAC & 21.9 \\
        \end{tabular}
	\begin{flushleft}
$a$: 5$\sigma$ point source completeness limit
\end{flushleft}
\label{tab:imaging}
\end{table}

\subsection{Optical Spectroscopy}
\label{optspec}

Imaging in the $r^{\prime}i^{\prime}z^{\prime}$ from LFC and $R_{C} I^{+} Z^{+}$ from Suprime-Cam were used to select spectroscopic targets
in the SG0023 and RXJ1716 fields, respectively, following the methods outlined in \citet{lub09}. Briefly, the two unique colours provided by
the three observed bands were used to prioritize spectroscopic targets, with the highest priority targets corresponding to galaxies with colours 
closest to the expected colours of quiescent galaxies at $z\sim0.8$ (see Table 2 of \citealt{lub09}). Objects with colours with progressively larger deviations
from the colour range which defined the highest priority targets were assigned progressively lower priorities. While galaxies were 
prioritized in such a way, due to the relative scarcity of the highest priority objects on the sky, the vast majority of spectroscopic targets in both fields 
($\sim$80\%) were objects with colours which deviated from those expected from $z\sim0.8$ quiescent galaxies. 
Observed-frame optical spectroscopy was 
performed with the DEep Imaging and Multi-Object Spectrometer (DEIMOS; \citealt{fab03}) at the Naysmith focus of the Keck \ion{}{II} telescope. All 
DEIMOS observations were performed using the 1200 l mm$^{-1}$ grating with slitmasks employing 1$\arcsec$ wide slits and the grism tilted 
to a $\lambda_{c}$ between 7500-7800\AA. This setup resulted a plate scale of 0.33\AA\ pix$^{-1}$, an $R\sim$5000 ($\lambda$/$\theta_{FWHM}$, 
where $\theta_{FWHM}$ is the full-width half-maximum resolution), and a wavelength coverage of $\Delta\lambda$$\sim$2600\AA. Spectroscopic 
targets were generally limited to $i^{\prime}/I^{+}<24.5$ with a tail extending to $i^{\prime}/I^{+}\sim25.5$. 

In the SG0023 field nine DEIMOS masks were observed between 
September 2005 and September 2010 under photometric conditions with seeing that ranged from 0.45$-$0.81$\arcsec$. Integration times per mask 
ranged from 5700s and 9400s, with $\tau_{int}$ scaled to roughly achieve a uniform distribution of continuum S/N per resolution element across all 
masks independent of both conditions (in this case seeing only) and the $i^{\prime}$ distribution of target objects. Data were reduced using a 
modified version of the Deep Evolutionary Extragalactic Probe 
2 (DEEP2; \citealt{davis03, new13}) \texttt{spec2d} pipeline. All objects, those targeted and those which serendipitously fell in a slit, were visually 
inspected and assigned a spectroscopic redshift (hereafter $z_{spec}$) and a redshift quality code ($Q$) 
in the \texttt{zspec} environment (see \citealt{new13}). A total of 1081 unique objects were targeted and/or detected, which resulted in 943 high-$Q$ $z_{spec}$
measurements ($Q$=-1,3,4, see \citealt{gal08,new13} for the meaning of these values) of which 213 are in the adopted redshift range of the supergroup,
$0.820\le z\le0.855$. 

In the RXJ1716 field six DEIMOS masks were observed between September 2010 to May 2015 with the grating tilted to $\lambda_c=7800$\AA\ for all masks. Exposure times ranged from 5400s 
to 9000s under seeing that ranged from 0.54$-$0.83$\arcsec$ and conditions that ranged from light cirrus to photometric. A total of 828 unique objects 
were targeted and/or detected, which resulted in 571 high-$Q$ $z_{spec}$ measurements of which 144 are in the adopted cluster redshift range 
$0.798\le z\le0.826$. These observations are sufficiently deep to determine continuum redshifts consistently to $I^{+}\le23.5-24$ 
or roughly 0.4$L^{\ast}$ at $z\sim0.8$\footnote{L$^{\ast}$ is adopted from \citet{depropris13} for $z=0.6$ cluster galaxies and translated to the redshift 
and filter of interest using EZGal, http://www.baryons.org/ezgal/} with secure redshifts based on emission line features obtainable to the limiting 
magnitude of the DEIMOS survey ($I^{+}\sim25.5$). 

Equivalent widths ($EW$s) measurements of the [OII] $\lambda$3726,3279\AA\ doublet and the H$\delta$ $\lambda$4101\AA\ Balmer series line were performed following
the bandpass method described in \citet{lem10} on all high-$Q$ extragalactic spectra where the wavelength coverage allowed for the possibility of the
features to be present. All measurements were visually inspected and bandpasses were tweaked when obvious reduction artifacts were present.  
These measurements would serve as the basis for the Multi-Object Spectrometer For Infra-Red Exploration (MOSFIRE; \citealt{mclean12})
campaign that followed. 
 
\subsection{Near-Infrared Spectroscopy} 
\label{NIRspec}

\begin{figure*}
\includegraphics[clip,angle=0,width=0.49\hsize]{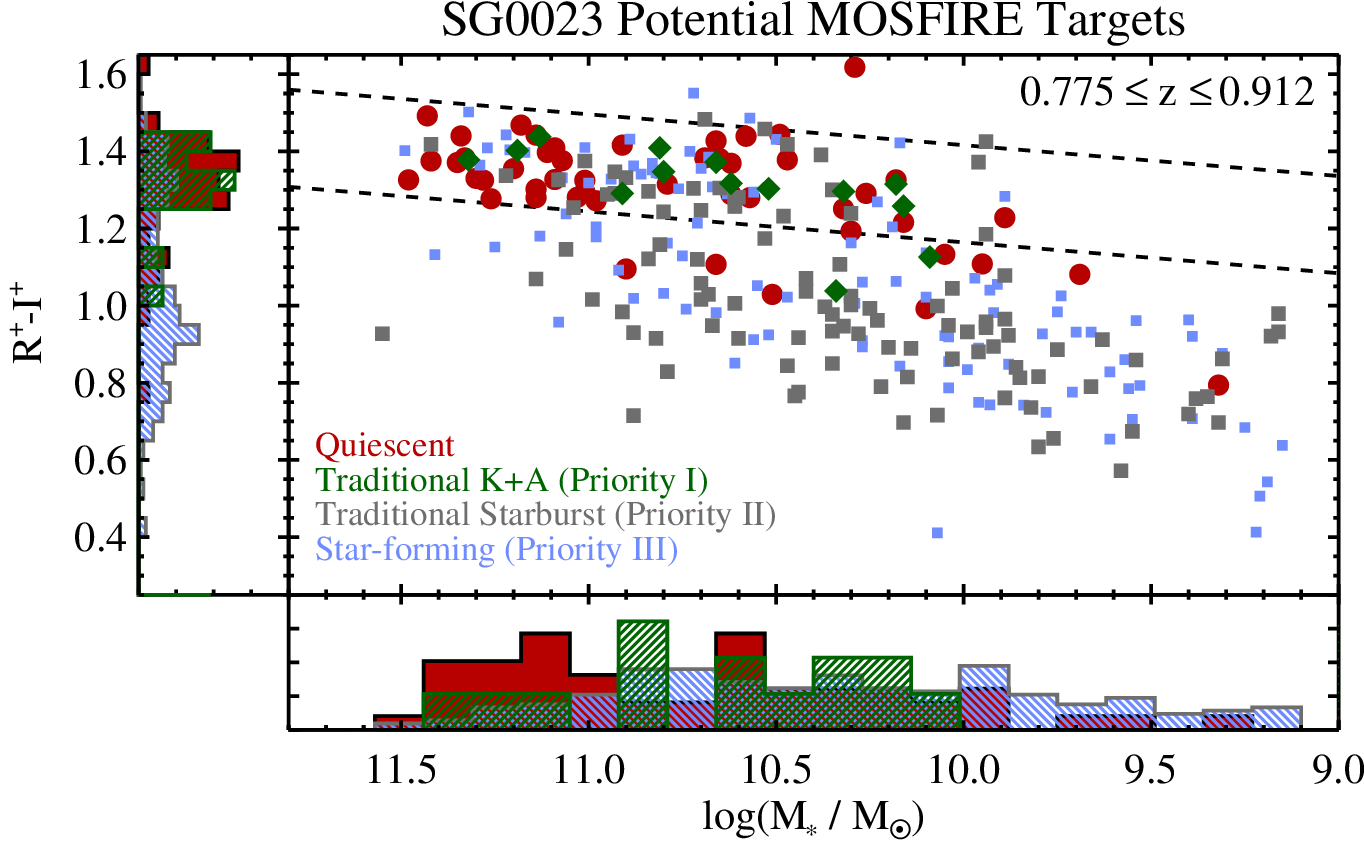}
\includegraphics[clip,angle=0,width=0.49\hsize]{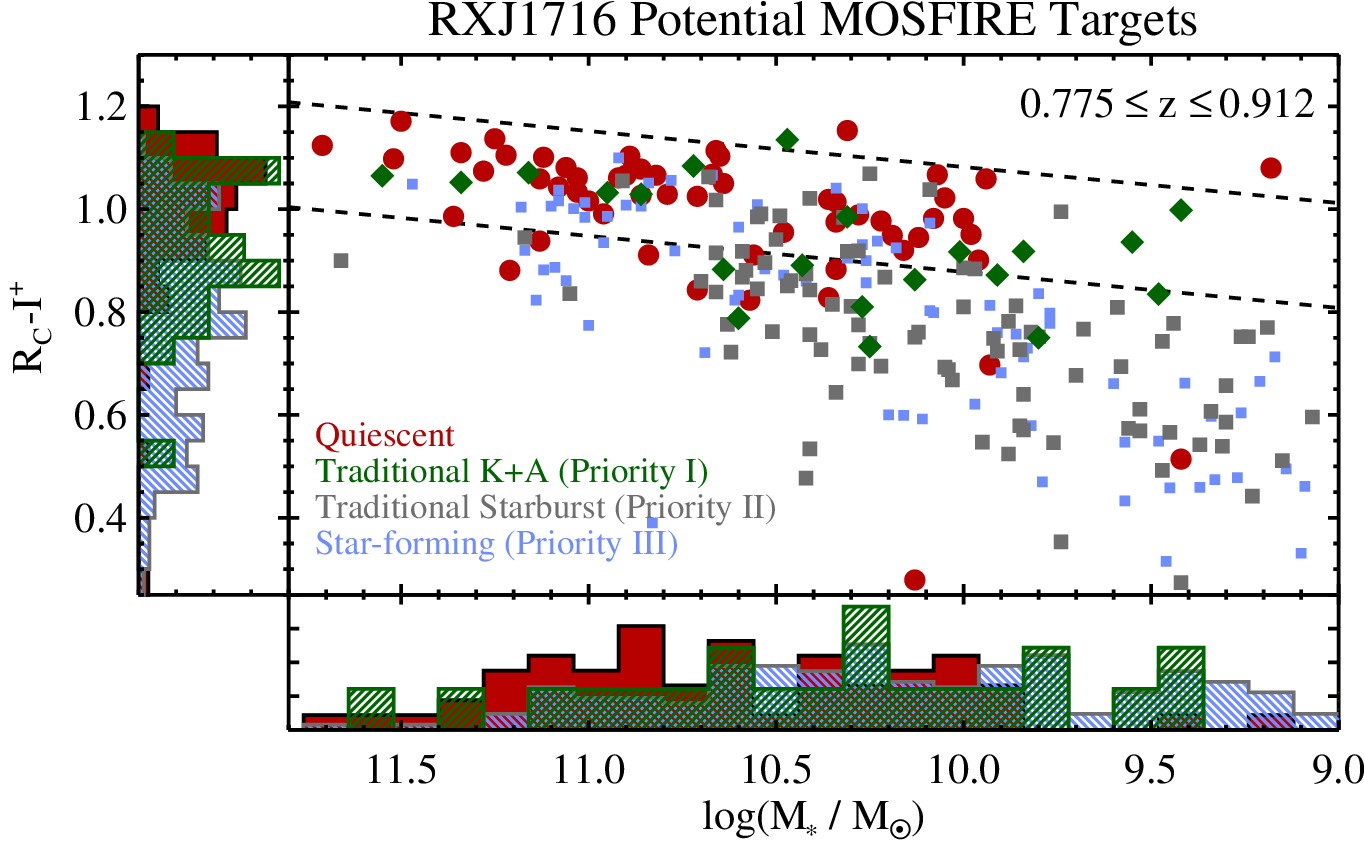}
\caption{\emph{Left}: Observed-frame $R^{+}-I^{+}$ colour-stellar mass (CSMD) for potential MOSFIRE target galaxies in the SG0023 field in the lower of the two target
redshift ranges ($0.775<z<0.912$). Only this redshift range is plotted as this is the redshift range adopted for all subsequent analysis presented in
this study. All galaxies shown in this plot have a secure spectroscopic redshift and measurements of EW([OII]) and EW(H$\delta$). Galaxies are classed
into quiescent (red circles), K+A (green diamonds), starburst (grey squares), and star-forming (light blue squares) galaxies using traditional
methods based on the strength of the [OII] and H$\delta$ features. Dashed lines indicate the bounds of the red sequence. Area normalized
histograms are plotted for quiescent (red filled), K+A (green hatched), and starburst+star-forming (grey/light blue hatched) galaxies for
both stellar mass and colour. \emph{Right:} Similar to the left panel except the population plotted is that of the RXJ1716 field and the $R^{+}$ filter
has been exchanged for the $R_{C}$ filter. Galaxies in an identical redshift range as those in the left panel are shown. In both SG0023 and RXJ1716 galaxies
classified as quiescent are generally redder and more massive in
their stellar content than any of the other spectal classes. K+A galaxies appear to be intermediate to the quiescent and star-forming/starbursting classes
both in terms of stellar mass and observed-frame colour. }
    \label{fig:priorities}
\end{figure*}

The DEIMOS $EW$ measurements for galaxies within the redshift ranges 0.775$<$$z$$<$0.912 and 0.975$<$$z$$<$1.020 were used to select targets for follow-up
NIR $J$-band spectroscopy with MOSFIRE. These redshift ranges were chosen so that [OII] and H$\delta$ are almost certain to fell within the wavelength 
coverage of DEIMOS for all slits\footnote{Variations of up to $\sim\pm$120\AA\ in $\lambda_c$ can occur depending on the placement of slit on the slitmask
along the direction parallel with the dispersion direction meaning, in some cases, we did not have the coverage to detect one of the two features.} and 
that both the H$\alpha$ $\lambda$6563\AA\ and the [NII] $\lambda$6583\AA\ feature fall within 
the wavelength coverage of the MOSFIRE $J$-band spectroscopy (1.150$\le \lambda \le $1.350$\mu m$) away from the strongest OH features in the $J$-band sky 
(1.260$\la\lambda\la$1.295$\mu$m). From these measurements, three main classes of MOSFIRE targets were formed. In order of priority these were
\textbf{I)} K+A [$EW$([OII])$>$-3\AA, $EW$(H$\delta$)>4\AA\ and, visually, the absence of H$\beta$ emission when the spectral coverage allowed for it], 
\textbf{II)} starbursts [$EW$([OII])$<$-3\AA, $EW$(H$\delta$)>4\AA], and \textbf{III)} star-forming galaxies [$EW$([OII])$<$-3\AA, $EW$(H$\delta$)<4\AA]. Quiescent galaxies 
[$EW$([OII])$>$-3\AA, $EW$(H$\delta$)<4\AA] were generally avoided as MOSFIRE
targets. 

\begin{table*}
        \centering
        \caption{MOSFIRE Targets in the SG0023 and RXJ1716 Fields}
        \label{tab:MOSFIREtargets}
        \begin{tabular}{lllll} 
                \hline
                Sample$^{a,b}$ & $N_{DEIMOS}$ & $N_{targeted}$ & $N_{KAIROS}$ & $N_{\rm{K+A-H}\alpha}$ \\
                \hline
                \textbf{Priority I:} Traditional K+A & 40 (29) & 20 (19) & --$^{c}$ & 15 (14) \\
                \textbf{Priority II:} Starburst/KAIROS & 188 (118) & 69 (60) & 15 (12) & --$^{c}$ \\
                \textbf{Priority III:} Star Forming & 178 (119) & 32 (31) & --$^{c}$ & --$^{c}$ \\
                \textbf{Quiescent} & 113 (82) & 15 (14) & --$^{c}$ & --$^{c}$ \\
                \hline
        \end{tabular}
\begin{flushleft}
a: Numbers given are for SG0023 and RXJ1716 combined and for galaxies in the redshift range $0.775\le z \le 0.912$ (see \S\ref{NIRspec})
b: Numbers in parentheses refer to large scale structure members galaxies only
c: Not possible by definition
\end{flushleft}
\end{table*}

Such a scheme is largely consistent with those employed by other K+A studies at $z\sim1$ (see the review of the logic behind a similar classification scheme
in \citealt{pog09}). Additionally, this metric of classification shows a high degree of congruence with classification which uses the 
observed-frame $R_{C}-I^{+}$ and $R^{+}-I^{+}$ colours. For example, in RXJ1716 and SG0023, spectroscopically-classified quiescent galaxies exhibit, by far, 
the reddest median colours 
(1.05 and 1.33, respectively), are the most massive in terms of their stellar content ($\langle\log{(\mathcal{M}_{\ast}/\mathcal{M}_{\odot})}\rangle$=10.67 and 10.79, respectively),
and have the highest incidence (87\% and 80\%, respectively) of galaxies with colours consistent with the cluster/group red sequences 
as measured by the methods defined in \citet{lem12} (see Figure \ref{fig:priorities}). The $EW$([OII]) threshold adopted here was set by fitting a half 
Gaussian to the positive portion of the distribution of $EW$([OII]) for all galaxies in the RXJ1716 and SG0023 field where this quantity was measured (i.e., where [OII] 
is observed to be in absorption). As such values are unphysical, this part of 
the distribution results purely from noise (instrumental or astrophysical). The threshold of $EW$([OII])=-3\AA\ was chosen as it is 1.5$\sigma$ of 
the resultant half-Gaussian fit, which, with 144 such galaxies in our sample, implies that $\le$10 galaxies with spuriously detected [OII] emission contaminate 
our star-forming and starbursting samples (i.e., $>$1.5$\sigma$ on the negative side of the distribution). 	
The true number is likely less as each [OII] feature is visually inspected in the 1d and 2d DEIMOS spectra. The H$\delta$ cut adopted here for K+A populations
is a compromise between those chosen by various other studies \citep{zabludoff96, balogh99, dressler99, bartho01, leborgne06, pog09, swinbank12} and, in conjunction 
with the cut on [OII], ensures, at least to the ability of the DEIMOS data to discriminate, that the last
major star-formation event ended within $\la$1.5 Gyr irrespective of star formation history (SFH) (see, e.g., discussion in \citealt{pog99}). While 
this may seem like an excessively long timescale constraint to use to select a transition population, this timescale is less ($\le$ 1 Gyr) both in synthetic models 
(e.g., \citealt{lem12}) and in hydrodynamic simulations 
(e.g., \citealt{snyder11}) when only considering SFHs which include, at some point during the history of a galaxy, a 
starburst. Such starbursts need not be strong, it is sufficient that they form $\sim$5-10\% of the stellar mass of the galaxy in the event, and, indeed, 
it is suggested that a starburst of at least this modest level is necessary to form K+A features (e.g., \citealt{balogh05, wild09, melnick13}, though see, e.g.,
\citealt{newberry90, pog99, leborgne06, yan09, falkenberg09} for an alternate view). However, a similarly tight age constraint also applies if K+A features are produced 
through rapidly quenched normal (or bursty) star formation \citep{yan09} making the distinction largely superfluous for this study. We will show 
later (see \S\ref{KAevolution}) that this cut selects K+A galaxies that have, on average, ended their star formation within $\la$1 Gyr. There were 194 (24/93/77) and 
257 (23/123/111) potential MOSFIRE targets of these three main classes in RXJ1716 and SG0023, respectively, bounded by the quoted redshift ranges, where values 
inside the parentheses indicate the number of priority \textbf{I}, \textbf{II}, and \textbf{III} targets, respectively.  

Three masks in RXJ1716 and three in SG0023 were observed on August 15th, 2014 with MOSFIRE under photometric conditions with seeing ranging from 
0.5$-$0.9$\arcsec$. Integration times for all masks was 7$\times$4$\times$120s (3360s), with an ABBA nod pattern employed for each block of 4$\times$120s 
exposures nod pattern. Slit widths were set to 0.7$\arcsec$ and the plate scale to 1.3\AA\ pix$^{-1}$ resulting in an $R\sim3500$. 
The \emph{python}-based \texttt{MOSFIRE DRP}\footnote{http://www2.keck.hawaii.edu/inst/mosfire/drp.html} was used to reduce the raw frames. This pipeline
provides dark-subtracted, flat-fielded, rectified, wavelength-calibrated, background-subtracted two-dimensional flux density and variance arrays for every 
slit. Each two-dimensional flux density spectrum output by the pipeline was collapsed along the dispersion axis and a Gaussian iteratively 
fit to the resulting collapsed profile with a mean location beginning with the expected spatial location of the targeted galaxy. The final parameters 
of the Gaussian fit, mean and $\pm$1.5$\sigma$, set the limits on the boxcar extraction used to generate the one-dimensional flux density and 
noise spectrum. In cases where the continuum was marginally detected in MOSFIRE or only emission lines were present, the dispersion axis would be 
collapsed over a limited wavelength range and the fit was done by hand. During this process it was noticed that the error arrays output by the 
pipeline (i.e., the square root of the variance arrays) appeared to be discordant with the RMS of the observed spectra away from bright OH lines 
in that the former appeared to overestimate the true noise present in the data\footnote{This was a known issue in the earlier version of the pipeline
employed for our data. According to the official documentation, there no longer remains issues with the output variance arrays for the newest version 
of the pipeline.}. In an attempt to rectify this discrepancy, all error spectra were 
scaled by the ratio of the RMS to the median error for that spectrum (roughly a factor of four in all cases). Absolute spectrophotometric calibration 
was performed by observations of standards throughout the night, though, in practice, this step was largely superfluous as we focus in this study
almost exclusively on relative quantities. 

In total, 78 galaxies were observed with MOSFIRE in each of the two fields to an average 3$\sigma$ line limit of 
$f_{line}\ge7\times10^{-18}$ ergs s$^{-1}$ cm$^{-2}$ including a slit loss correction appropriate for the average targeted galaxy, equivalent, 
for H$\alpha$, to an unobscured star formation rate of $\sim0.1$ M$_{\odot}$ yr$^{-1}$ at $z=0.83$ adopting the conversion of \citet{kenn98} 
and scaling to a \citet{chab03} initial mass function (IMF). 
Of these 156 galaxies, 61 (12/36/13) and 65 (8/37/20) in RXJ1716 and SG0023, respectively, were comprised of the three main classes described above and fell within the 
redshift ranges adopted earlier in this section (0.775$<$$z$$<$0.912 and 0.975$<$$z$$<$1.020). The remaining 17 and 13 MOSFIRE targets in RXJ1716
and SG0023, respectively, were at different redshifts, were spectrally classed as quiescent, or were $z_{phot}$ members with no corresponding DEIMOS spectra and thus
excluded from the remainder of our analysis. The above numbers for the three main classes result in a 42.6\%, 32.0\%, and 17.5\% MOSFIRE 
targeting completeness for K+As, starbursts, and star-forming galaxies across the redshift ranges of interest. 
As only a small number of priority \textbf{I}, \textbf{II}, and \textbf{III} galaxies were observed across the two fields in the higher redshift bin 
(i.e., four priority \textbf{II} targets), in this paper we will focus 
exclusively on the lower redshift bin for which the MOSFIRE targeting
completeness was higher, 50.0\%, 36.7\%, and 18.0\% for priority \textbf{I}, \textbf{II}, and \textbf{III} galaxies, respectively.
This redshift range contains the vast majority ($\sim95$\%) of galaxies targeted with MOSFIRE across the two fields, is roughly centreed at the redshift range of the two LSSs, 
and retains a large number of both DEIMOS- and MOSFIRE-targeted galaxies while not appreciably compromising completeness. While some portion of the galaxies in this
redshift range are found in the coeval field, a majority ($\sim70$\%) are members of the two LSSs.
Line flux measurements were made following the bandpass method of \citet{lem10} 
with 3$\sigma$ upper limits imposed on any line not significantly detected. Because the MOSFIRE observations did not generally go to sufficient depth to 
significantly detect stellar continua, the H$\alpha$ emission line flux (or limits thereon) calculated after applying a slit loss correction was used in conjunction with the 
$J$-band magnitudes, transformed to the rest-frame, to calculate EW(H$\alpha$).

\section{Methods}
\label{methods}

\begin{figure*}
\includegraphics[angle=0,width=0.487\hsize]{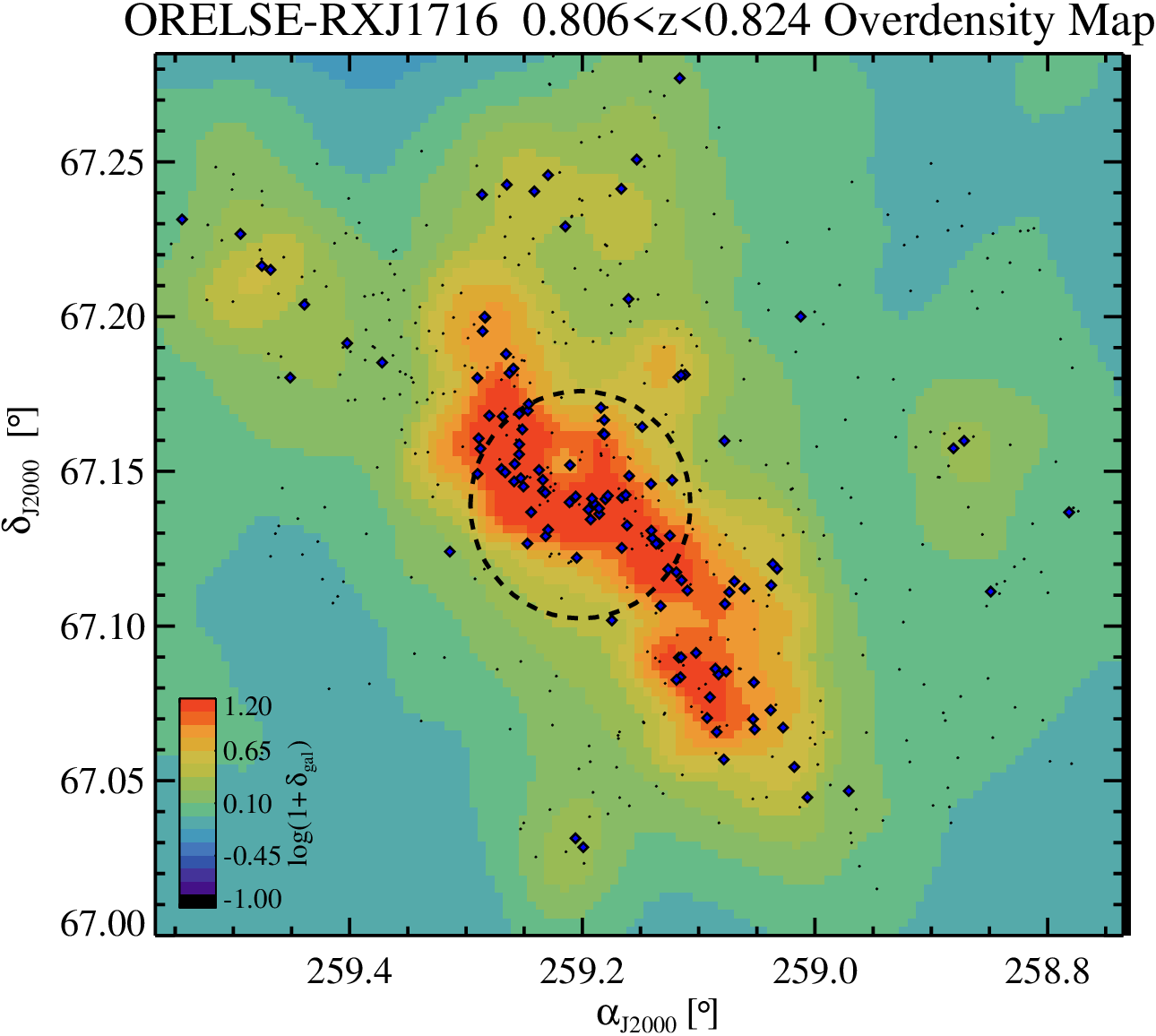}
\includegraphics[angle=0,width=0.474\hsize]{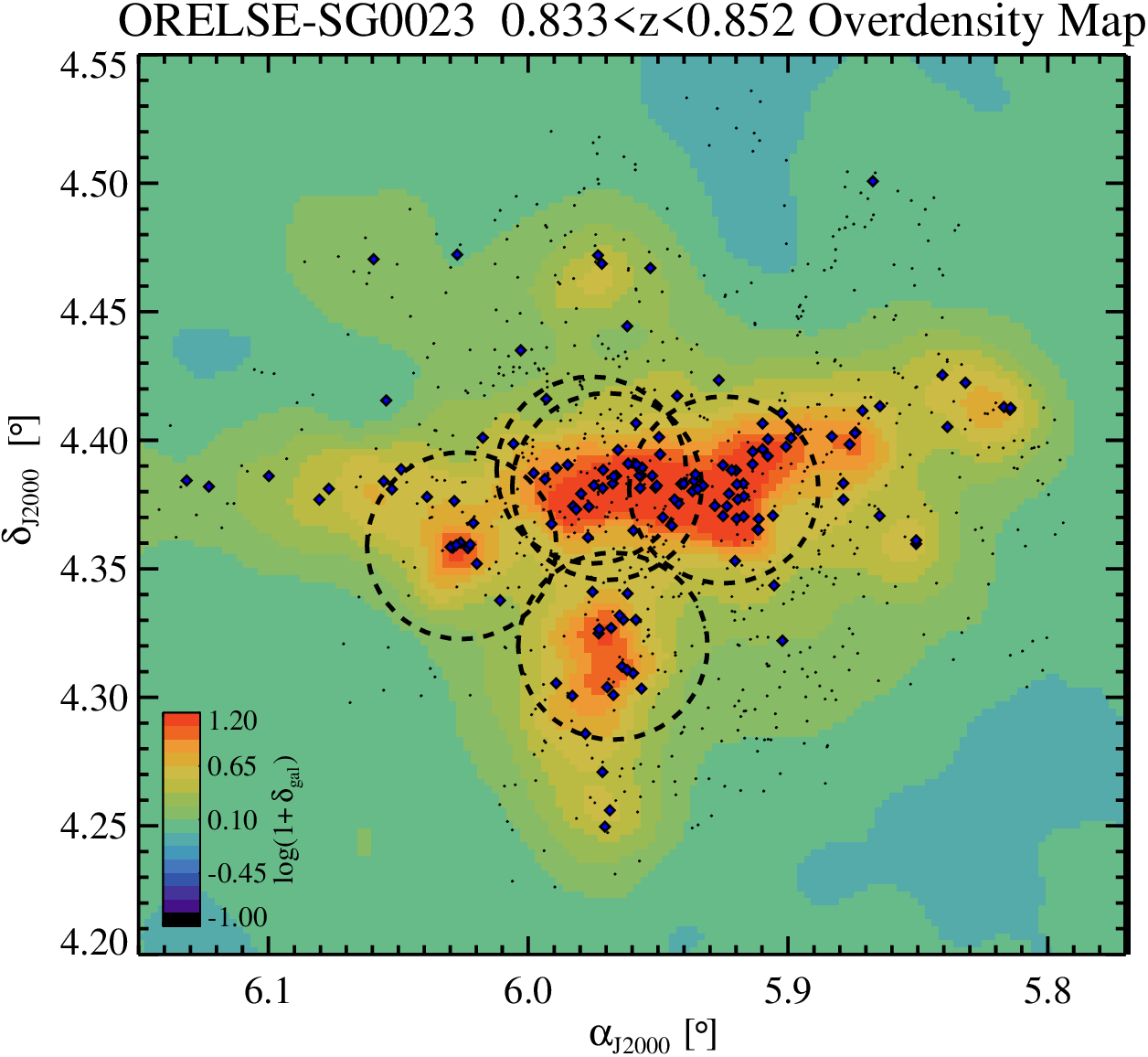}

\caption{Voronoi Monte-Carlo overdensity maps (see \S\ref{voronoi}) of one redshift slice containing part (along the line of sight) of the RXJ1716 cluster (\emph{left}) 
and the SG0023 supergroup (\emph{right}). Redshift slices are bounded by $\pm$1500 km s$^{-1}$ from the central redshift of each slice. 
The scale bar shown in the bottom left of each panel gives maps colours to their associated level of overdensity. Blue diamonds show the sky distribution
of galaxies with high-quality spectroscopic redshifts that fall within the redshift range of each LSS (see \S\ref{optspec}), while small black points show that 
of all objects with high-quality spectroscopic redshifts. Dashed circles of radius $R_{proj}=1 $ $h^{-1}_{70}$ Mpc originating at the $I^{+}/i^{\prime}$
luminosity-weighted centres of the RXJ1716 cluster and five SG0023 member groups are overplotted. Only those objects with $18\le I^{+} \le 24.5$ 
were used to generate the maps. Both maps were smoothed by a Gaussian kernel with a FWHM of 4.5 pixels ($\sim$300 $h^{-1}_{70}$ kpc) for visualization purposes.
Actual overdensity measurements are made on unsmoothed maps.} 
    \label{fig:voronoi}
\end{figure*}

\subsection{Broadband Spectral Energy Distribution Fitting}
\label{SEDfitting}

Spectral energy distribution (SED) fitting was performed on the imaging in three stages. First, aperture magnitudes were input to the code 
Easy and Accurate $z_{phot}$ from Yale (\texttt{EAZY}, \citealt{brammer08}) for the purposes of estimating photometric redshifts (hereafter $z_{phot}$).		
For each object, the $z_{phot}$ is estimated from from a probability distribution function (PDF) generated by minimizing the $\chi^2$ of the observed flux 
densities and a set of seven basis templates at each redshift. These seven basis templates were generated from a large library of P\'{E}GASE models (\citealt{fioc97}, 
see also \citealt{grazian06}) following the methodology given in \S2.2 of \citet{brammer08}, though several changes in the input models were made since this time 
including the implementation of emission lines \citep{brammer11}. The parameter ``$z_{\rm{peak}}$'' was adopted as the measure of $z_{phot}$, with the uncertainties 
on this parameter estimated from the PDF of each source. Also at this stage, a second round of fitting was performed separately  
for which we exclusively employed stellar templates drawn from the \citet{pickles98} library. The reduced $\chi^{2}$ values between the two set of fits were compared to separate, 
in conjunction with other criteria, stars from galaxies. A use flag was generated for all objects in the photometric catalogue which allowed for the removal
of those objects that were likely stellar, had a S/N$<$3 in the detection band, were covered in less than five of the broadband images, had
significant saturation ($>20$\% of the segmentation map pixels), or were poorly fit in the galaxy portion of the SED fitting 
(see Straatman et al.\ \emph{in prep.}). Such objects totaled $\sim4$ and $\sim9$\% of the photometric objects in the SG0023 and RXJ1716 fields, respectively,
over the area bounded by the DEIMOS coverage. 
These objects were removed from our analysis.  

The precision and accuracy of the photometric redshifts were estimated from fitting
a Gaussian to the distribution of $(z_{spec}-z_{phot})/(1+z_{spec})\def\Delta z/(1+z)$ for those galaxies with high-$Q$ $z_{spec}$ measurements in the range 
0.5$\le z \le$1.2 and was found to be $\sigma_{\Delta z/(1+z)}=0.025$ with a catastrophic outlier rate ($\Delta z/(1+z)>0.15$) of $\eta\sim4$\% for both
fields to a limit of $I^{+}\le24.5$. At this point, a slight systematic offset from zero ($\Delta z/(1+z_{phot})\sim0.01$) was noticed for both fields. The value 
of this offset, multiplied by (1+$z_{phot}$), was applied to all raw $z_{phot}$ values. The spectroscopic sample was also used during the initial fitting
to iteratively correct the photometric zero points of each filter following the methodology of \citet{brammer11}. These corrections assumes that the properties and 
statistics of the spectroscopic sample, a sample which constitutes only $1.4-2.8$\% of the usable photometric objects in the region bounded by the DEIMOS coverage, can 
be applied to the underlying photometric sample. For the single facet of this analysis where we rely at all on $z_{phot}$ measurements (\S\ref{voronoi}), we cut the photometric 
sample in a magnitude range matched to the spectroscopic sample such that this assumption likely holds. 

For the second stage of the SED fitting process, the \texttt{EAZY} code is again run, but this time setting either the high-$Q$ $z_{spec}$, when available, or the $z_{phot}$ 
from \texttt{EAZY} as a redshift prior. Identical models are employed at this stage as were employed at the first stage. Identical photometric zero points are 
also applied. 
At the conclusion of this fitting, rest-frame magnitudes are directly estimated from the best-fit template following the methodology of \citet{brammer11}. These 
extinction-uncorrected rest-frame aperture magnitudes are corrected to total magnitudes by the scaling method described in \S\ref{phot}. 

For the final stage of the fitting process, 
the code Fitting and Assessment of Synthetic Templates (\texttt{FAST}, \citealt{kriek09}) was used to perform SED fitting on the aperture-corrected magnitudes 
using the same redshift priors as were used in the second stage, again with identical photometric zero points applied. Exponentially declining stellar population 
synthesis (SPS) \citealt{bc03} models (hereafter BC03) were adopted with 
a \citealt{chab03} IMF and a \citet{calz00} extinction law. The ranges of allowed parameters are comparable to those of \citet{AA16}.
For each fit, the maximum age bounded by the age of the universe at the $z_{spec}$ or $z_{phot}$ redshift. Stellar-phase metallicity was fixed 
to $Z=Z_{\odot}$. For this paper only stellar mass derived from this fitting are used. Each parameter is estimated from the best-fit value and uncertainties are derived through 
100 realizations of re-fitting to an SED with photometry that has been tweaked by a Gaussian random multiple of its photometric errors for each band (as in, e.g., \citealt{Rusl13}). 

\subsection{Local Overdensity}
\label{voronoi}

For certain aspects of this analysis we will focus on the environmental distribution of various types of galaxies. While environmentally-driven evolution within 
LSSs is certainly a complex function of redshift, smaller-scale galaxy density, and the properties of the LSS in which a galaxy resides, we choose in this 
paper to focus almost exclusively on local overdensity as a proxy for environment. In future studies that will include all ORELSE fields this analysis will 
extend to separately binning the distribution of different classes of galaxies varying local density, halo mass, dynamical and spatial distribution within the LSS, 
and redshift while holding the other quantities fixed. In order to estimate the local environment of the galaxies in our sample, we employ 
the Voronoi Monte-Carlo technique described in \citep{lem17}. The method that we employ broadly follows that employed in \citet{darvish15}, in
which it was found, through comparisons to simulated density fields, that the ``weighted Voronoi tessellation estimator" computed in that study matched or exceeded the 
accuracy and precision of all other metrics of density estimation. The one metric with comparable performance to the Voronoi approach, weighted adaptive kernel estimation, 
is sensitive to both the form and size of the kernel and, generally, employs a spatially symmetric kernel (along the transverse dimensions) which is not ideal for the complex LSSs 
studied here. Further quantitative measures of the accuracy and precision of our implementation derived from (over)density field reconstructions on mock catalogs will be described in
Tomczak et al.\ (\emph{in prep}) and Lemaux et al.\ (\emph{in prep}). We discuss the version of Voronoi Monte-Carlo technique employed here briefly. 

For each Monte-Carlo realization, Gaussian sampling is performed for each 
object without a high quality $z_{spec}$ (but with a good use flag, see \S\ref{SEDfitting}). The sampled value, in units of $\sigma$, is then multiplied by either 
the effective $1\sigma$ lower or upper uncertainty on $z_{phot}$ for that object depending on which side of the peak of the Gaussian sample fell. For each object, this 
value is then either subtracted from or added to its original $z_{phot}$ to create a new 
$z_{phot,\ MC_{i}}$ for that realization. These objects along with all galaxies with high quality extragalactic $z_{spec}$ are sliced into 85  
redshift bins running from $0.55 \le z\le1.4$, and Voronoi tessellation is performed on each realization of redshift slice on all objects that fall within that redshift bin. For 
each realization of each slice, a grid of 75$\times$75 kpc is created to sample the underlying local density distribution. The local density 
at each grid value for each realization and slice is set equal to the inverse of the Voronoi cell area (multiplied by $D_{A}^2$) of the cell that encloses the 
central point of that grid. Final local densities, $\Sigma_{VMC}$, for each grid point in each redshift slice are then computed by median combining the values
of 100 realizations of the Voronoi maps for that slice. The local overdensity value for each grid point is then computed as 
$\log(1+\delta_{gal}) \equiv \log(1+ (\Sigma_{VMC}-\tilde{\Sigma}_{VMC})/\tilde{\Sigma}_{VMC})$, where $\tilde{\Sigma}_{VMC}$ is the median $\Sigma_{VMC}$
for all grid points over which the map is defined (i.e., where there is coverage in a sufficient number of bands). By adopting local overdensity rather 
than local density as a proxy of evironment, we largely mitigate issues of sample selection and differential bias as a function of redshift.

Two main changes were made with respect to the method detailed in \citep{lem17} to adapt this technique to the ORELSE data. The first is that 
redshift slices were defined not to be constant in line of sight proper distance but rather in velocity space encompassing
$\pm$1500 km s$^{-1}$ from the central redshift of each bin. This velocity width was imposed to roughly match the average $\pm$3$\sigma_{v}$ of the constituent 
sub-structures of the LSSs presented in this paper and ranges from $\pm$1.3$\sigma_{v}$ to $\pm$14.2$\sigma_{v}$ for individual sub-structures (see Table \ref{tab:LSSs}). 
The second change made was to treat $z_{spec}$ measurements in a binary fashion in that galaxies with high-$Q$ $z_{spec}$ measurements always had 
their redshifts fixed to $z_{spec}$ rather than have some chance in each realization of being treated in the probabilistic manner described above. 
Voronoi Monte-Carlo maps were generated in half steps of 1500 km s$^{-1}$ with central redshift bins running from $0.55\le z \le 1.4$. For all Voronoi Monte-Carlo
maps, the photometric and spectroscopic catalogues were cut at $18\le I^{+} \le 24.5$, a magnitude range which encompasses nearly all 
high-$Q$ ORELSE objects. For the redshift slices used in this study (0.775$\le z \le$0.912, see \S\ref{NIRspec}), the median fraction of $z_{spec}$ to $z_{phot}$ 
objects for all realizations across the entire area over which the maps were defined ($\sim$0.25$\Box^{\circ}$) varies from 3-37\% across the two fields, with a 
similar distribution in each field. Figure \ref{fig:voronoi} shows example slices of the Voronoi Monte-Carlo maps centreed at the average 
systemic redshifts of SG0023 and RXJ1716. While the redshift slices shown in each panel of Figure \ref{fig:voronoi} do not span the entire redshift range of
each LSS, it can be seen that the estimated density field largely traces the LSS members and peaks near the central regions of the constituent cluster and
groups.

\section{Post-Starburst Galaxies in High-Redshift Large Scale Structures}
\label{PSB}

With these observations and measurements in place, we now begin a preliminary census of the true K+A population in RXJ1716 and SG0023. In this study we focus our attention on 
two main aspects of this census. First, we investigate the incompleteness and impurity of K+A populations selected purely by observed-frame optical 
spectroscopy (e.g., DEIMOS) at $z\sim1$ as revealed by our MOSFIRE observations and estimate the fraction of true K+A galaxies. Secondly, we 
investigate differences between the DEIMOS-selected K+A population and the true population of K+As in terms of their stacked broadband properties
and distribution across different environments. 

\subsection{Revealing the True Post-Starburst Population}
\label{revealKA}

Using the traditional scheme to select post-starburst galaxies at high redshift, strong Balmer features, proxied by $EW$(H$\delta$)$>$4\AA, and 
the absence of emission lines traditionally associated with ongoing star formation, proxied by $EW$([OII])$>$-3\AA\ (i.e., priority \textbf{I} galaxies, 
see \S\ref{NIRspec}), 
results in a K+A fraction of 7.7$\pm$1.2\% (40/519). This fraction is calculated for all galaxies in our sample in the redshift range 0.775$\le z \le$0.912
with $I^{+}\le24.5$ for which a reliable EW(H$\delta$) and EW([OII]) measurement could be made (see \S\ref{NIRspec}). The galaxies selected using this method will be referred to as 
``traditional K+As" throughout the remainder of the paper. The fraction of traditional K+As does not change significantly if the redshift range 
is restricted to that of the two LSSs (8.3$\pm$1.5\%, 29/348) and is broadly consistent with those found in other LSSs at similar redshifts 
\citep{tran03, tran07, pog09, muz12, pwu14}. The former point is likely a reflection
of the majority of the spectroscopic sample in our adopted redshift range being associated with the LSSs. This fraction also does not change significantly 
(7.7$\pm$1.3\%, 31/405) if we instead calculate it after imposing the stellar mass limit of the spectroscopic sample (see \S\ref{tarnobs}). In a nearly
equivalent exercise, we find the traditional K+A fraction also remains statistically unchanged if we impose a stricter $I^{+}$-band cut, e.g., 
imposing $I^{+}\le23.5$ results in a traditional K+A fraction of $6.9\pm1.2$\% (29/422).

As dicussed in \S\ref{NIRspec}, with MOSFIRE we have targeted two populations that have the potential to be true K+A galaxies, defined as those galaxies with no ongoing 
star-formation activity but which still fulfill our $EW$(H$\delta$)$>$4\AA\ requirement. The first are the traditional K+As for which we have no
evidence of ongoing star formation in our DEIMOS spectra (priority \textbf{I} galaxies). The second are the galaxies with K+A features that would have 
been classified as such but for the presence of formally significant [OII] emission (priority \textbf{II} galaxies). With the former we will attempt here to 
quantify the level of purity attained using the observed-frame optical selection, with the latter the level of completeness of this selection. 

\begin{figure*}
\includegraphics[clip,angle=0,width=0.49\hsize]{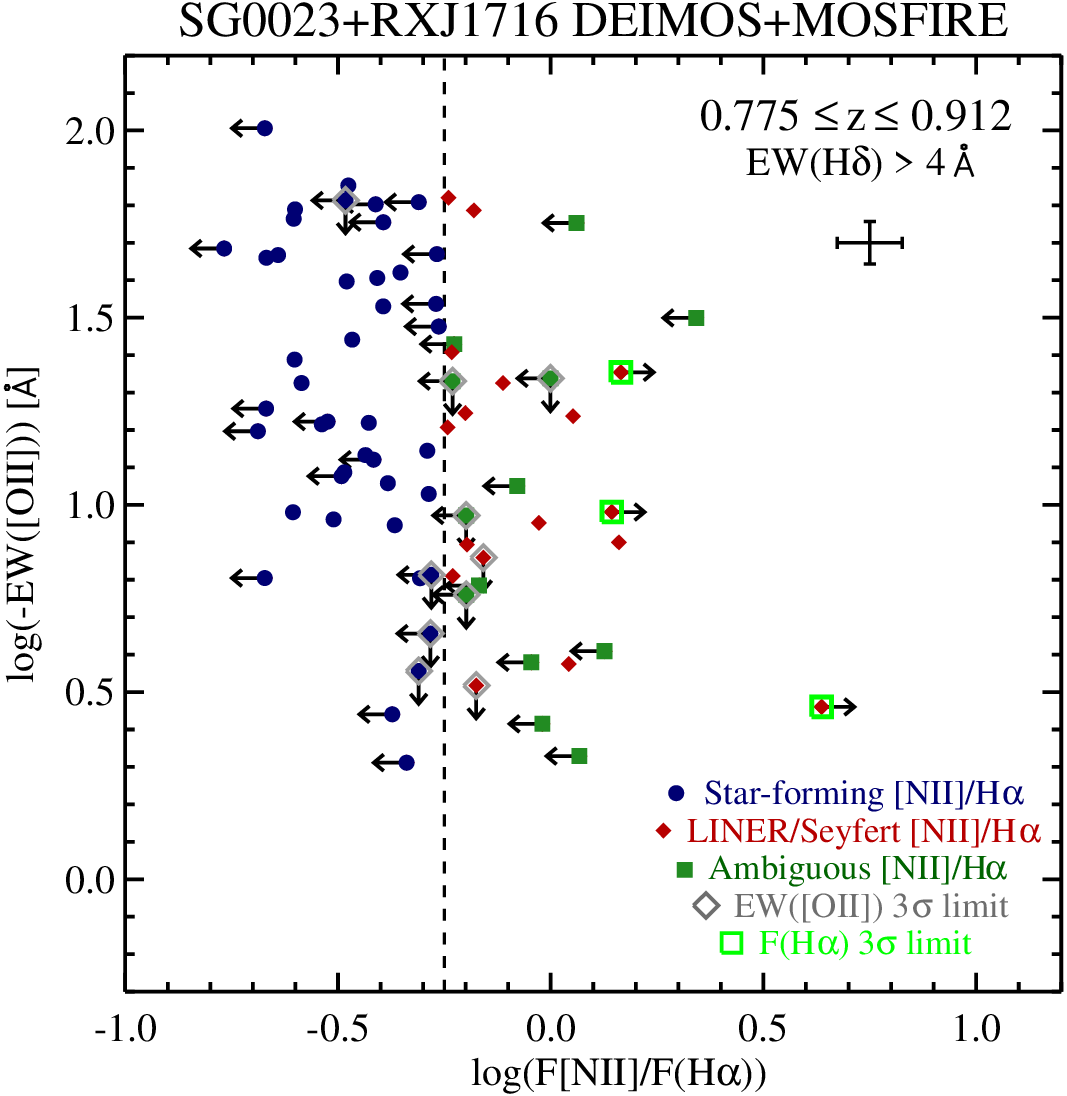}
\includegraphics[clip,angle=0,width=0.49\hsize]{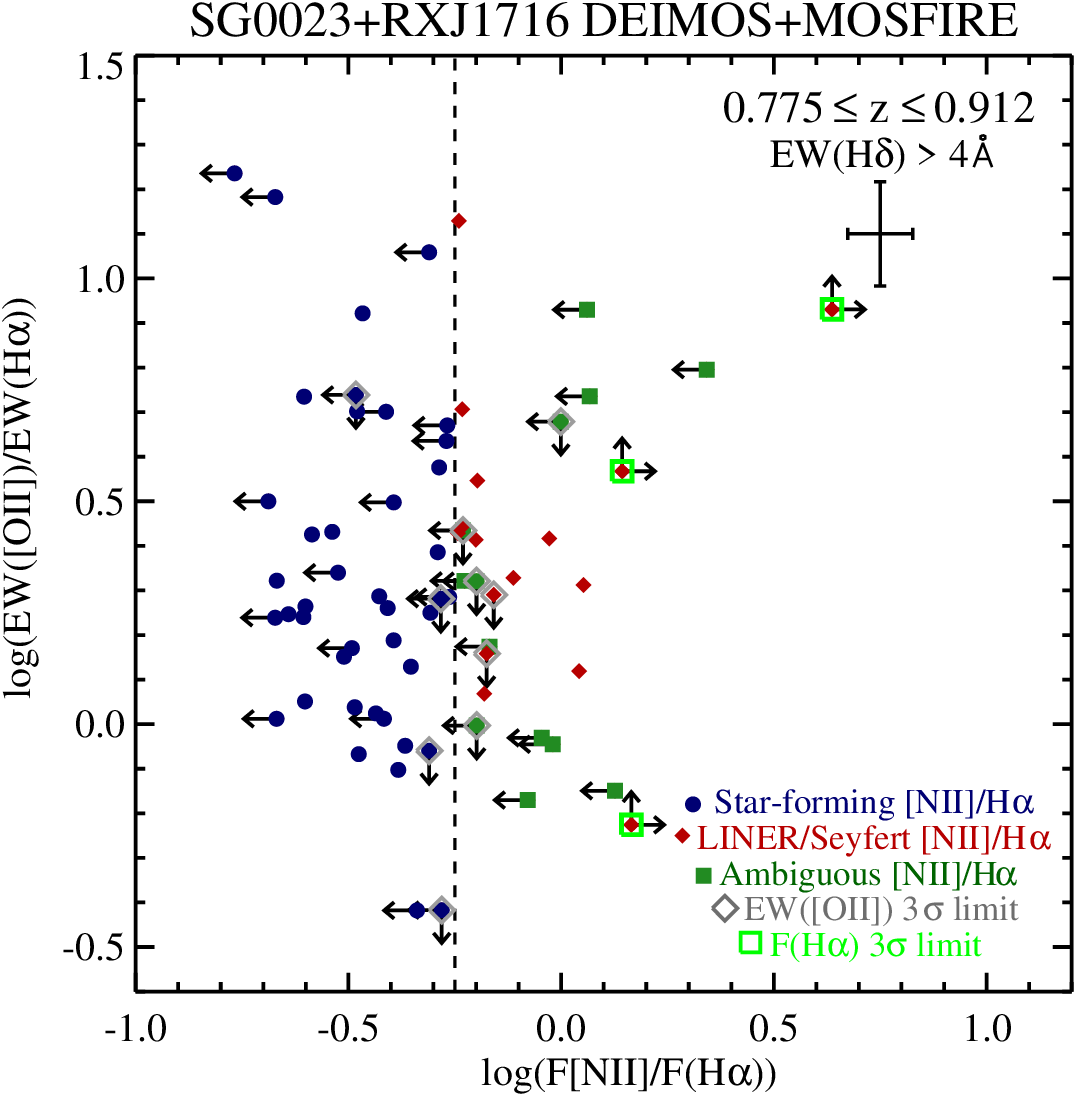}
\caption{\emph{Left}: Plot of (inverted) [OII] equivalent width (EW) vs. the emission line flux ratio of [NII]/H$\alpha$ for all MOSFIRE targets in the redshift range indicated 
with EW(H$\delta$)$>4$ \AA\ and for which at least H$\alpha$ \emph{or} [NII] were detected at a significant ($>$$3\sigma$) level. The dashed line demarcates
galaxies dominated by LINER/Seyfert emission ($\log([\rm{NII]}/\rm{H}\alpha)$$\ge$-0.25) from those galaxies consistent with having some residual ongoing star formation activity 
($\log([\rm{NII}]/\rm{H}\alpha)$$<$-0.25). All galaxies with EW([OII]>-3\AA\ (see \S\ref{NIRspec}) are plotted by adopting their 3$\sigma$ EW([OII]) limits and are shown with downward
facing arrows and are circumscribed grey diamonds. These galaxies comprise the subset of traditional K+A galaxies targeted by MOSFIRE where H$\alpha$ or [NII] was significantly detected.
Galaxies with a 3$\sigma$ limit on [NII] are shown with leftward facing arrows, those with a 3$\sigma$ limit on H$\alpha$ are shown with rightward facing arrows 
and are circumscribed by green boxes. Galaxies definitively classified as star-forming or LINER/Seyfert galaxies are shown as filled blue circles and red diamonds, 
respectively. Those galaxies whose limits on [NII]/H$\alpha$ were not sufficiently constraining are shown as darker green filled squares. The full
sample of KAIROS galaxies are visible on this plot as those red diamonds not circumscribed by grey diamonds. Typical 
(median) errors for individual measurements are shown in the upper right portion of the panel. \emph{Right:} Same as left, but with the EW ratio of [OII] and H$\alpha$ plotted
on the ordinate. The meaning of symbols, arrows, and lines are the same. Only those galaxies for which we measured significant detection in both [NII] and [OII] and/or [NII] and 
H$\alpha$ are shown. The generally lower values of EW([OII])/EW(H$\alpha$) exhibited by the KAIROS galaxies suggests that most have their emission powered by Seyfert or 
hybrid activity.} 
    \label{fig:spectraldiagnostic}
\end{figure*}

From the MOSFIRE observations of each galaxy we measured the flux ratio of the H$\alpha$ and [NII] emission features or placed limits thereon (see \S\ref{NIRspec}). 
This measure of (or limit on) the [NII]/H$\alpha$ is used in an attempt to determine the 
source of the emission for each galaxy. We classify those galaxies for which this ratio (or lower limit) is in excess of $\log(\rm{[NII]}/\rm{H}\alpha)$$\ge$-0.25 
as being \emph{dominated} by emission from a process other than star formation, i.e., the emission originates from either that of a Low-Ionization 
Nuclear Emission-line Region (LINER) or a Seyfert (hereafter LINER/Seyfert). This value was adopted as it excludes 100\% of galaxies within the Sloan 
Digital Sky Survey (SDSS) star-forming locus \citep{kauff03} as seen in the incarnation of the \citet{BPT81} (hereafter BPT) diagram which employs [NII]. This 
[NII]/H$\alpha$ ratio is also higher than that exhibited by most SDSS composite objects and includes nearly all objects within the regions populated by 
LINERs and Seyferts \citep{kew06}. While this classification scheme is calibrated for a galaxy sample at $z\sim0.1$, little to no evolution is 
observed or predicted in various versions of the BPT diagram up to $z\sim1$ \citep{kew13b,kew13a,jones15}. While systematically elevated 
[NII]/H$\alpha$ ratios have been observed for other star-forming populations observed with MOSFIRE (e.g., \citealt{sanders15}) 
such an offset may only apply to particular subsets of galaxies \citep{shapley15} and are observed in samples appearing 4 Gyr earlier in cosmic time 
($z\sim2$) when physical conditions in star-forming galaxies appear to be appreciably different than those at lower redshift (see \citealt{sanders15} and 
references therein). Such a cut also appears, from modeling, to 
be sufficient to select the vast majority of galaxies whose emission features are powered by both fast and slow shocks \citep{alatalo16}. Galaxies with an 
upper limit or a measured ratio below $\log(\rm{[NII]}/\rm{H}\alpha)$$<$-0.25 we classify here as star forming. While the emission lines of such galaxies can still
be partially powered by LINER/Seyfert emission, it is enough for our purposes that at least some of the emission can potentially come from star formation. In the left panel 
of Figure \ref{fig:spectraldiagnostic} we show EW([OII]) vs. $\log(\rm{[NII]}/\rm{H}\alpha)$ for all galaxies with EW(H$\delta$)$>$4\AA\
that were targeted with MOSFIRE for which we measured a significant detection for at least [NII] or H$\alpha$. In addition, we show the emission class of
the galaxies when a definitive classification could be made. Traditional K+As are shown as galaxies with downward-facing arrows. The remaining galaxies are 
potential K+As that have [OII] emission significantly detected in their DEIMOS spectra (i.e., priority \textbf{II} galaxies). In the right panel of Figure \ref{fig:spectraldiagnostic} 
we plot the EW ratio of [OII] and H$\alpha$ vs. $\log(\rm{[NII]}/\rm{H}\alpha)$ for a subset of these galaxies (see Figure \ref{fig:spectraldiagnostic} caption). 

Of the 20 traditional K+As targeted with MOSFIRE for which we could make meaningful measurements, five exhibited significant ($>3\sigma$) H$\alpha$ emission 
in tandem with [NII]/H$\alpha$ ratios (or upper
limits) consistent with originating from regions of ongoing star formation. Such galaxies are likely the product of heavy dust obscuration in which the [OII] 
(and potentially H$\beta$ when it was possible to observe) is heavily differentially attenuated within \ion{H}{II} regions strong enough to suppress even 
the $EW$([OII]) measurement, but from which appreciable numbers of H$\alpha$ photons can escape (see, e.g., \citealt{oem09}). As post-starburst galaxies cannot,
by definition, house ongoing star formation, traditional classification of $z\sim1$ K+A galaxies based solely on observed-frame optical spectroscopy results in sample 
25\% contaminated by galaxies with active star formation. Such a fraction is statistically consistent with that seen in \citet{pwu14} from a study of 
traditionally-selected K+As in the SC1604 supercluster at $z\sim0.9$ based on the prevalence of significant mid-infrared (MIR) emission. The remaining 15 
galaxies for which either no H$\alpha$ emission was significantly detected or whose [NII]/H$\alpha$ ratios (or limits thereon) did not definitively point to 
star-formation processes as the origin of the H$\alpha$ emission (hereafter termed K+A-H$\alpha$ galaxies) we adopt as part of the true K+A population for 
the remainder of the paper. 

Of the 69 priority \textbf{II} (starburst) galaxies at $0.775\le z \le 0.912$ targeted with MOSFIRE, [NII]/H$\alpha$ classifications, either through significant detections 
of both lines or constraining limits, were made of 54. Of these, 39 had $\log(\rm{[NII]}/\rm{H}\alpha)<$-0.25 consistent with emission originating from star formation, the 
null hypothesis for these galaxies. However, 15 of these galaxies had [NII]/H$\alpha$ ratios (or lower limits) consistent with dominant LINER/Seyfert emission 
($\log(\rm{[NII]}/\rm{H}\alpha)\ge$-0.25). In such cases, the [OII] emission observed in the DEIMOS spectrum can also be attributed to a LINER/Seyfert process 
rather than star formation \citep{lem10}. Such galaxies are therefore spuriously classified by our DEIMOS spectra, as they have ceased star formation activity 
but show signs of a recently truncated ($\le$1 Gyr) episode of star formation, the defining criteria of a K+A galaxy. For the remainder of the paper, such galaxies, 
which would be classified as K+A but for their relatively strong [OII] emission which is powered by a dominant LINER/Seyfert source (i.e., false negative K+As) are
referred to as K+A with ImposteR [OII]-derived Star formation (KAIROS) galaxies. Applying these statistics to those priority \textbf{II} galaxies that either went 
untargeted with MOSFIRE or were unclassifiable 
to the depth of our observations results in $\sim50$ KAIROS galaxies over the redshift range $0.775\le z \le 0.912$, a value which \emph{exceeds} the full 
traditional K+A sample in the same redshift range. Accounting for the impurity of the traditional K+A sample and including the estimated number of KAIROS 
galaxies in our full sample results in a true K+A fraction of 15.7$\pm$1.7\%. Thus, not only does traditional selection of K+As at $z\sim1$ result in a 
sample in which a moderate fraction of galaxies have ongoing star formation, but \emph{such a selection also misses more than half of the true K+A 
population}. This fraction also does not change significantly if computed to the stellar mass completeness limit of our spectroscopic sample 
(see Table \ref{tab:kafractions}) nor if a different $I^{+}$ cut is imposed on the sample.

\subsubsection{Environmental Dependence of the Post-Starburst Fraction}
\label{envPSB}

While we defer a complete analysis of the dependence of the two types of K+A fractions across different large scale environments (i.e., field, group, and
cluster) and its dependence on various properties of LSSs to future work, we attempt here a cursory look. For this analysis we apply the statistics of the full sample 
and calculate fractions using all galaxies in the redshift range $0.775 \le z \le 0.912$. This choice was made to maximise the number 
of galaxies in each sub-sample we will define, though we note that results presented here largely hold if fractions are instead calculated only to
the stellar mass completeness limit of our spectroscopic sample and/or by applying the statistics of each sub-sample separately. 
Group and cluster galaxies are defined broadly as those in the adopted SG0023 and RXJ1716 redshift ranges (see \S\ref{optspec}), with field galaxies 
defined as all galaxies which do not fall in the previous two categories. Only those galaxies which were spectrally classifiable were used.
Adopting the traditional K+A classification scheme results in K+A fractions of 4.4$\pm$2.0\%, 7.2$\pm$1.8\%, and 10.1$\pm$2.6\% amongst field, group, 
and cluster galaxies, respectively. Such fractions and the trend of increasing traditional K+A fraction with increasing mass of the average central 
halo hosting the galaxy population is consistent with those observed in a variety of 
other $z\sim1$ K+A studies (e.g., \citealt{tran03, tran04, pog09, muz12, pwu14}). The observed trend is flattened considerably when comparing 
the combined KAIROS/K+A-H$\alpha$\footnote{This term will be used interchangeably with the term ``true K+A" throughout the paper.} fractions, 
which are 13.2$\pm$3.4\%, 14.3$\pm$2.7\%, and 15.9$\pm$3.3\% for the same three sub-samples, respectively. At first glance, this lack of dependence seems to imply that 
more massive central halos do not preferentially induce a K+A phase, contrary to results obtained for the vast majority of K+A studies at $z\sim1$. 

However, a slightly more nuanced approach can be taken. 
The number of K+A galaxies relative to the total number of combined starbursting and star-forming galaxies, hereafter K+A/(SB+SF), is a quantity more 
intimately linked with the efficacy of quenching processes. This fraction has been shown to be markedly higher amongst members of massive clusters and 
groups whose galaxy populations resemble those of more relaxed clusters \citep{pog09} than the field or groups with a dominant 
star-forming population. Despite the similar true K+A
fractions across the three sub-samples, a suggestive excess is observed in the true K+A/(SB+SF) fraction for cluster galaxies 
(29.3$\pm$6.6\%) as compared to group or field galaxies (21.5$\pm$3.9\% and 20.6$\pm$5.4, respectively). As the vast majority of the 
SG0023 galaxy population is undergoing active star formation ($\sim$70\%) and the groups show no large scale X-ray emission 
indicative of an excessively harsh medium for the average group halo mass, the similarity of this fraction between the SG0023 members 
and in the field is perhaps not surprising. A similar trend was seen a study of 11 groups at $z\sim1$ \citep{Mok13} in which colour-selected, 
green ``transition" galaxies were found to comprise a consistent fraction of SB+SF galaxies relative to a coeval field sample at all 
stellar masses, though this fraction was seen to change dramatically as a function
of stellar mass. The elevated fraction amongst the RXJ1716 members points to more efficient quenching within the 
bounds of the cluster, which, perhaps not coincidentally, is the only region in our sample definitively known to contain a hot medium. 
This trend is qualitatively similar to results found in the SC1604 supercluster, in which the only two constituent structures
with detectable ICM emission (clusters A and B) exhibited both higher traditional K+A fractions and lower fractions of star-forming galaxies than all
other clusters/groups within the supercluster as well as the coeval field \citep{lem12, pwu14}. Though differing spatial and spectral selections 
and different breadth and depth of NIR spectroscopic coverage preclude a rigorous quantitative comparison between our sample and that presented in
\citet{pwu14}, the completeness/purity-corrected, LINER/Seyfert-corrected true K+A/(SB+SF) fractions in clusters A and B range from 
$\sim25-30$\% as contrasted with $\sim$15\% for the galaxy populations of the SC1604 groups. These values are broadly consistent with 
those found amongst the galaxy population of SG0023 and RXJ1716. While this trend holds in our own data if 
we instead calculate these fractions for traditional K+As, these fractions decrease by a factor of 2-2.5$\times$ across all 
sub-samples. As the relative abundance of transitional populations is used to place constraints on quenching timescales 
(e.g., \citealt{balogh11, balogh16, mok14, muz14}), such a decrease can lead to drastically different conclusions on both the relative and 
absolute effectiveness of quenching mechanisms \citep{pog09}. In the following sections we investigate whether the shortcomings of traditional K+A selection are 
limited to lower purity and completeness or whether traditional K+As are comprised of a distinct population from the sample selected here.  

\begin{table*}
        \centering
        \caption{K+A Fractions} 
        \label{tab:kafractions}
        \begin{tabular}{llll} 
                \hline
                Sample & Traditional K+A Fraction$^{a}$ & KAIROS/K+A-H$\alpha$ Fraction$^{a}$ & (KAIROS/K+A-H$\alpha$)/(SB+SF)\\ 
                \hline
                Full  & 7.7$\pm$1.2\% (7.7$\pm$1.3\%) & 15.7$\pm$1.7\% (15.3$\pm$1.9\%) & -- \\ 
                LSSs Only &  8.3$\pm$1.5\% (8.5$\pm$1.7\%) & 15.6$\pm$2.0\% (14.4$\pm$2.2\%) & -- \\ 
                Field Only & 4.4$\pm$2.0\%  & 13.2$\pm$3.4\% & 20.6$\pm$5.4\% \\	
                SG0023 & 7.2$\pm$1.8\% & 14.3$\pm$2.7\% & 21.5$\pm$3.9\% \\ 
                RXJ1716 &  10.1$\pm$2.6\% & 15.9$\pm$3.3\% &29.3$\pm$6.6\%\\ 
                \hline
        \end{tabular}
\begin{flushleft}
a: Where calculated, the numbers in parentheses gives fractions for the stellar-mass-limited sample ($\log{(\mathcal{M}_{\ast}/\mathcal{M}_{\odot})}\ge10$)
\end{flushleft} 
\end{table*}

\subsection{The Evolutionary Stages of K+A and KAIROS Galaxies}
\label{KAevolution}

\begin{figure}
\includegraphics[clip,angle=0,width=0.98\hsize]{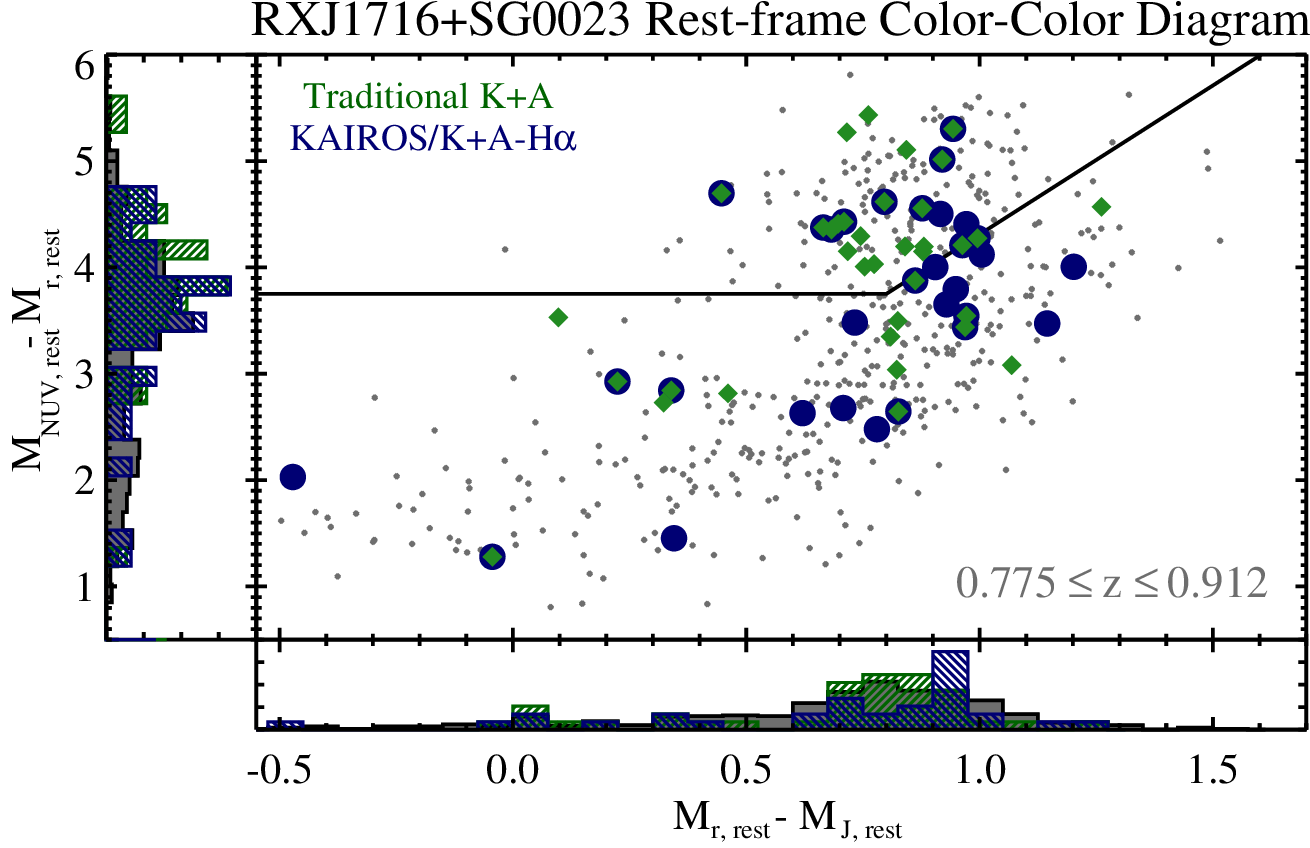}
\caption{Rest-frame $M_{NUV}-M_{r^{\prime}}$ vs. $M_{r^{\prime}}-M_{J}$ distribution of all galaxies (small grey points) in SG0023 and RXJ1716 
spectroscopically confirmed in the redshift range $0.775\le z \le 0.912$. Overplotted are those galaxies which meet our criteria for traditional 
K+A (green diamonds) or KAIROS/K+A-H$\alpha$ (blue circles). The black line shows the delination in this phase space between star-forming and quiescent 
galaxies adopted from \citet{lem14}. Area normalized histograms are shown for all galaxies (grey filled), traditional K+As 
(green hatched), and KAIROS/K+A-H$\alpha$ galaxies (blue hatched). The histograms in the left panel are plotted as ``effective colour" defined as the 
distance away from the quiescent/star-forming dividing line. The two flavours of K+A galaxies have similar colours in this phase space, with 
KAIROS/K+A-H$\alpha$ preferring slightly bluer $M_{NUV}-M_{r^{\prime}}$ and slightly redder $M_{r^{\prime}}-M_{J}$ colours, possibly indicating
a younger population on average (see \S\ref{KAevolution}).}
\label{fig:K+A_CCD}
\end{figure}

We begin this investigation by comparing the average properties of traditional K+As and those K+As selected from the MOSFIRE+DEIMOS data. The latter sample
is comprised both of known KAIROS galaxies and the pure population of traditional K+As, i.e., K+A-H$\alpha$ galaxies (see \S\ref{revealKA}). Galaxies selected 
using traditional K+A techniques appear, on average, both more massive in their 
stellar content ($\log(\langle \mathcal{M}_{\ast} \rangle/\mathcal{M}_{\odot})=10.65$ vs. $\log(\langle \mathcal{M}_{\ast} \rangle/\mathcal{M}_{\odot})=10.35$) 
and redder ($M_{NUV}-M_{r^{\prime}}=4.2$ vs. $M_{NUV}-M_{r^{\prime}}=4.0$) than the combined population of KAIROS/K+A-H$\alpha$ galaxies. The two populations
also appear at different positions in $M_{NUV}-M_{r^{\prime}}$ vs. $M_{r^{\prime}}-M_{J}$ phase space, a phase space which is commonly employed to separate star-forming 
from quiescent populations. Traditional K+As, with average colours of 4.2 and 0.80, respectively, fall comfortably into the region of this phase space 
indicating quiescence (at $0.5<z<1$, see \citealt{lem14}), though in the region of this phase space is potentially also populated by recently ($\la$1 Gyr) rapidly quenched 
galaxies (see, e.g., \citealt{thibaud16}). In moderate contrast, the average KAIROS/K+A-H$\alpha$ galaxy, with colours of 4.0 and 0.85, respectively, appears 
in the liminal region of this phase space between star-forming and quiescent populations where younger transitional populations are likely to lie 
\citep{ilbert10,lem14, thibaud16}. In Figure \ref{fig:K+A_CCD} we plot $M_{NUV}-M_{r^{\prime}}$ vs. $M_{r^{\prime}}-M_{J}$ for all traditional 
K+A and KAIROS/K+A-H$\alpha$ galaxies.  

Such differences suggest that the true K+A population is predominantly younger than the K+A population selected using traditional
means. While we, in principle, have estimates of the mean luminosity-weighted stellar age output by the SED fitting described in \S\ref{SEDfitting}, 
age estimates from applying traditional SED-fitting techniques to broadband photometry alone are highly uncertain (e.g., \citealt{lee09, janine12, romain16}). 
Here, however, we have the luxury of high S/N DEIMOS spectra
that contain several age sensitive features (e.g., $D_{n}(4000)$, H$\delta$, G-band $\lambda$4305\AA). These spectra in conjunction with the broad 
rest-frame wavelength coverage of our photometry allow us to place much stronger constraints on internal extinctions than is possible with the limited 
wavelength range of our DEIMOS spectroscopy. These constraints, in turn, largely allows for the breaking of the degeneracy between the stellar age of a galaxy and 
its dust content (as in \citealt{romain16}, though at higher redshift), which subsequently allows for at least the potential of 
precision measurements on ages\footnote{Here and throughout the remainder 
of the paper ``age" is defined as the time since the onset of star formation and is denoted $t_{SB}$}. 

\begin{figure*}
\includegraphics[clip,angle=0,width=0.49\hsize]{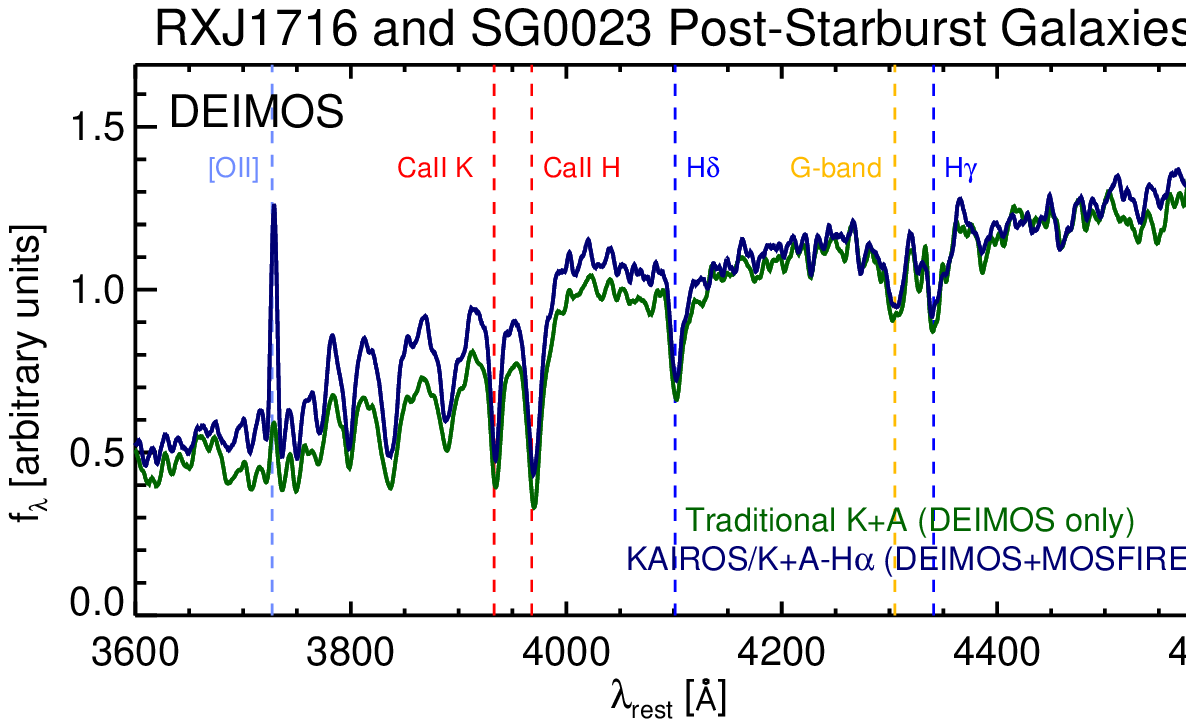}
\includegraphics[clip,angle=0,width=0.49\hsize]{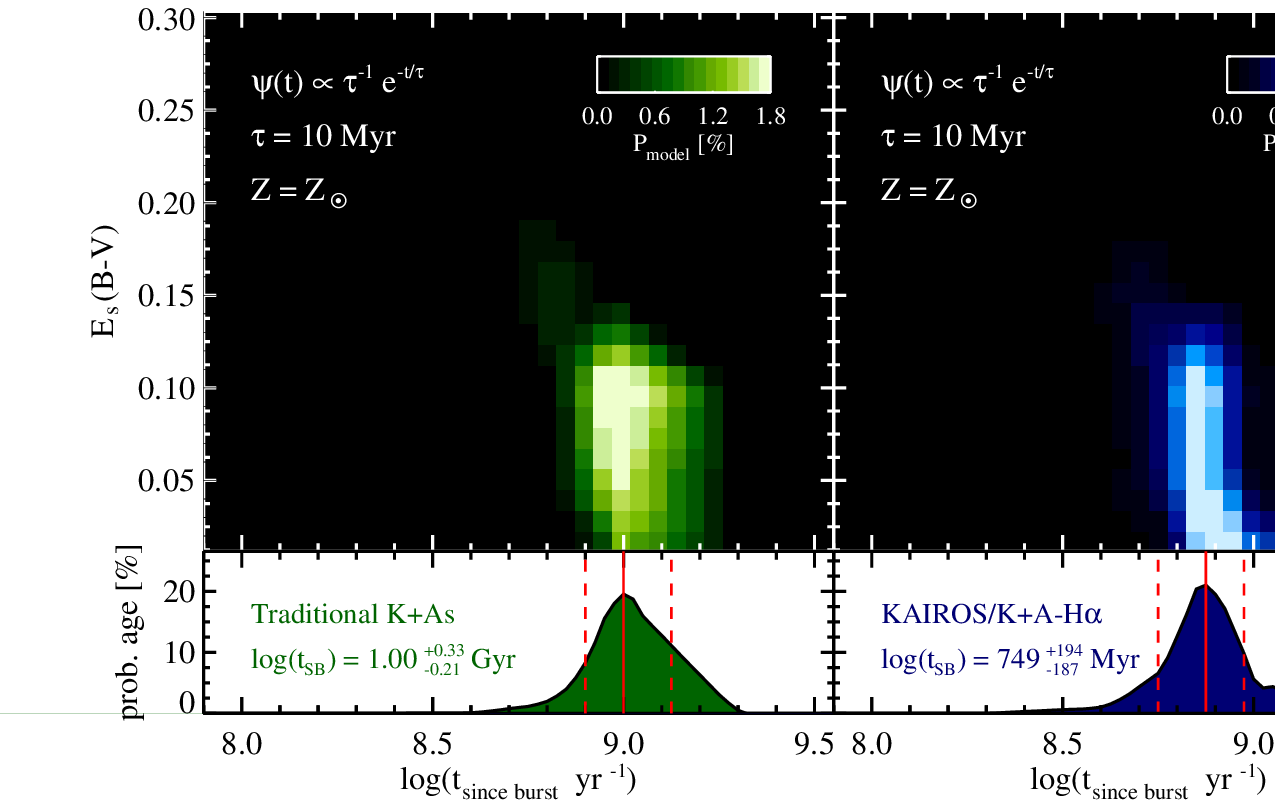}

\caption{\emph{Left:} Inverse-variance, unit-weighted coadded DEIMOS spectra of the traditional K+A (green) and KAIROS/K+A-H$\alpha$ (blue)
galaxies. Important spectral features are marked by vertical dashed lines and labeled. The MOSFIRE spectra for the latter population was
coadded in a similar manner (not shown) and the logarithm of the measured [NII]/H$\alpha$ ratio is given at the bottom of the figure. While the
two DEIMOS coadded spectra appear to have similar strength Balmer features (H$\gamma$, H$\delta$, and higher order features blueward of \ion{Ca}{II} K),
strong [OII] emission attributable to a dominant LINER/Seyfert source (see \S\ref{revealKA}) is present in the average KAIROS/K+A-H$\alpha$ galaxy
spectrum. \emph{Right:} Zoom in of the probability density maps (PDMs)
for luminosity-weighted stellar ages and stellar extinctions of the average traditional K+A (upper left panel) and KAIROS/K+A-H$\alpha$
(upper right panel) galaxy estimated from fitting stellar synthesis models simultaneously to the stacked DEIMOS spectra and stacked photometry.
In both panels, $>>99.9$\% of the probability density of the full PDMs are contained within the displayed area.
The form of the SFH, the e-folding time of the exponential decline, and the stellar-phase metallicity used for the fitting are given in the top
left of each panel. A scale bar is shown in the top right of each panel and maps the colours to their associated probabilities. The bottom panels
show the (extinction) marginalized one-dimensional probability distribution functions (PDFs) of the luminosity-weighted stellar age for each sample.
The median value of each PDF along with the associated effective 1$\sigma$ uncertainties is reported in the left of each panel.}
    \label{fig:spectranPDM}
\end{figure*}

\begin{figure}
\includegraphics[clip,angle=0,width=0.98\hsize]{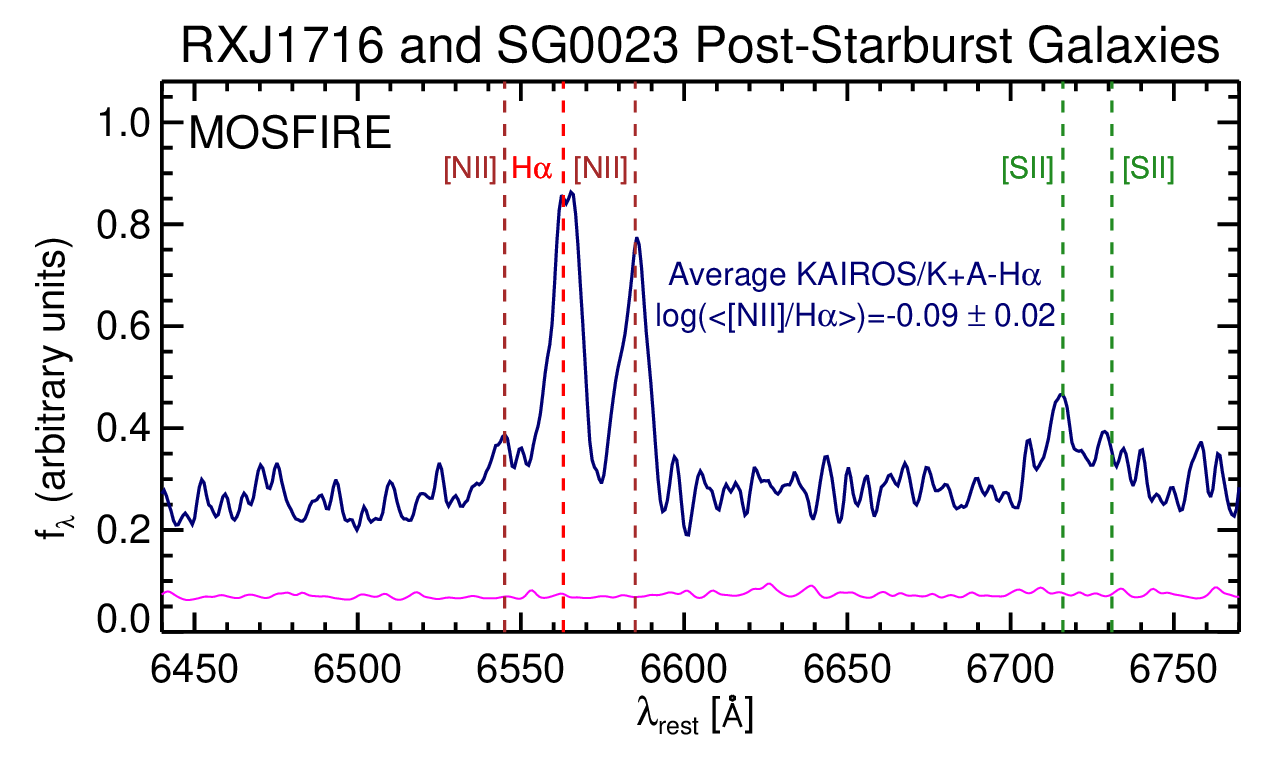}
\caption{Inverse-variance, unit-weighted coadded MOSFIRE spectrum of the KAIROS/K+A-H$\alpha$ sample. The flux density
spectrum is shown in blue and the error spectrum is shown in magenta. The location of H$\alpha$ and the [NII] $\lambda$6548\AA, 
$\lambda$6583\AA\ features is marked by dashed lines. Also marked are the locations of the \ion{S}{II} $\lambda$6716\AA\ and 6731\AA\ features, both
of which are significantly detected although appear to be relatively weak ($\log(\langle \rm{[SII]}/\rm{H}\alpha \rangle)=-0.65\pm0.04$ for the 
former feature). The average $\log(\rm{[NII]}/\rm{H}\alpha)$ measurement derived from bandpass measurements 
on the spectrum is given. Such a value firmly excludes star formation as the dominant source of emission in these galaxies.} 
\label{fig:MOSFIREcoadd}
\end{figure}

The DEIMOS spectra of the two populations was combined (hereafter ``coadded") through an inverse variance-weighted average after shifting each 
individual spectrum to the rest frame, interpolating onto a standard grid with constant plate scale of $\Delta\lambda=0.33/(1+z_{min})$ (where $z_{min}$ is the 
minimum $z_{spec}$ for each sample), and normalizing each spectrum to an average flux density of unity (e.g., unit weighted) following the methodology 
described in \citet{lem12}. The resulting coadded spectra of the traditional K+A and the combined KAIROS and K+A-H$\alpha$ population are shown in the 
left panel of Figure \ref{fig:spectranPDM}. The MOSFIRE spectra of the KAIROS/K+A-H$\alpha$ population was also coadded in an identical manner 
(see Figure \ref{fig:MOSFIREcoadd}). While this coadded MOSFIRE spectrum has continuum emission which is too faint to lend itself appreciably to the SED 
fitting process described here, the measure of the average [NII]/H$\alpha$ of the resulting coadded spectra, 
$\log(\langle \rm{[NII]}/\rm{H}\alpha \rangle)=$-0.09$\pm$0.02, clearly establishes LINER/Seyfert activity as the dominant source of 
emission in these galaxies. The strength of the [SII] $\lambda$6716\AA\ line relative to H$\alpha$ in the same spectrum was measured 
employing an identical method and custom bandpasses and found to be $\log(\langle \rm{[SII]}/\rm{H}\alpha \rangle)=-0.65\pm0.04$, a value 
constraining in its own right (see below). 

It is important to pause at this point to consider several features of the coadded DEIMOS spectra presented in Figure \ref{fig:spectranPDM}. 
The most striking difference is the 
strong [OII] feature of the average KAIROS/K+A-H$\alpha$ galaxy ($EW$([OII])=-7.1$\pm$0.2\AA), which is absent in the average traditional K+A spectrum.
While some galaxies without strong [OII] emission are contained within the former sample (the K+A-H$\alpha$ galaxies), these galaxies are subdominant 
to the KAIROS galaxies, which results in the observed strong [OII] emission. In addition, the high S/N of the coadded KAIROS/K+A-H$\alpha$ spectrum allows  
us the luxury of significantly detecting the [NeIII] $\lambda$3868\AA\ line ($EW$([NeIII]=-0.7$\pm$0.1\AA). This measurement when combined with that of
[OII] and various colour measurements strongly indicate the presence of activity other than star formation \citep{Stasinska06,trouille11, Marocco11},
though it is difficult to say definitively how much that activity dominates the emission profile based on these diagnostics alone. 
The most striking similarity between the two coadded spectra is the very strong Balmer series absorption observed both visually and 
quantitatively ($EW$(H$\delta$)$\sim$6.0$\pm$0.2\AA\ in both cases) 
highlighting the large fractional population of A and B stars contained within the average galaxy in both populations (or, more precisely, the fraction of 
those stars not selectively affected by dust extinction, see, e.g., \citealt{pog99}). Though less striking visually, 
the spectra exhibit significantly different strengths in the continuum break seen at 4000\AA\, $D_{n}(4000)$ (adopting the method of \citealt{balogh99}), 
1.33$\pm$0.01 vs. 1.47$\pm$0.01 for the average KAIROS/K+A-H$\alpha$ galaxy and traditional K+A, respectively. Such a difference again strongly points to a 
younger galaxies comprising the KAIROS/K+A-H$\alpha$ population. 

In parallel with this spectral coaddition, observed-frame broadband photometry was coadded following the manner described in Appendix A. 
The coadded spectrum and photometry for each sample was then fit simultaneously to synthetic models (see Appendix A) to further 
investigate the evolutionary states of the two populations. 
In the right panels of Figure \ref{fig:spectranPDM} we show a visualization of the results of this fitting process for a \citet{CB07} model with the model parameters 
set to the values shown in each panel (for justification on these choices, see Appendix A). To make these visualizations, hereafter referred to as 
probability density maps (PDMs), the probability at 
each step in $t_{SB}$ and $E_{s}(B-V)$ is calculated from the formula given in Appendix A, with $E_{s}(B-V)$ allowed to vary between 0 and 0.7 in steps of 0.05.  
It can be clearly seen in the PDMs that the dust-age degeneracy has been largely broken by this analysis, as there exists no anti-correlated behavior 
between age and extinction in the observed PDMs. The vertical extent of the non-trivial values observed in the PDM in both cases is a result of the inability of our data 
to discriminate between various low levels of dust content. Below each panel shows the one-dimensional $t_{SB}$ PDF generated by adding probabilities of all 
values of $E_{s}(B-V)$ for each age step. The median value of the PDF is denoted by a solid red vertical line. The effective $\pm$1$\sigma$ values are taken from the 
84th and 16th percentiles of the PDF, respectively, such that 68.3\% of the PDF is contained within their bounds. These effective $\pm$1$\sigma$ values are shown as 
vertical dashed lines. The KAIROS/K+A-H$\alpha$ population 
has a median $t_{SB}=0.75\pm0.19$ Gyr, younger than the corresponding value for the traditional K+A population of 
$t_{SB}=1.00^{+0.33}_{-0.21}$ Gyr.            
The distributions are not Gaussian, and the $t_{SB}$ PDF of the average KAIROS/K+A-H$\alpha$ galaxy likely excludes 
the possibility ($<$10\%) that the average galaxy is older than the median $t_{SB}$ of the traditional K+A population. These relative differences
persist, at the same level of statistical significance, if we instead choose a different stellar synthesis prescription (\citealt{maraston05}, BC03), 
SFH delayed $\tau$, multi-burst, metallicity (0.2$Z_{\odot}$, 0.4$Z_{\odot}$), or extinction scheme \citep{prevot84} for both sets of galaxies. 

The combination of all analysis presented in this section strongly indicates that the true K+As are at an earlier evolutionary stage   
than those K+As selected using traditional methods. Adopting this possibility as truth, we then have a scenario in which a population 
of galaxies with a recently truncated massive star formation event with, on average, strong emission powered by a dominant LINER/Seyfert 
component evolve within $\sim$300 Myr to a population largely devoid of such emission. Such a scenario lends itself to the intriguing 
possibility that the source powering the LINER/Seyfert emission comes from residual AGN activity that was incited coevally (or nearly so) with 
the starburst. Other processes are known to give rise to LINER emission, galactic shocks \citep{dopita95, veilleux95}, cooling flows		
\citep{heckman81, heckman89}, photoionization by hot stars \citep{terlevich85, shields92}, and post-asymptotic giant branch stars 
\citep{binette94,taniguchi00}, and have been favored over a central engine as powering LINER activity in recent results from 
spatially resolved spectroscopy \citep{singh13,bel16} leading to a suggested change in terminology for such sources (Low Ionization Emission Regions).  
In this study, we do not have the luxury of spatially resolved spectroscopy. However, since such processes operate at different evolutionary phases and over different 
timescales than AGN activity the source of emission can, in principle, be uncovered through timing arguments or through other more direct observational means 
(see, e.g., the discussion in \citealt{alatalo16}). Such timing arguments have been used on large samples of local post-starburst galaxies to argue that 
AGN activity is the dominant source of quenching for the most massive of such galaxies ($\log(\mathcal{M}_{\ast}/\mathcal{M}_{\odot})>10$, \citealt{kaviraj07}). 
While we make no serious effort to uncover the source of the emission
in this paper, we note that most (80\%) of the galaxies definitively classified LINER/Seyfert have $\log$(EW([OII])/EW(H$\alpha$))$<$0.6 (see Figure 
\ref{fig:spectraldiagnostic}), with a median value of $\log$(EW([OII])/EW(H$\alpha$))=0.4. Such values are more typical of emission powered by Seyfert 
or hybrid activity rather than LINER-powered emission \citep{yan06,lem10}, which suggests that the narrow line emission in the KAIROS/K+A-H$\alpha$ sample 
originates from an AGN. Such a suggestion is in line with the relatively strong [NeIII] emission observed in the coadded KAIROS/K+A-H$\alpha$ DEIMOS spectrum 
(see, e.g., \citealt{greg15} and references therein). Additionally, the relatively weak [SII] $\lambda$6716\AA\ emission observed in the coadded 
KAIROS/K+A-H$\alpha$ MOSFIRE spectrum (see Figure \ref{fig:MOSFIREcoadd}) appears to disfavor alternatives
to AGN activity (see, e.g., \citealt{kew06,alatalo16}). More detailed modeling will be required to explore this suggestion further and will be attempted 
in future studies. Here, we limit ourselves to searching for residual AGN acitivity through other more overt means, though any such activity is likely to be 
at a low level $\sim$700 Myr after the cessation of star formation (e.g., \citealt{hopkins08}).

We searched the $\sim$50 ks \emph{Chandra} ACIS-I 
\citep{garmire03} images obtained in both fields (Obs. IDs 7194 \& 548 for SG0023 \& RXJ1716, respectively; \citealt{vikhlinin02, rum12}) 
for individual detections of galaxies in both the traditional K+A and KAIROS samples. These images reach a 
50\% completeness limit of $L_{X,\ 0.5-7 \rm{keV}}\ga10^{42.6}$ 
ergs s$^{-1}$ at $z=0.83$ over the area spanned by our sample as derived from Monte Carlo simulations (see Rumbaugh et al.\ \emph{in prep.}). 
No traditional K+A and only one galaxy from the KAIROS sample was individually matched to an X-ray source ($L_{X,\ 0.5-7 \rm{keV}}=10^{42.7}$ ergs s$^{-1}$, 
$\log(\rm{[NII]}/\rm{H}\alpha)=0.68$). We also preformed a stack of the X-ray data at the optical positions of the galaxies 
in two samples, separately, after removing the one X-ray detected KAIROS galaxy, sources at spatial locations coincident with 
the RXJ1716 ICM emission, and those at large ($>7\arcmin$) off-axis angles. No detection was found in either sample to a 
3$\sigma$ limit of $L_{X,\ 0.5-7 \rm{keV}}\sim41.8$ at $\langle z \rangle=0.83$. 

It appears, that any residual X-ray AGN activity in the  
KAIROS galaxies, if it exists, is generally too faint, both individually or on average, to observe with our data. Such activity, however, is likely extremely 
difficult to detect given our current data. In a study of traditionally-selected K+As at $0.10<z<0.35$, 
no optically-fainter K+As ($M_{R}>-22$) were observed with X-ray emission to a limit of $L_{X,\ 0.5-7 \rm{keV}}<10^{41}-10^{42}$ ergs s$^{-1}$ \citep{brown09}. 
Galaxies at these luminosities comprise the bulk ($\sim75$\%) of both our traditional K+A and KAIROS samples. While one third of optically-brighter K+As ($M_{R}<-22$) 
are found to have nuclear X-ray activity in \citet{brown09}, this activity was measured exclusively at levels below the 10\% completeness limit of our \emph{Chandra} 
imaging. Such a paucity of nuclear activity was also observed by \citet{depropris14} in an extensive study of 10 local ($z\sim0.03$) similarly optically-faint 
traditionally-selected K+As. Similarly weak X-ray emission was also observed in a sample of traditionally-selected K+As drawn from the zCOSMOS-bright survey 
\citep{lilly07} at redshifts more comparable to our sample. Using \emph{Chandra} imaging considerably deeper than our own \citep{elvis09}, \citet{vergani10} 
found that only $\sim10$\% of their K+A sample exhibited detectable individual X-ray emission, all at $L_{X, \ 0.5-10 \rm{keV}}<10^{43}$ ergs s$^{-1}$,
and a stacking analysis of the remaining population revealed no significant detection to a limit comparable to that of our data \citep{vergani10}.
Thus, the lack of coincident X-ray AGN activity in our current samples is likely not sufficiently constraining. 
It is also entirely possible that such activity is manifest in another mode, as strong X-ray and narrow-line activity from AGN are 
often times observed distinctly \citep{renbin11, trouille11}. We defer further investigation
of this scenario to the full population of ORELSE K+As for which we will be able to place extremely stringent limits on AGN activity, both
for individual sources and for stacked data, from deeper X-ray, Very Large Array 1.4 GHz, NIR, and MIR imaging, and more complex
photoionization modelling.  

\begin{figure*}
\includegraphics[clip,angle=0,width=0.48\hsize]{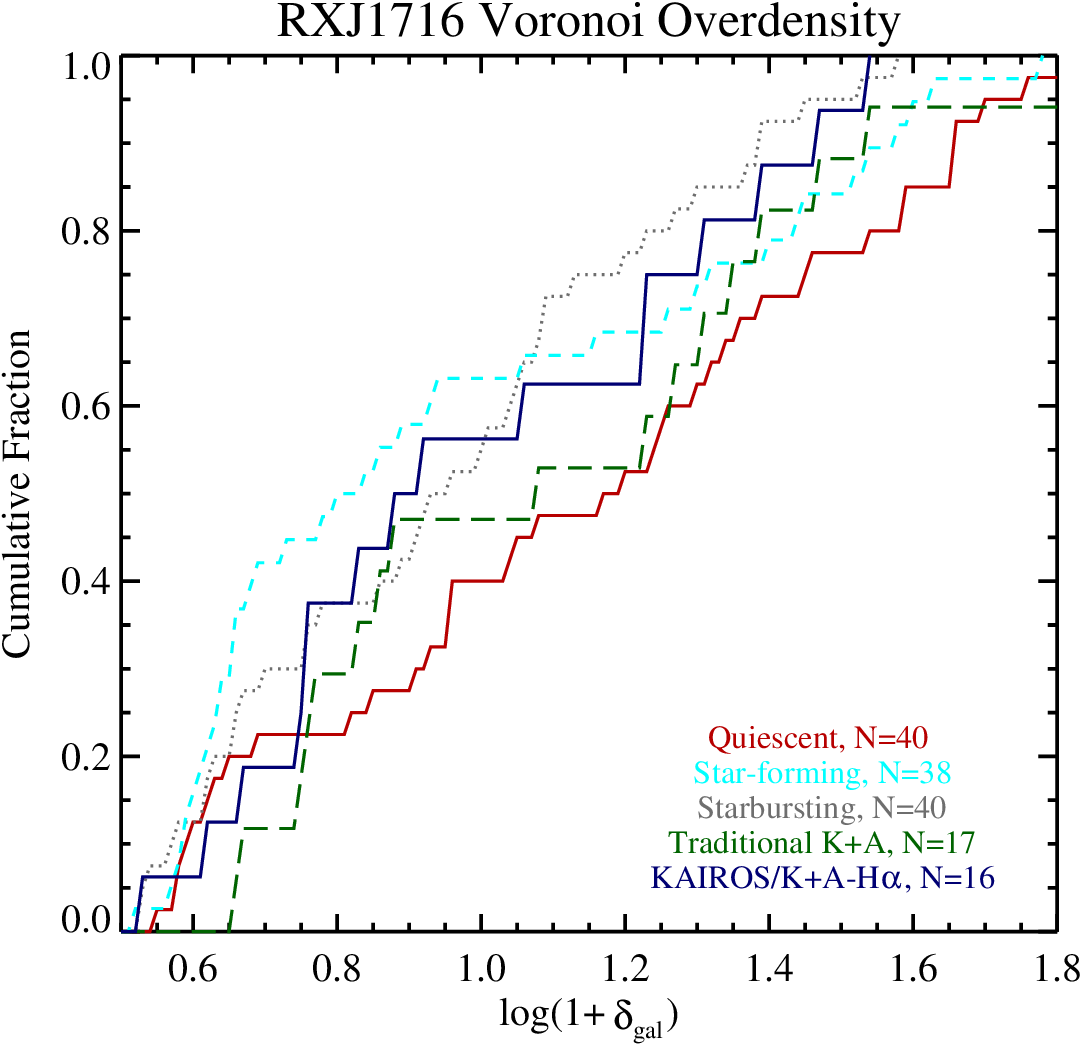}
\includegraphics[clip,angle=0,width=0.48\hsize]{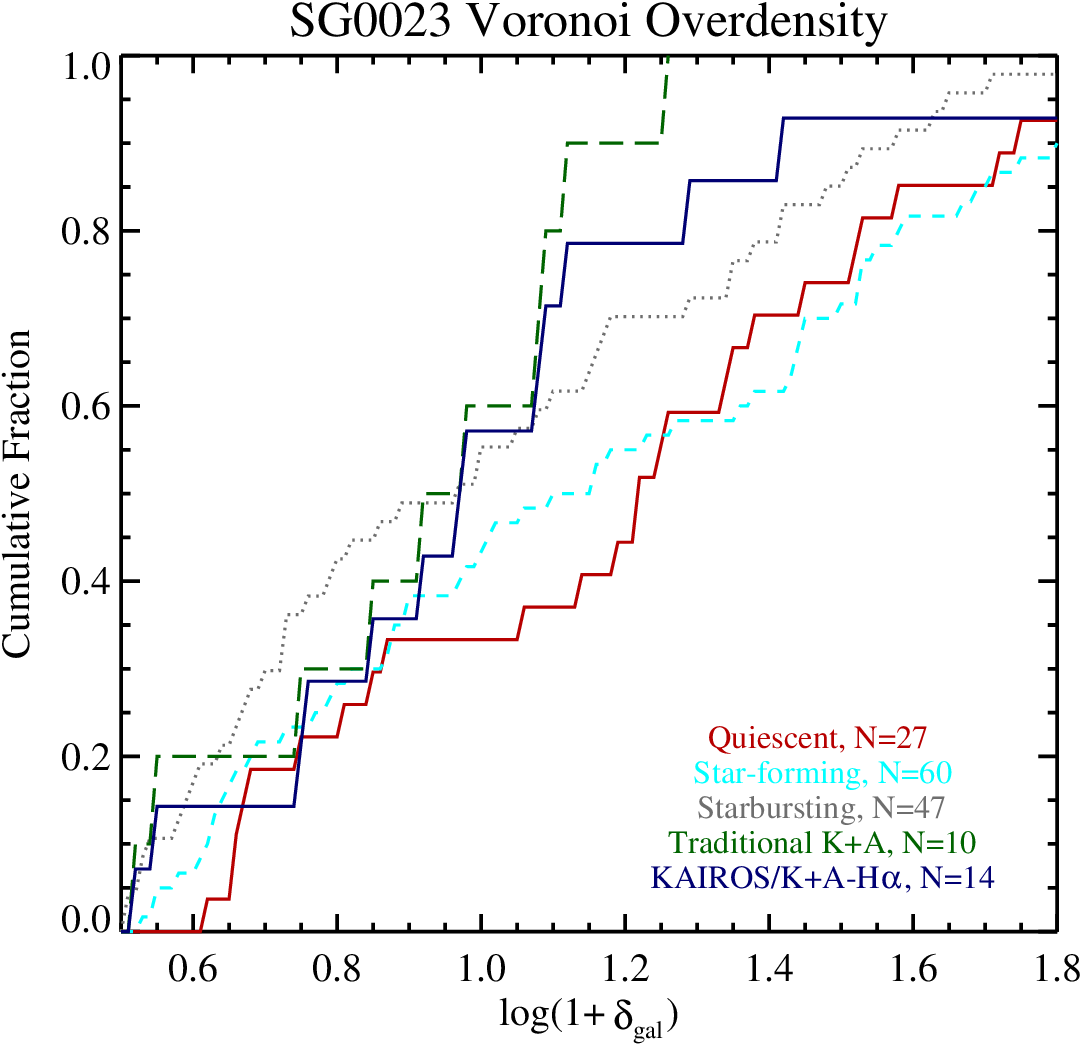}

\caption{Overdensity cumulative distribution functions (CDFs) of various spectrally classed galaxies in the RXJ1716 (\emph{left}) and SG0023 (\emph{right}) fields
in the redshift range $0.775\le z \le 0.912$. The CDFs of quiescent, starbursting, and star-forming galaxies are plotted with red solid, grey dotted, and cyan
dashed lines, respectively. The CDFs of the two flavours of K+As, traditional K+As and KAIROS/K+A-H$\alpha$, are shown by green long-dashed and blue solid lines,
respectively. The majority of galaxies comprising each CDF are member galaxies of each LSS. Only galaxies with $I^{+}<24.5$ and 
at $\log(1+\delta_{gal})>0.5$ are used to generate the CDFs as our DEIMOS observations are highly complete for such galaxies (see \S\ref{environKA}). MOSFIRE observations 
are also highly complete for the relevant galaxy populations (traditional K+A and starburst) when applying these cuts. 
The CDF of each galaxy type appears different in each field highlighting the effect of the differing LSS environment at common (local) 
overdensities. In each field, the CDFs of the traditional K+A and KAIROS/K+A-H$\alpha$ galaxies are significantly different, with the latter largely 
tracing that of the starbursting population. In RXJ1716, the CDF of traditional K+As shows consistency with that of quiescent galaxies, a consistency
not observed in SG0023, suggestive of a different evolutionary path.
}
    \label{fig:CDFz}
\end{figure*}

\subsection{Environments of Post-Starburst Galaxies}
\label{environKA}
We have shown in the previous sections that the inclusion of galaxies exhibiting K+A features with moderately strong [OII] emission powered 
by a LINER/Seyfert results in large changes in the number and type of galaxies selected as K+A. While we made a cursory attempt to investigate
the effect of environment, defined in a broad sense, on the various post-starburst fractions in \S\ref{envPSB}, in this section we investigate the 
distribution of the true K+A population across different environments and compare this distribution to traditional K+As and galaxies of other spectral types.
As mentioned in \S\ref{voronoi}, we chose here to use local overdensity as the sole metric for estimating environment. However, we note that, within
the limits of our data, none of the results presented in this section are sensitive to this choice. Identical results are obtained 
if we had instead chosen to perform the analysis based on group-/clusto-centric distance or in $R_{proj}/R_{vir}-|\Delta_{v,\ LOS}|/\sigma_{v}$ 
phase space (see, e.g., \citealt{carlberg97, balogh99, biviano02, haines12, Noble13}). Since completeness now becomes an issue for our analysis, for this section, we limit 
the distributions for each galaxy sample to 
$\log(1+\delta_{gal})\ge0.5$. Above this value of overdensity we have obtained high-$Q$ spectral measurements for which we can measure both  
$EW$([OII]) and $EW$(H$\delta$) reliably for $\ge$40\% of \emph{all} objects in both fields (as estimated by our $z_{phot}$ measurements), with 
an average completeness of 64\%, over the redshift range $0.775\le z \le 0.912$, at $I^{+}\le24.5$, and within the spatial extent covered by our 
DEIMOS masks coverage. With such a high level of spectroscopic 
completeness it is reasonable to assume the distributions observed in our data are reflective of the true underlying distributions. In these environments 
MOSFIRE observations for which [NII]/H$\alpha$ measurements were obtained (or meaningful limits placed) of traditional K+A and potential KAIROS 
galaxies is also high, 55\% and 45\%, respectively. Thus, such trends observed in the data are likely robust to sample variance. 

In Figure \ref{fig:CDFz} we plot the cumulative $\log(1+\delta_{gal})$ distribution functions (CDFs) of quiescent, star-forming, and 
starbursting galaxies along with the two flavours of K+A galaxies over roughly a decade and a half in local overdensity. A few broad initial 
observations can be made. While the distribution of quiescent galaxies appears 
largely similar, with quiescent galaxies preferring the densest environments in both fields and a Kolmogorov-Smirnov (KS) test finding no significant difference
between the two populations, the environments of galaxies 	
forming (relatively) modest amounts of stars (i.e., star-forming galaxies) are very different between the two fields. Star-forming galaxies strongly
avoid the densest regions in the massive RXJ1716 cluster, while such galaxies are heavily clustered in the central regions of the SG0023 groups.
In both cases starbursting galaxies generally avoid the densest regions, though with a stronger aversion to such regions seen in RXJ1716. These 
trends highlight the large range of evolutionary states of the LSSs in the two fields, and, underscore the necessity of accounting for both local 
(over)density and LSS properties (e.g., halo mass) when investigating environmentally-driven galaxy evolution. 
  
The environments of traditional K+A populations appear markedly different than the quiescent galaxies in SG0023. Such a trend is also observed
for traditionally-selected K+A member galaxies of the five groups in the SC1604 supercluster \citep{pwu14} using a different metric of environment 
(projected radius). We have independently confirmed that this trend holds in the SC1604 groups if $\log(1+\delta_{gal})$ (calculated as in 
\S\ref{voronoi}) or projected positional-differential velocity phase space is instead 
used as an environmental metric and the same K+A selection techniques employed in this study are instead adopted. This concordance 
across ten groups in two different fields strongly indicates that this is a general trend
of K+A galaxies within groups. Conversely, traditional K+As and quiescent galaxies appear to show some overlap in their environments 
in RXJ1716, however, and a KS test cannot reject the possibility that the two distributions are drawn from the 
same underlying parent sample with any significance. Such overlap is also seen in the K+As housed in the three clusters of the SC1604 supercluster 
as well in preliminary results for other ORELSE clusters (RXJ1757, RXJ1821, SC1324, RCS0224) indicating that this is, again, a general trend of traditionally-selected cluster K+As. 	
The differences in the observed overlap between traditionally selected K+As and quiescent galaxies may have frustrated proper interpretation 
in previous studies which conflate group and cluster K+A populations or simply measure K+A properties and prevalence as a function of 
local density (or some other equally blunt environmental metric) without regard to the properties of the parent structure. No such 
ambiguity exists for the environments of the KAIROS/K+A-H$\alpha$ galaxies, as a KS test firmly rejects ($>>3\sigma$) the hypothesis that 
their distribution is compatible with that of the quiescent population in both SG0023 and RXJ1716. This 		
dissimilarity immediately precludes the possibility that, at least in the LSSs and their surrounding fields studied here, K+As are solely the products of 
rejuvenated quiescent galaxies (as in, e.g., \citealt{dressler13}). Instead, the KAIROS/K+A-H$\alpha$ distributions appear to, more faithfully than
those of the traditional K+As, track the environments of starbursts. In both RXJ1716 and SG0023 the environments of the traditional K+As are inconsistent
with those of the starbursting population at $>$3$\sigma$, while those of the KAIROS/K+A-H$\alpha$ populations are statistically indistinguishable in both 
cases. 

\subsubsection{Possible Evolutionary Scenarios for Cluster and Group Post-Starburst Galaxies}

The results in the previous section brought a dramatic shift in the interpretation of the possible progenitors of K+A galaxies. From 
a traditional K+A population moderately consistent with descending from quiescent galaxies, this study revealed the true K+A population 
within RXJ1716 and SG0023 almost certainly descended predominantly from bluer starbursting galaxies. 
As most of the quiescent galaxies are harbored, in both
fields, within the cluster/group cores, this shifting interpretation also has dramatic consequences on the prevalence and efficacy of mechanisms
acting on the K+A populations throughout their transformation. The more moderate density environments (i.e., the outskirts of groups and
clusters) inhabited by the true K+A population, where differential velocities remain relatively low, and their large environmental overlap with the 
starbursting population, a population known to exhibit a high fraction of disturbed morphologies \citep{dirtydale11b, kartaltepe12, pawlik16}, points to 
galaxy-galaxy interactions or merging activity as the primary formative events leading to the K+A phase. The latter activity in particular have been shown
in simulations to effectively induce a K+A phase under certain conditions \citep{bekki05, johansson09, wild09, snyder11}. 

While this result is seemingly at odds with the elevated true K+A/(SB+SF) fraction found for the galaxy population of the RXJ1716 cluster found in 
\S\ref{envPSB}, the two results can be reconciled in the following manner. Interactions occurring near the outskirts (i.e., 1-2 $R_{vir}$) of LSSs 
cause a starburst which decays over 10-100 Myr due either to gas exhaustion, stellar feedback (including supernovae), or feedback from an AGN. 
The massive reservoirs of \ion{H}{I} or CO gas (the latter being a proxy for H$_{2}$) discovered surrounding $\sim$50\% of both field and 
group/cluster K+As in the local universe 
\citep{chang01, zwaan13, french15} points to the latter two mechanisms as both mechanisms can inject large amounts of kinetic energy into any 
remaining gas, disrupting star formation without, necessarily, completely removing the fuel supply. More recently, a study of SDSS K+A galaxies which were
selected in a similar manner to the KAIROS selection employed for this study found tentative evidence of
widespread galactic winds in such galaxies driven either by AGN activity or stellar feedback processes \citep{alatalo16}. 
Such feedback is also consistent with that inferred from large surveys of dusty starbursting galaxies in both the local and distant universe 
in which the feedback is required to be relatively rapid (see \citealt{juneau13, lem14} and references therein). These mechanisms also provide a 
natural explanation to explain the small values of 
$E_{s}(B-V)$ estimated in \S\ref{KAevolution} as the same processes would also excise or diffuse the dust content near the inception of the K+A phase 
(as in, e.g., \citealt{yesuf14}). In field and group environments devoid of the presence of a well-formed hot medium, galaxies are largely allowed 
to retain this reservoir of hot gas, which is used to fuel rejuvenated star formation in $\sim$1-2 Gyr (e.g., \citealt{bahe15}). Such a picture is consistent with the similar fraction of galaxies with active star formation
($\sim70$\%) and those classified as starburst ($\sim28$\%) in both environments as well as their similar fraction of true K+A/(SB+SF).
In such a picture, the environmental distributions of KAIROS galaxies, galaxies which have residual signs of some form of this feedback 
(see \S\ref{KAevolution}), would appear more similar to starbursting galaxies, while traditional K+As, being further removed from the 
starbursting event, would start to diverge from the starbursting population. In neither case, however, should  
K+As follow the distributions of the quiescent populations, a restriction which holds for SG0023 K+As. Such fractions and distributions
can be used to place constraints on the K+A duty cycle and will be attempted in future works.

The end stages of such a picture within a cluster environment would look considerably different. An initial starburst 
occurs near the cluster outskirts through galaxy interactions or merging events. Appealing to this method of inducement for RXJ1716 members is
reasonable given that MIR-detected members of RXJ1716 (i.e., dusty starbursting galaxies) are seen to preferentially inhabit 
moderate local density environments within the overall LSS, roughly equivalent to our $\log(1+\delta_{gal})\la1$, and strongly avoid the region 
corresponding to the projected cluster core \citep{koyama08}. Such interactions lower the binding energy of the remaining gas through 
kinematic effects and stellar or AGN feedback, which allow it to be more easily stripped when entering the radius where strangulation and ram pressure stripping
are effective ($\sim$$R_{vir}$ and $\la$0.5$R_{vir}$, respectively, \citealt{moran07}, assuming for the latter a spherically symmetric 
ICM centreed at the cluster optical centre, as is observed in RXJ1716). In a study of simulated cluster and group galaxies performed by
\citet{bahe15} it was found that the inclusion of stellar feedback doubles the effectiveness of ram pressure stripping on member galaxies 
of massive halos even without including the (likely additive) effects of AGN feedback, which lends credence to this scenario. These 
interactions or merging events need not occur prior to the first pericentre passage of galaxies accreting into the cluster environment and, indeed, are 
perhaps more likely after this or several passages (see discussion in \citealt{struck06}). Thus, such a scenario is still possible to 
reconcile with the ``delayed-then-rapid" quenching inferred by comparisons of observations of local cluster and group galaxies with 
numerical simulations \citep{wetzel13} and the slight-less-delayed then rapid quenching inferred from observations of other samples of
$z\sim1$ K+A galaxies \citep{muz14} or other types of transitional populations observed at a variety of redshifts \citep{thibaud16, kschawin14}.  

In this scenario the K+A phase experienced by a cluster galaxy is largely not cyclic, 
but rather marks the end of a galaxy building up its stellar mass through \emph{in-situ} star formation. As a consequence,  
both the fraction of quiescent galaxies and the K+A/(SB+SF) fraction would elevate relative to those in less dense environments, though 
the exact value of this increase depends on the length of the duty cycle of the K+A phase in the latter environments. Additionally, 
older (traditional) K+As would be observed to inhabit environments intermediate to those of starbursting and quiescent populations. 
All three trends are observed in the RXJ1716 galaxy population. Further supporting this scenario, $\sim$50\% of the traditional K+A 
and KAIROS/K+A-H$\alpha$ galaxies which are members of the RXJ1716 cluster lie within $R_{proj}<0.5 R_{vir}$
from the cluster centre. From the properties of its ICM emission, this radius is equivalent to the radius at which ram pressure stripping 
effectively acts on a Milky Way analog RXJ1716 member travelling at a radial velocity relative to the ICM equivalent to $\sigma_{v}$
(see appendix B of \citealt{treu03} for details on the calculation). Such a distribution is consistent with that observed in an exhaustive
spectroscopic search for K+A galaxies in the massive $z\sim0.55$ cluster MACS J0717.5+3745 \citep{ma08} as well as published \citep{pwu14}
and preliminary investigations of cluster K+A distributions in other ORELSE fields. In contrast, such a (projected) concentration for the
two flavours of member K+As in SG0023 is not observed, as a majority of both types lie at larger ($R_{proj}>0.5R_{vir}$ $h_{70}^{-1}$ Mpc) radii from 
the group centres. In this scenario, strangulation or ram pressure stripping acts in a maintenance role that forces galaxies in the K+A phase
to persist in their quiescence after being acted upon by an (or several) initial, separate quenching mechanism(s).

Such cursory comparisons emphasise the importance of proper K+A selection in the task of 
determining their progenitors and the conditions necessary to invoke their presence. In this study we largely chose to proxy environment based on 
local overdensity. Since distributions of various populations, and, subsequently, inferences drawn from them, can be markedly different for different 
metrics of environment, and since our limited sample size precludes the possibility of further sub-dividing the samples presented here,  
we refrain here from attempting to compare this naive scenario to other more sophisticated scenarios for K+A
evolution (e.g., \citealt{muz14}). The lines of investigation presented in this study, in concert with other metrics of 
environment and measures of morphological properties, will be followed further upon completion of the full sample.  

\section{Conclusions}
\label{concl}

In this study we used MOSFIRE to target a large number ($\sim$100) of DEIMOS-selected traditional and potential K+A galaxies in and around two LSSs, SG0023 and RXJ1716, 
at $z\sim0.83$ drawn from the ORELSE survey. Through measuring or placing constraining limits on the H$\alpha$ emission and the [NII]/H$\alpha$ emission ratio, these 
observations 
were used to quantify the fraction of galaxies included in traditional ($z\ga0.3$) K+A selection which are actively forming stars and the number of galaxies 
missed by traditional K+A selection due to spuriously ascribing [OII] emission to star-formation processes. A sample of true K+A galaxies was formed comprised of 
two populations, those traditional K+As without significant H$\alpha$ emission or whose [NII]/H$\alpha$ ratios did not indicate ongoing star formation 
(K+A-H$\alpha$ galaxies) and those galaxies selected by our DEIMOS data to be starbursting for which we determined the dominant source powering the [OII] emission
was either LINER or Seyfert activity (KAIROS galaxies). This sample of true K+A galaxies was used to compare to a variety of different aspects of the 
traditional K+A populations. Here we list the most important results of these comparisons. 

\begin{itemize}
\renewcommand{\labelitemi}{$\bullet$}

\item The sample of traditional K+A selected in and around SG0023 and RXJ1716 was found to have 25\% impurity, with contamination coming from dusty starbursting 
galaxies. Based on scaling the statistics of our MOSFIRE sample to the entire DEIMOS sample at $0.775\le z \le 0.912$ we estimated that \emph{traditional K+A selection 
misses more than half of the true K+A population}. 

\item The traditional K+A fraction of our entire sample was found to be 7.7$\pm$1.2\% as compared to the purity/completeness-corrected KAIROS/K+A-H$\alpha$ fraction
of 15.7$\pm$1.7\%, with neither of these numbers changing significantly when calculating fractions to the stellar mass completeness limit of our DEIMOS sample. 

\item While the traditional K+A fraction was found to vary considerably across different large-scale environments, 4.4$\pm$2.0\%, 7.2$\pm$1.8\%, and 10.1$\pm$2.6\%, for 
field galaxies, group, and cluster members, respectively, the true K+A fraction was consistent with being constant across all large scale environments. However,
the number of KAIROS/K+A-H$\alpha$ galaxies relative to the number of galaxies actively forming stars, K+A/(SB+SF), though constant amongst field galaxies and 
group members (20.6$\pm$5.4 and 21.5$\pm$3.9\%, respectively), was found to be considerably higher for cluster members (29.3$\pm$6.6\%). 

\item Traditional K+A galaxies were found on average to both contain more stellar mass and exhibit a moderately redder rest-frame $M_{NUV}-M_{r^{\prime}}$ colour 
than KAIROS/K+A-H$\alpha$ galaxies. A combined fit of stacked DEIMOS spectra and optical/NIR photometry with a variety of stellar synthesis models revealed that 
the onset of the last major star-formation event in the average traditional K+A was appreciably earlier than that of the average KAIROS/K+A-H$\alpha$ galaxy 
($1.00^{+0.33}_{-0.21}$ Gyr vs. $0.75\pm0.19$ Gyr, respectively) indicating that the former population possibly descended from the latter. 

\item Relatively strong [OII] and [NeIII] emission observed in the stacked DEIMOS spectrum of KAIROS/K+A-H$\alpha$ galaxies (EW([OII])=-7.1$\pm$0.2\AA, 
$EW$([NeIII]=-0.7$\pm$0.1\AA) as well as the relatively high $\log(\rm{[NII]}/\rm{H}\alpha)$ ratio (-0.09$\pm$0.02) observed in the stacked MOSFIRE spectrum 
strongly indicated the presence of either stellar or active galactic nuclei feedback. The relatively low values of EW([OII])/EW(H$\alpha$) observed
for the bulk of the KAIROS sample along with the relatively high EW([NeIII])/EW([OII]) measured in their stacked DEIMOS spectra favored the latter possibility. 
While X-ray emission was generally not detected in the KAIROS/K+A-H$\alpha$
population either individually or in a stacked analysis, the size of our current sample, the depth of our X-ray imaging, and the cursory nature of this analysis 
did not allow us to rule out pervasive X-ray AGN activity in these galaxies.

\item When analysing the local overdensity distributions of (predominantly) member galaxies of the SG0023 supergroup and the massive RXJ1716 X-ray cluster we
found that the distributions of KAIROS/K+A-H$\alpha$ were consistent with tracing the starbursting population and generally avoided the regions of highest (local)
overdensity in both LSSs. In SG0023 the overdensity distribution of both the traditional K+As and the KAIROS/K+A-H$\alpha$ galaxies were inconsistent with that
of quiescent galaxies, while in RXJ1716 this distribution of K+As was indistinguishable from that of the quiescent population. 

\end{itemize}

These lines of evidence were used to formulate a scenario in which true K+A galaxies evolve in a different manner in lower density large scale environments 
(field and groups) than in cluster environments. In all cases, the K+A phase appears in galaxies inhabiting regions of moderate local overdensity relatively 
far removed from the cores of the groups or cluster. Such a distribution points to galaxy-galaxy interactions or mergers as \emph{inducers} of the K+A phase rather
than cluster- or massive group-specific processes such as ram pressure stripping. However, there does appear to be a signature of the latter processes at work on
K+A galaxies in our data which is more subtle. Feedback from stellar or other processes is maintained for a considerable time after the cessation of star 
formation in the KAIROS/K+A-H$\alpha$ galaxies ($\sim700$ Myr). Such feedback along with any initial stronger feedback associated with this cessation serves to increase the 
thermal and kinetic energy in any remaining gas reservoir. In the field and in groups, those KAIROS/K+A-H$\alpha$ galaxies whose feedback processes combined 
with any kinematic effects associated with the merger/interaction do not accelerate the gas to the escape velocity are largely allowed to retain their 
diffuse reservoirs of gas, as the effects of ram pressure stripping and strangulation are minimal or nonexistent in such environments. Such a scenario 
allows for the possibility of a cyclic K+A phase, a scenario consistent with the concordant fraction of starbursting, star-forming, and true K+A galaxies
in these environments as well as the disparate overdensity distributions of traditional K+As and quiescent galaxies. In cluster
environments, however, the effectiveness of mechanisms related to the stripping of hot or cold gas are enhanced by the decrease in the binding energy of that gas. As such, 
the K+A phase is rather a precursor to quiescence, with ram pressure stripping and/or strangulation acting in a \emph{preventative} role that precludes the possibility reignited star 
formation. Such a scenario consistent both with the observed overdensity distribution of traditional K+As and the elevated K+A/(SB+SF)
fraction in RXJ1716. Future planned MOSFIRE observations of these fields and other LSSs in ORELSE along with cross-correlations of the 
wealth of multiwavelength data in the ORELSE survey will seek to calcify or reject this scenario. Deep, high resolution imaging from the Atacama Large 
Millimeter/submillimeter Array (ALMA; \citealt{wootten09}) as well as observations of large sample of K+A galaxies from adaptive-optics-fed 
integral field unit (IFU) spectrometers, such as the one being developed for the Multi-Unit Spectroscopic Explorer (MUSE; \citealt{bacon10}) 
can also be extremely useful in challenging K+A formation scenarios (as in, e.g., \citealt{bekki05}) and how they evolve across different environments. 

\section*{Acknowledgements}

{\footnotesize This material is based upon work supported by the National Science Foundation under Grant No. 1411943. Part of the work 
presented herein is supported by NASA Grant Number NNX15AK92G. BCL thanks Nelson
Cheung and Stephen Lampa for laying the foundation for this study and for looking through thousands of DEIMOS spectra for little glory and even less 
pay. BCL gratefully acknowledges Romain Thomas for discussions and guidance related to age estimates and Alison Mansheim and Lu Shen for discussions
that improved the paper. We also thank the anonymous referee for helpful and careful guidance. This study is based, in part, on data collected 
at the Subaru Telescope and obtained from the SMOKA, which is operated by the Astronomy Data Center, National Astronomical Observatory of 
Japan. This work is based, in part, on observations made with the Spitzer Space Telescope, which is operated by the Jet Propulsion Laboratory, California 
Institute of Technology under a contract with NASA. UKIRT is supported by NASA and operated under an agreement among the University of Hawaii, the 
University of Arizona, and Lockheed Martin Advanced Technology Center; operations are enabled through the cooperation of the East Asian Observatory.  
When the data reported here were acquired, UKIRT was operated by the Joint Astronomy Centre on behalf of the Science and Technology Facilities Council 
of the U.K. This study is also based, in part, on observations obtained with WIRCam, a joint project of CFHT, Taiwan, Korea, Canada, France, and the 
Canada-France-Hawaii Telescope which is operated by the National Research Council (NRC) of Canada, the Institut National des Sciences de l'Univers 
of the Centre National de la Recherche Scientifique of France, and the University of Hawai'i. The scientific results reported in this article are based in part 
on observations made by the Chandra X-ray Observatory and data obtained from the Chandra Data Archive. The spectrographic data presented herein were 
obtained at the W.M. Keck Observatory, which is operated as a scientific partnership among the California Institute of Technology, the University of 
California, and the National Aeronautics and Space Administration. The Observatory was made possible by the generous financial support of the W.M. Keck 
Foundation. We wish to thank the indigenous Hawaiian community for allowing us to be guests on their sacred mountain, a privilege, without which, this 
work would not have been possible. We are most fortunate to be able to conduct observations from this site.}





\bibliographystyle{mnras}
\bibliography{chronosnKAIROS} 

\begin{thebibliography}{}
\makeatletter
\relax
\def\mn@urlcharsother{\let\do\@makeother \do\$\do\&\do\#\do\^\do\_\do\%\do\~}
\def\mn@doi{\begingroup\mn@urlcharsother \@ifnextchar [ {\mn@doi@}
  {\mn@doi@[]}}
\def\mn@doi@[#1]#2{\def\@tempa{#1}\ifx\@tempa\@empty \href
  {http://dx.doi.org/#2} {doi:#2}\else \href {http://dx.doi.org/#2} {#1}\fi
  \endgroup}
\def\mn@eprint#1#2{\mn@eprint@#1:#2::\@nil}
\def\mn@eprint@arXiv#1{\href {http://arxiv.org/abs/#1} {{\tt arXiv:#1}}}
\def\mn@eprint@dblp#1{\href {http://dblp.uni-trier.de/rec/bibtex/#1.xml}
  {dblp:#1}}
\def\mn@eprint@#1:#2:#3:#4\@nil{\def\@tempa {#1}\def\@tempb {#2}\def\@tempc
  {#3}\ifx \@tempc \@empty \let \@tempc \@tempb \let \@tempb \@tempa \fi \ifx
  \@tempb \@empty \def\@tempb {arXiv}\fi \@ifundefined
  {mn@eprint@\@tempb}{\@tempb:\@tempc}{\expandafter \expandafter \csname
  mn@eprint@\@tempb\endcsname \expandafter{\@tempc}}}

\bibitem[\protect\citeauthoryear{{Alatalo} et~al.,}{{Alatalo}
  et~al.}{2016}]{alatalo16}
{Alatalo} K.,  et~al., 2016, \mn@doi [\apjs] {10.3847/0067-0049/224/2/38},
  \href {http://adsabs.harvard.edu/abs/2016ApJS..224...38A} {224, 38}

\bibitem[\protect\citeauthoryear{{Alberts} et~al.,}{{Alberts}
  et~al.}{2016}]{alberts16}
{Alberts} S.,  et~al., 2016, \mn@doi [\apj] {10.3847/0004-637X/825/1/72}, \href
  {http://adsabs.harvard.edu/abs/2016ApJ...825...72A} {825, 72}

\bibitem[\protect\citeauthoryear{{Ascaso}, {Lemaux}, {Lubin}, {Gal},
  {Kocevski}, {Rumbaugh}  \& {Squires}}{{Ascaso} et~al.}{2014}]{begona14}
{Ascaso} B.,  {Lemaux} B.~C.,  {Lubin} L.~M.,  {Gal} R.~R.,  {Kocevski} D.~D.,
  {Rumbaugh} N.,   {Squires} G.,  2014, \mn@doi [\mnras]
  {10.1093/mnras/stu877}, \href
  {http://adsabs.harvard.edu/abs/2014MNRAS.442..589A} {442, 589}

\bibitem[\protect\citeauthoryear{{Bacon} et~al.,}{{Bacon}
  et~al.}{2010}]{bacon10}
{Bacon} R.,  et~al., 2010, in Ground-based and Airborne Instrumentation for
  Astronomy III. p. 773508, \mn@doi{10.1117/12.856027}

\bibitem[\protect\citeauthoryear{{Bah{\'e}} \& {McCarthy}}{{Bah{\'e}} \&
  {McCarthy}}{2015}]{bahe15}
{Bah{\'e}} Y.~M.,  {McCarthy} I.~G.,  2015, \mn@doi [\mnras]
  {10.1093/mnras/stu2293}, \href
  {http://adsabs.harvard.edu/abs/2015MNRAS.447..969B} {447, 969}

\bibitem[\protect\citeauthoryear{{Baldwin}, {Phillips}  \&
  {Terlevich}}{{Baldwin} et~al.}{1981}]{BPT81}
{Baldwin} J.~A.,  {Phillips} M.~M.,   {Terlevich} R.,  1981, \mn@doi [\pasp]
  {10.1086/130766}, \href {http://adsabs.harvard.edu/abs/1981PASP...93....5B}
  {93, 5}

\bibitem[\protect\citeauthoryear{{Balogh}, {Morris}, {Yee}, {Carlberg}  \&
  {Ellingson}}{{Balogh} et~al.}{1999}]{balogh99}
{Balogh} M.~L.,  {Morris} S.~L.,  {Yee} H.~K.~C.,  {Carlberg} R.~G.,
  {Ellingson} E.,  1999, \mn@doi [\apj] {10.1086/308056}, \href
  {http://adsabs.harvard.edu/abs/1999ApJ...527...54B} {527, 54}

\bibitem[\protect\citeauthoryear{{Balogh}, {Miller}, {Nichol}, {Zabludoff}  \&
  {Goto}}{{Balogh} et~al.}{2005}]{balogh05}
{Balogh} M.~L.,  {Miller} C.,  {Nichol} R.,  {Zabludoff} A.,   {Goto} T.,
  2005, \mn@doi [\mnras] {10.1111/j.1365-2966.2005.09047.x}, \href
  {http://adsabs.harvard.edu/abs/2005MNRAS.360..587B} {360, 587}

\bibitem[\protect\citeauthoryear{{Balogh} et~al.,}{{Balogh}
  et~al.}{2011}]{balogh11}
{Balogh} M.~L.,  et~al., 2011, \mn@doi [\mnras]
  {10.1111/j.1365-2966.2010.18052.x}, \href
  {http://adsabs.harvard.edu/abs/2011MNRAS.412.2303B} {412, 2303}

\bibitem[\protect\citeauthoryear{{Balogh} et~al.,}{{Balogh}
  et~al.}{2016}]{balogh16}
{Balogh} M.~L.,  et~al., 2016, \mn@doi [\mnras] {10.1093/mnras/stv2949}, \href
  {http://adsabs.harvard.edu/abs/2016MNRAS.456.4364B} {456, 4364}

\bibitem[\protect\citeauthoryear{{Bartholomew}, {Rose}, {Gaba}  \&
  {Caldwell}}{{Bartholomew} et~al.}{2001}]{bartho01}
{Bartholomew} L.~J.,  {Rose} J.~A.,  {Gaba} A.~E.,   {Caldwell} N.,  2001,
  \mn@doi [\aj] {10.1086/324229}, \href
  {http://adsabs.harvard.edu/abs/2001AJ....122.2913B} {122, 2913}

\bibitem[\protect\citeauthoryear{{Bekki}, {Couch}, {Shioya}  \&
  {Vazdekis}}{{Bekki} et~al.}{2005}]{bekki05}
{Bekki} K.,  {Couch} W.~J.,  {Shioya} Y.,   {Vazdekis} A.,  2005, \mn@doi
  [\mnras] {10.1111/j.1365-2966.2005.08932.x}, \href
  {http://adsabs.harvard.edu/abs/2005MNRAS.359..949B} {359, 949}

\bibitem[\protect\citeauthoryear{{Belfiore} et~al.,}{{Belfiore}
  et~al.}{2016}]{bel16}
{Belfiore} F.,  et~al., 2016, \mn@doi [\mnras] {10.1093/mnras/stw1234}, \href
  {http://adsabs.harvard.edu/abs/2016MNRAS.tmp..899B} {}

\bibitem[\protect\citeauthoryear{{Belloni}, {Bruzual}, {Thimm}  \&
  {Roser}}{{Belloni} et~al.}{1995}]{belloni95}
{Belloni} P.,  {Bruzual} A.~G.,  {Thimm} G.~J.,   {Roser} H.-J.,  1995, \aap,
  \href {http://adsabs.harvard.edu/abs/1995A%26A...297...61B} {297, 61}

\bibitem[\protect\citeauthoryear{{Bertin} \& {Arnouts}}{{Bertin} \&
  {Arnouts}}{1996}]{BertinArn96}
{Bertin} E.,  {Arnouts} S.,  1996, \mn@doi [\aaps] {10.1051/aas:1996164}, \href
  {http://adsabs.harvard.edu/abs/1996A%26AS..117..393B} {117, 393}

\bibitem[\protect\citeauthoryear{{Binette}, {Magris}, {Stasi{\'n}ska}  \&
  {Bruzual}}{{Binette} et~al.}{1994}]{binette94}
{Binette} L.,  {Magris} C.~G.,  {Stasi{\'n}ska} G.,   {Bruzual} A.~G.,  1994,
  \aap, \href {http://adsabs.harvard.edu/abs/1994A%26A...292...13B} {292, 13}

\bibitem[\protect\citeauthoryear{{Biviano}, {Katgert}, {Thomas}  \&
  {Adami}}{{Biviano} et~al.}{2002}]{biviano02}
{Biviano} A.,  {Katgert} P.,  {Thomas} T.,   {Adami} C.,  2002, \mn@doi [\aap]
  {10.1051/0004-6361:20020340}, \href
  {http://adsabs.harvard.edu/abs/2002A%26A...387....8B} {387, 8}

\bibitem[\protect\citeauthoryear{{Brammer}, {van Dokkum}  \& {Coppi}}{{Brammer}
  et~al.}{2008}]{brammer08}
{Brammer} G.~B.,  {van Dokkum} P.~G.,   {Coppi} P.,  2008, \mn@doi [\apj]
  {10.1086/591786}, \href {http://adsabs.harvard.edu/abs/2008ApJ...686.1503B}
  {686, 1503}

\bibitem[\protect\citeauthoryear{{Brammer} et~al.,}{{Brammer}
  et~al.}{2011}]{brammer11}
{Brammer} G.~B.,  et~al., 2011, \mn@doi [\apj] {10.1088/0004-637X/739/1/24},
  \href {http://adsabs.harvard.edu/abs/2011ApJ...739...24B} {739, 24}

\bibitem[\protect\citeauthoryear{{Brown} et~al.,}{{Brown}
  et~al.}{2009}]{brown09}
{Brown} M.~J.~I.,  et~al., 2009, \mn@doi [\apj] {10.1088/0004-637X/703/1/150},
  \href {http://adsabs.harvard.edu/abs/2009ApJ...703..150B} {703, 150}

\bibitem[\protect\citeauthoryear{{Bruzual}}{{Bruzual}}{2007}]{CB07}
{Bruzual} G.,  2007, in {Vallenari} A.,  {Tantalo} R.,  {Portinari} L.,
  {Moretti} A.,  eds,  Astronomical Society of the Pacific Conference Series
  Vol. 374, From Stars to Galaxies: Building the Pieces to Build Up the
  Universe. p.~303 (\mn@eprint {} {astro-ph/0702091})

\bibitem[\protect\citeauthoryear{{Bruzual} \& {Charlot}}{{Bruzual} \&
  {Charlot}}{2003}]{bc03}
{Bruzual} G.,  {Charlot} S.,  2003, \mn@doi [\mnras]
  {10.1046/j.1365-8711.2003.06897.x}, \href
  {http://adsabs.harvard.edu/abs/2003MNRAS.344.1000B} {344, 1000}

\bibitem[\protect\citeauthoryear{{Bundy} et~al.,}{{Bundy}
  et~al.}{2006}]{bundy06}
{Bundy} K.,  et~al., 2006, \mn@doi [\apj] {10.1086/507456}, \href
  {http://adsabs.harvard.edu/abs/2006ApJ...651..120B} {651, 120}

\bibitem[\protect\citeauthoryear{{Bundy} et~al.,}{{Bundy}
  et~al.}{2010}]{bundy10}
{Bundy} K.,  et~al., 2010, \mn@doi [\apj] {10.1088/0004-637X/719/2/1969}, \href
  {http://adsabs.harvard.edu/abs/2010ApJ...719.1969B} {719, 1969}

\bibitem[\protect\citeauthoryear{{Butcher} \& {Oemler}}{{Butcher} \&
  {Oemler}}{1978}]{bo1978}
{Butcher} H.,  {Oemler} Jr. A.,  1978, \mn@doi [\apj] {10.1086/156640}, \href
  {http://adsabs.harvard.edu/abs/1978ApJ...226..559B} {226, 559}

\bibitem[\protect\citeauthoryear{{Calzetti}, {Armus}, {Bohlin}, {Kinney},
  {Koornneef}  \& {Storchi-Bergmann}}{{Calzetti} et~al.}{2000}]{calz00}
{Calzetti} D.,  {Armus} L.,  {Bohlin} R.~C.,  {Kinney} A.~L.,  {Koornneef} J.,
   {Storchi-Bergmann} T.,  2000, \mn@doi [\apj] {10.1086/308692}, \href
  {http://adsabs.harvard.edu/abs/2000ApJ...533..682C} {533, 682}

\bibitem[\protect\citeauthoryear{{Carlberg}, {Morris}, {Yee}  \&
  {Ellingson}}{{Carlberg} et~al.}{1997}]{carlberg97}
{Carlberg} R.~G.,  {Morris} S.~L.,  {Yee} H.~K.~C.,   {Ellingson} E.,  1997,
  \mn@doi [\apjl] {10.1086/310577}, \href
  {http://adsabs.harvard.edu/abs/1997ApJ...479L..19C} {479, L19}

\bibitem[\protect\citeauthoryear{{Casali} et~al.,}{{Casali}
  et~al.}{2007}]{casali07}
{Casali} M.,  et~al., 2007, \mn@doi [\aap] {10.1051/0004-6361:20066514}, \href
  {http://adsabs.harvard.edu/abs/2007A%26A...467..777C} {467, 777}

\bibitem[\protect\citeauthoryear{{Chabrier}}{{Chabrier}}{2003}]{chab03}
{Chabrier} G.,  2003, \mn@doi [\pasp] {10.1086/376392}, \href
  {http://adsabs.harvard.edu/abs/2003PASP..115..763C} {115, 763}

\bibitem[\protect\citeauthoryear{{Chang}, {van Gorkom}, {Zabludoff}, {Zaritsky}
   \& {Mihos}}{{Chang} et~al.}{2001}]{chang01}
{Chang} T.-C.,  {van Gorkom} J.~H.,  {Zabludoff} A.~I.,  {Zaritsky} D.,
  {Mihos} J.~C.,  2001, \mn@doi [\aj] {10.1086/319959}, \href
  {http://adsabs.harvard.edu/abs/2001AJ....121.1965C} {121, 1965}

\bibitem[\protect\citeauthoryear{{Clowe}, {Luppino}, {Kaiser}, {Henry}  \&
  {Gioia}}{{Clowe} et~al.}{1998}]{clowe98}
{Clowe} D.,  {Luppino} G.~A.,  {Kaiser} N.,  {Henry} J.~P.,   {Gioia} I.~M.,
  1998, \mn@doi [\apjl] {10.1086/311285}, \href
  {http://adsabs.harvard.edu/abs/1998ApJ...497L..61C} {497, L61}

\bibitem[\protect\citeauthoryear{{Cooke} et~al.,}{{Cooke}
  et~al.}{2016}]{cooke16}
{Cooke} E.~A.,  et~al., 2016, \mn@doi [\apj] {10.3847/0004-637X/816/2/83},
  \href {http://adsabs.harvard.edu/abs/2016ApJ...816...83C} {816, 83}

\bibitem[\protect\citeauthoryear{{Cooper} et~al.,}{{Cooper}
  et~al.}{2007}]{mcoopz07}
{Cooper} M.~C.,  et~al., 2007, \mn@doi [\mnras]
  {10.1111/j.1365-2966.2007.11534.x}, \href
  {http://adsabs.harvard.edu/abs/2007MNRAS.376.1445C} {376, 1445}

\bibitem[\protect\citeauthoryear{{Couch} \& {Sharples}}{{Couch} \&
  {Sharples}}{1987}]{couch1987}
{Couch} W.~J.,  {Sharples} R.~M.,  1987, \mn@doi [\mnras]
  {10.1093/mnras/229.3.423}, \href
  {http://adsabs.harvard.edu/abs/1987MNRAS.229..423C} {229, 423}

\bibitem[\protect\citeauthoryear{{Cucciati} et~al.,}{{Cucciati}
  et~al.}{2016}]{olga16}
{Cucciati} O.,  et~al., 2016, preprint, \href
  {http://adsabs.harvard.edu/abs/2016arXiv161107049C} {} (\mn@eprint {arXiv}
  {1611.07049})

\bibitem[\protect\citeauthoryear{{Darvish}, {Mobasher}, {Sobral}, {Scoville}
  \& {Aragon-Calvo}}{{Darvish} et~al.}{2015}]{darvish15}
{Darvish} B.,  {Mobasher} B.,  {Sobral} D.,  {Scoville} N.,   {Aragon-Calvo}
  M.,  2015, \mn@doi [\apj] {10.1088/0004-637X/805/2/121}, \href
  {http://adsabs.harvard.edu/abs/2015ApJ...805..121D} {805, 121}

\bibitem[\protect\citeauthoryear{{Darvish}, {Mobasher}, {Sobral}, {Rettura},
  {Scoville}, {Faisst}  \& {Capak}}{{Darvish} et~al.}{2016}]{darvish16}
{Darvish} B.,  {Mobasher} B.,  {Sobral} D.,  {Rettura} A.,  {Scoville} N.,
  {Faisst} A.,   {Capak} P.,  2016, \mn@doi [\apj]
  {10.3847/0004-637X/825/2/113}, \href
  {http://adsabs.harvard.edu/abs/2016ApJ...825..113D} {825, 113}

\bibitem[\protect\citeauthoryear{{Davidzon} et~al.,}{{Davidzon}
  et~al.}{2013}]{iary13}
{Davidzon} I.,  et~al., 2013, \mn@doi [\aap] {10.1051/0004-6361/201321511},
  \href {http://adsabs.harvard.edu/abs/2013A%26A...558A..23D} {558, A23}

\bibitem[\protect\citeauthoryear{{Davis} et~al.,}{{Davis}
  et~al.}{2003}]{davis03}
{Davis} M.,  et~al., 2003, in {Guhathakurta} P.,  ed.,  \procspie Vol. 4834,
  Discoveries and Research Prospects from 6- to 10-Meter-Class Telescopes II.
  pp 161--172 (\mn@eprint {} {astro-ph/0209419}), \mn@doi{10.1117/12.457897}

\bibitem[\protect\citeauthoryear{{De Propris} \& {Melnick}}{{De Propris} \&
  {Melnick}}{2014}]{depropris14}
{De Propris} R.,  {Melnick} J.,  2014, \mn@doi [\mnras] {10.1093/mnras/stu141},
  \href {http://adsabs.harvard.edu/abs/2014MNRAS.439.2837D} {439, 2837}

\bibitem[\protect\citeauthoryear{{De Propris}, {Phillipps}  \& {Bremer}}{{De
  Propris} et~al.}{2013}]{depropris13}
{De Propris} R.,  {Phillipps} S.,   {Bremer} M.~N.,  2013, \mn@doi [\mnras]
  {10.1093/mnras/stt1262}, \href
  {http://adsabs.harvard.edu/abs/2013MNRAS.434.3469D} {434, 3469}

\bibitem[\protect\citeauthoryear{{Deng}, {Chen}  \& {Jiang}}{{Deng}
  et~al.}{2011}]{deng11}
{Deng} X.-F.,  {Chen} Y.-Q.,   {Jiang} P.,  2011, \mn@doi [\mnras]
  {10.1111/j.1365-2966.2011.19277.x}, \href
  {http://adsabs.harvard.edu/abs/2011MNRAS.417..453D} {417, 453}

\bibitem[\protect\citeauthoryear{{Dey}, {Lee}, {Reddy}, {Cooper}, {Inami},
  {Hong}, {Gonzalez}  \& {Jannuzi}}{{Dey} et~al.}{2016}]{dey16}
{Dey} A.,  {Lee} K.-S.,  {Reddy} N.,  {Cooper} M.,  {Inami} H.,  {Hong} S.,
  {Gonzalez} A.~H.,   {Jannuzi} B.~T.,  2016, \mn@doi [\apj]
  {10.3847/0004-637X/823/1/11}, \href
  {http://adsabs.harvard.edu/abs/2016ApJ...823...11D} {823, 11}

\bibitem[\protect\citeauthoryear{{Dopita} \& {Sutherland}}{{Dopita} \&
  {Sutherland}}{1995}]{dopita95}
{Dopita} M.~A.,  {Sutherland} R.~S.,  1995, \mn@doi [\apj] {10.1086/176596},
  \href {http://adsabs.harvard.edu/abs/1995ApJ...455..468D} {455, 468}

\bibitem[\protect\citeauthoryear{{Dressler}}{{Dressler}}{1980}]{dressler80}
{Dressler} A.,  1980, \mn@doi [\apj] {10.1086/157753}, \href
  {http://adsabs.harvard.edu/abs/1980ApJ...236..351D} {236, 351}

\bibitem[\protect\citeauthoryear{{Dressler} \& {Gunn}}{{Dressler} \&
  {Gunn}}{1983}]{dressler1983}
{Dressler} A.,  {Gunn} J.~E.,  1983, \mn@doi [\apj] {10.1086/161093}, \href
  {http://adsabs.harvard.edu/abs/1983ApJ...270....7D} {270, 7}

\bibitem[\protect\citeauthoryear{{Dressler} \& {Gunn}}{{Dressler} \&
  {Gunn}}{1992}]{dressler1992}
{Dressler} A.,  {Gunn} J.~E.,  1992, \mn@doi [\apjs] {10.1086/191620}, \href
  {http://adsabs.harvard.edu/abs/1992ApJS...78....1D} {78, 1}

\bibitem[\protect\citeauthoryear{{Dressler}, {Smail}, {Poggianti}, {Butcher},
  {Couch}, {Ellis}  \& {Oemler}}{{Dressler} et~al.}{1999}]{dressler99}
{Dressler} A.,  {Smail} I.,  {Poggianti} B.~M.,  {Butcher} H.,  {Couch} W.~J.,
  {Ellis} R.~S.,   {Oemler} Jr. A.,  1999, \mn@doi [\apjs] {10.1086/313213},
  \href {http://adsabs.harvard.edu/abs/1999ApJS..122...51D} {122, 51}

\bibitem[\protect\citeauthoryear{{Dressler}, {Oemler}, {Poggianti}, {Gladders},
  {Abramson}  \& {Vulcani}}{{Dressler} et~al.}{2013}]{dressler13}
{Dressler} A.,  {Oemler} Jr. A.,  {Poggianti} B.~M.,  {Gladders} M.~D.,
  {Abramson} L.,   {Vulcani} B.,  2013, \mn@doi [\apj]
  {10.1088/0004-637X/770/1/62}, \href
  {http://adsabs.harvard.edu/abs/2013ApJ...770...62D} {770, 62}

\bibitem[\protect\citeauthoryear{{Elvis} et~al.,}{{Elvis}
  et~al.}{2009}]{elvis09}
{Elvis} M.,  et~al., 2009, \mn@doi [\apjs] {10.1088/0067-0049/184/1/158}, \href
  {http://adsabs.harvard.edu/abs/2009ApJS..184..158E} {184, 158}

\bibitem[\protect\citeauthoryear{{Ettori}, {Tozzi}, {Borgani}  \&
  {Rosati}}{{Ettori} et~al.}{2004}]{ettori04}
{Ettori} S.,  {Tozzi} P.,  {Borgani} S.,   {Rosati} P.,  2004, \mn@doi [\aap]
  {10.1051/0004-6361:20034119}, \href
  {http://adsabs.harvard.edu/abs/2004A%26A...417...13E} {417, 13}

\bibitem[\protect\citeauthoryear{{Faber} et~al.,}{{Faber} et~al.}{2003}]{fab03}
{Faber} S.~M.,  et~al., 2003, in {Iye} M.,  {Moorwood} A.~F.~M.,  eds,  Society
  of Photo-Optical Instrumentation Engineers (SPIE) Conference Series Vol.
  4841, Instrument Design and Performance for Optical/Infrared Ground-based
  Telescopes. pp 1657--1669, \mn@doi{10.1117/12.460346}

\bibitem[\protect\citeauthoryear{{Faber} et~al.,}{{Faber}
  et~al.}{2007}]{faber07}
{Faber} S.~M.,  et~al., 2007, \mn@doi [\apj] {10.1086/519294}, \href
  {http://adsabs.harvard.edu/abs/2007ApJ...665..265F} {665, 265}

\bibitem[\protect\citeauthoryear{{Falkenberg}, {Kotulla}  \&
  {Fritze}}{{Falkenberg} et~al.}{2009}]{falkenberg09}
{Falkenberg} M.~A.,  {Kotulla} R.,   {Fritze} U.,  2009, \mn@doi [\mnras]
  {10.1111/j.1365-2966.2009.14416.x}, \href
  {http://adsabs.harvard.edu/abs/2009MNRAS.397.1940F} {397, 1940}

\bibitem[\protect\citeauthoryear{{Fazio} et~al.,}{{Fazio}
  et~al.}{2004}]{fazio04}
{Fazio} G.~G.,  et~al., 2004, \mn@doi [\apjs] {10.1086/422843}, \href
  {http://adsabs.harvard.edu/abs/2004ApJS..154...10F} {154, 10}

\bibitem[\protect\citeauthoryear{{Fioc} \& {Rocca-Volmerange}}{{Fioc} \&
  {Rocca-Volmerange}}{1997}]{fioc97}
{Fioc} M.,  {Rocca-Volmerange} B.,  1997, \aap, \href
  {http://adsabs.harvard.edu/abs/1997A%26A...326..950F} {326, 950}

\bibitem[\protect\citeauthoryear{{French}, {Yang}, {Zabludoff}, {Narayanan},
  {Shirley}, {Walter}, {Smith}  \& {Tremonti}}{{French}
  et~al.}{2015}]{french15}
{French} K.~D.,  {Yang} Y.,  {Zabludoff} A.,  {Narayanan} D.,  {Shirley} Y.,
  {Walter} F.,  {Smith} J.-D.,   {Tremonti} C.~A.,  2015, \mn@doi [\apj]
  {10.1088/0004-637X/801/1/1}, \href
  {http://adsabs.harvard.edu/abs/2015ApJ...801....1F} {801, 1}

\bibitem[\protect\citeauthoryear{{Fukugita}, {Ichikawa}, {Gunn}, {Doi},
  {Shimasaku}  \& {Schneider}}{{Fukugita} et~al.}{1996}]{fukugita96}
{Fukugita} M.,  {Ichikawa} T.,  {Gunn} J.~E.,  {Doi} M.,  {Shimasaku} K.,
  {Schneider} D.~P.,  1996, \mn@doi [\aj] {10.1086/117915}, \href
  {http://adsabs.harvard.edu/abs/1996AJ....111.1748F} {111, 1748}

\bibitem[\protect\citeauthoryear{{Fumagalli} et~al.,}{{Fumagalli}
  et~al.}{2016}]{fumagalli16}
{Fumagalli} M.,  et~al., 2016, \mn@doi [\apj] {10.3847/0004-637X/822/1/1},
  \href {http://adsabs.harvard.edu/abs/2016ApJ...822....1F} {822, 1}

\bibitem[\protect\citeauthoryear{{Gal}, {Lemaux}, {Lubin}, {Kocevski}  \&
  {Squires}}{{Gal} et~al.}{2008}]{gal08}
{Gal} R.~R.,  {Lemaux} B.~C.,  {Lubin} L.~M.,  {Kocevski} D.,   {Squires}
  G.~K.,  2008, \mn@doi [\apj] {10.1086/590416}, \href
  {http://adsabs.harvard.edu/abs/2008ApJ...684..933G} {684, 933}

\bibitem[\protect\citeauthoryear{{Garmire}, {Bautz}, {Ford}, {Nousek}  \&
  {Ricker}}{{Garmire} et~al.}{2003}]{garmire03}
{Garmire} G.~P.,  {Bautz} M.~W.,  {Ford} P.~G.,  {Nousek} J.~A.,   {Ricker} Jr.
  G.~R.,  2003, in {Truemper} J.~E.,  {Tananbaum} H.~D.,  eds,  \procspie Vol.
  4851, X-Ray and Gamma-Ray Telescopes and Instruments for Astronomy.. pp
  28--44, \mn@doi{10.1117/12.461599}

\bibitem[\protect\citeauthoryear{{G{\'o}mez} et~al.,}{{G{\'o}mez}
  et~al.}{2003}]{gomez03}
{G{\'o}mez} P.~L.,  et~al., 2003, \mn@doi [\apj] {10.1086/345593}, \href
  {http://adsabs.harvard.edu/abs/2003ApJ...584..210G} {584, 210}

\bibitem[\protect\citeauthoryear{{Goto} et~al.,}{{Goto}
  et~al.}{2003a}]{goto03b}
{Goto} T.,  et~al., 2003a, \mn@doi [\pasj] {10.1093/pasj/55.4.739}, \href
  {http://adsabs.harvard.edu/abs/2003PASJ...55..739G} {55, 739}

\bibitem[\protect\citeauthoryear{{Goto}, {Yamauchi}, {Fujita}, {Okamura},
  {Sekiguchi}, {Smail}, {Bernardi}  \& {Gomez}}{{Goto} et~al.}{2003b}]{goto03a}
{Goto} T.,  {Yamauchi} C.,  {Fujita} Y.,  {Okamura} S.,  {Sekiguchi} M.,
  {Smail} I.,  {Bernardi} M.,   {Gomez} P.~L.,  2003b, \mn@doi [\mnras]
  {10.1046/j.1365-2966.2003.07114.x}, \href
  {http://adsabs.harvard.edu/abs/2003MNRAS.346..601G} {346, 601}

\bibitem[\protect\citeauthoryear{{Grazian} et~al.,}{{Grazian}
  et~al.}{2006}]{grazian06}
{Grazian} A.,  et~al., 2006, \mn@doi [\aap] {10.1051/0004-6361:20053979}, \href
  {http://adsabs.harvard.edu/abs/2006A%26A...449..951G} {449, 951}

\bibitem[\protect\citeauthoryear{{Gunn} \& {Gott}}{{Gunn} \&
  {Gott}}{1972}]{gunn72}
{Gunn} J.~E.,  {Gott} III J.~R.,  1972, \mn@doi [\apj] {10.1086/151605}, \href
  {http://adsabs.harvard.edu/abs/1972ApJ...176....1G} {176, 1}

\bibitem[\protect\citeauthoryear{{Haines} et~al.,}{{Haines}
  et~al.}{2012}]{haines12}
{Haines} C.~P.,  et~al., 2012, \mn@doi [\apj] {10.1088/0004-637X/754/2/97},
  \href {http://adsabs.harvard.edu/abs/2012ApJ...754...97H} {754, 97}

\bibitem[\protect\citeauthoryear{{Hansen}, {Sheldon}, {Wechsler}  \&
  {Koester}}{{Hansen} et~al.}{2009}]{hansen09}
{Hansen} S.~M.,  {Sheldon} E.~S.,  {Wechsler} R.~H.,   {Koester} B.~P.,  2009,
  \mn@doi [\apj] {10.1088/0004-637X/699/2/1333}, \href
  {http://adsabs.harvard.edu/abs/2009ApJ...699.1333H} {699, 1333}

\bibitem[\protect\citeauthoryear{{Heckman}}{{Heckman}}{1981}]{heckman81}
{Heckman} T.~M.,  1981, \mn@doi [\apjl] {10.1086/183674}, \href
  {http://adsabs.harvard.edu/abs/1981ApJ...250L..59H} {250, L59}

\bibitem[\protect\citeauthoryear{{Heckman}, {Baum}, {van Breugel}  \&
  {McCarthy}}{{Heckman} et~al.}{1989}]{heckman89}
{Heckman} T.~M.,  {Baum} S.~A.,  {van Breugel} W.~J.~M.,   {McCarthy} P.,
  1989, \mn@doi [\apj] {10.1086/167181}, \href
  {http://adsabs.harvard.edu/abs/1989ApJ...338...48H} {338, 48}

\bibitem[\protect\citeauthoryear{{Hopkins}, {Hernquist}, {Cox}  \& {Kere{\v
  s}}}{{Hopkins} et~al.}{2008}]{hopkins08}
{Hopkins} P.~F.,  {Hernquist} L.,  {Cox} T.~J.,   {Kere{\v s}} D.,  2008,
  \mn@doi [\apjs] {10.1086/524362}, \href
  {http://adsabs.harvard.edu/abs/2008ApJS..175..356H} {175, 356}

\bibitem[\protect\citeauthoryear{{Ilbert} et~al.,}{{Ilbert}
  et~al.}{2006}]{dreadOlivier06}
{Ilbert} O.,  et~al., 2006, \mn@doi [\aap] {10.1051/0004-6361:20065138}, \href
  {http://adsabs.harvard.edu/abs/2006A%26A...457..841I} {457, 841}

\bibitem[\protect\citeauthoryear{{Ilbert} et~al.,}{{Ilbert}
  et~al.}{2010}]{ilbert10}
{Ilbert} O.,  et~al., 2010, \mn@doi [\apj] {10.1088/0004-637X/709/2/644}, \href
  {http://adsabs.harvard.edu/abs/2010ApJ...709..644I} {709, 644}

\bibitem[\protect\citeauthoryear{{Ilbert} et~al.,}{{Ilbert}
  et~al.}{2013}]{ilbert13}
{Ilbert} O.,  et~al., 2013, \mn@doi [\aap] {10.1051/0004-6361/201321100}, \href
  {http://adsabs.harvard.edu/abs/2013A%26A...556A..55I} {556, A55}

\bibitem[\protect\citeauthoryear{{Johansson}, {Naab}  \& {Burkert}}{{Johansson}
  et~al.}{2009}]{johansson09}
{Johansson} P.~H.,  {Naab} T.,   {Burkert} A.,  2009, \mn@doi [\apj]
  {10.1088/0004-637X/690/1/802}, \href
  {http://adsabs.harvard.edu/abs/2009ApJ...690..802J} {690, 802}

\bibitem[\protect\citeauthoryear{{Jones}, {Martin}  \& {Cooper}}{{Jones}
  et~al.}{2015}]{jones15}
{Jones} T.,  {Martin} C.,   {Cooper} M.~C.,  2015, \mn@doi [\apj]
  {10.1088/0004-637X/813/2/126}, \href
  {http://adsabs.harvard.edu/abs/2015ApJ...813..126J} {813, 126}

\bibitem[\protect\citeauthoryear{{Juneau} et~al.,}{{Juneau}
  et~al.}{2013}]{juneau13}
{Juneau} S.,  et~al., 2013, \mn@doi [\apj] {10.1088/0004-637X/764/2/176}, \href
  {http://adsabs.harvard.edu/abs/2013ApJ...764..176J} {764, 176}

\bibitem[\protect\citeauthoryear{{Kannappan}, {Guie}  \& {Baker}}{{Kannappan}
  et~al.}{2009}]{sheilaK09}
{Kannappan} S.~J.,  {Guie} J.~M.,   {Baker} A.~J.,  2009, \mn@doi [\aj]
  {10.1088/0004-6256/138/2/579}, \href
  {http://adsabs.harvard.edu/abs/2009AJ....138..579K} {138, 579}

\bibitem[\protect\citeauthoryear{{Kartaltepe} et~al.,}{{Kartaltepe}
  et~al.}{2012}]{kartaltepe12}
{Kartaltepe} J.~S.,  et~al., 2012, \mn@doi [\apj] {10.1088/0004-637X/757/1/23},
  \href {http://adsabs.harvard.edu/abs/2012ApJ...757...23K} {757, 23}

\bibitem[\protect\citeauthoryear{{Kauffmann} et~al.,}{{Kauffmann}
  et~al.}{2003}]{kauff03}
{Kauffmann} G.,  et~al., 2003, \mn@doi [\mnras]
  {10.1111/j.1365-2966.2003.07154.x}, \href
  {http://adsabs.harvard.edu/abs/2003MNRAS.346.1055K} {346, 1055}

\bibitem[\protect\citeauthoryear{{Kaviraj}, {Kirkby}, {Silk}  \&
  {Sarzi}}{{Kaviraj} et~al.}{2007}]{kaviraj07}
{Kaviraj} S.,  {Kirkby} L.~A.,  {Silk} J.,   {Sarzi} M.,  2007, \mn@doi
  [\mnras] {10.1111/j.1365-2966.2007.12475.x}, \href
  {http://adsabs.harvard.edu/abs/2007MNRAS.382..960K} {382, 960}

\bibitem[\protect\citeauthoryear{{Kennicutt}}{{Kennicutt}}{1998}]{kenn98}
{Kennicutt} Jr. R.~C.,  1998, \mn@doi [\araa] {10.1146/annurev.astro.36.1.189},
  \href {http://adsabs.harvard.edu/abs/1998ARA%26A..36..189K} {36, 189}

\bibitem[\protect\citeauthoryear{{Kewley}, {Groves}, {Kauffmann}  \&
  {Heckman}}{{Kewley} et~al.}{2006}]{kew06}
{Kewley} L.~J.,  {Groves} B.,  {Kauffmann} G.,   {Heckman} T.,  2006, \mn@doi
  [\mnras] {10.1111/j.1365-2966.2006.10859.x}, \href
  {http://adsabs.harvard.edu/abs/2006MNRAS.372..961K} {372, 961}

\bibitem[\protect\citeauthoryear{{Kewley}, {Dopita}, {Leitherer}, {Dav{\'e}},
  {Yuan}, {Allen}, {Groves}  \& {Sutherland}}{{Kewley} et~al.}{2013a}]{kew13b}
{Kewley} L.~J.,  {Dopita} M.~A.,  {Leitherer} C.,  {Dav{\'e}} R.,  {Yuan} T.,
  {Allen} M.,  {Groves} B.,   {Sutherland} R.,  2013a, \mn@doi [\apj]
  {10.1088/0004-637X/774/2/100}, \href
  {http://adsabs.harvard.edu/abs/2013ApJ...774..100K} {774, 100}

\bibitem[\protect\citeauthoryear{{Kewley}, {Maier}, {Yabe}, {Ohta}, {Akiyama},
  {Dopita}  \& {Yuan}}{{Kewley} et~al.}{2013b}]{kew13a}
{Kewley} L.~J.,  {Maier} C.,  {Yabe} K.,  {Ohta} K.,  {Akiyama} M.,  {Dopita}
  M.~A.,   {Yuan} T.,  2013b, \mn@doi [\apjl] {10.1088/2041-8205/774/1/L10},
  \href {http://adsabs.harvard.edu/abs/2013ApJ...774L..10K} {774, L10}

\bibitem[\protect\citeauthoryear{{Kocevski} et~al.,}{{Kocevski}
  et~al.}{2011}]{dirtydale11b}
{Kocevski} D.~D.,  et~al., 2011, \mn@doi [\apj] {10.1088/0004-637X/736/1/38},
  \href {http://adsabs.harvard.edu/abs/2011ApJ...736...38K} {736, 38}

\bibitem[\protect\citeauthoryear{{Koyama} et~al.,}{{Koyama}
  et~al.}{2008}]{koyama08}
{Koyama} Y.,  et~al., 2008, \mn@doi [\mnras]
  {10.1111/j.1365-2966.2008.13931.x}, \href
  {http://adsabs.harvard.edu/abs/2008MNRAS.391.1758K} {391, 1758}

\bibitem[\protect\citeauthoryear{{Kriek}, {van Dokkum}, {Labb{\'e}}, {Franx},
  {Illingworth}, {Marchesini}  \& {Quadri}}{{Kriek} et~al.}{2009}]{kriek09}
{Kriek} M.,  {van Dokkum} P.~G.,  {Labb{\'e}} I.,  {Franx} M.,  {Illingworth}
  G.~D.,  {Marchesini} D.,   {Quadri} R.~F.,  2009, \mn@doi [\apj]
  {10.1088/0004-637X/700/1/221}, \href
  {http://adsabs.harvard.edu/abs/2009ApJ...700..221K} {700, 221}

\bibitem[\protect\citeauthoryear{{Laigle} et~al.,}{{Laigle}
  et~al.}{2016}]{laigle16}
{Laigle} C.,  et~al., 2016, \mn@doi [\apjs] {10.3847/0067-0049/224/2/24}, \href
  {http://adsabs.harvard.edu/abs/2016ApJS..224...24L} {224, 24}

\bibitem[\protect\citeauthoryear{{Lamareille} et~al.,}{{Lamareille}
  et~al.}{2009}]{Lamareille2009}
{Lamareille} F.,  et~al., 2009, \mn@doi [\aap] {10.1051/0004-6361:200810397},
  \href {http://adsabs.harvard.edu/abs/2009A%26A...495...53L} {495, 53}

\bibitem[\protect\citeauthoryear{{Landolt}}{{Landolt}}{1992}]{landolt92}
{Landolt} A.~U.,  1992, \mn@doi [\aj] {10.1086/116242}, \href
  {http://adsabs.harvard.edu/abs/1992AJ....104..340L} {104, 340}

\bibitem[\protect\citeauthoryear{{Le Borgne} et~al.,}{{Le Borgne}
  et~al.}{2006}]{leborgne06}
{Le Borgne} D.,  et~al., 2006, \mn@doi [\apj] {10.1086/500005}, \href
  {http://adsabs.harvard.edu/abs/2006ApJ...642...48L} {642, 48}

\bibitem[\protect\citeauthoryear{{Le F{\`e}vre} et~al.,}{{Le F{\`e}vre}
  et~al.}{2013}]{dong13}
{Le F{\`e}vre} O.,  et~al., 2013, \mn@doi [\aap] {10.1051/0004-6361/201322179},
  \href {http://adsabs.harvard.edu/abs/2013A%26A...559A..14L} {559, A14}

\bibitem[\protect\citeauthoryear{{Lee}, {Idzi}, {Ferguson}, {Somerville},
  {Wiklind}  \& {Giavalisco}}{{Lee} et~al.}{2009}]{lee09}
{Lee} S.-K.,  {Idzi} R.,  {Ferguson} H.~C.,  {Somerville} R.~S.,  {Wiklind} T.,
    {Giavalisco} M.,  2009, \mn@doi [\apjs] {10.1088/0067-0049/184/1/100},
  \href {http://adsabs.harvard.edu/abs/2009ApJS..184..100L} {184, 100}

\bibitem[\protect\citeauthoryear{{Lemaux}, {Lubin}, {Shapley}, {Kocevski},
  {Gal}  \& {Squires}}{{Lemaux} et~al.}{2010}]{lem10}
{Lemaux} B.~C.,  {Lubin} L.~M.,  {Shapley} A.,  {Kocevski} D.,  {Gal} R.~R.,
  {Squires} G.~K.,  2010, \mn@doi [\apj] {10.1088/0004-637X/716/2/970}, \href
  {http://adsabs.harvard.edu/abs/2010ApJ...716..970L} {716, 970}

\bibitem[\protect\citeauthoryear{{Lemaux} et~al.,}{{Lemaux}
  et~al.}{2012}]{lem12}
{Lemaux} B.~C.,  et~al., 2012, \mn@doi [\apj] {10.1088/0004-637X/745/2/106},
  \href {http://adsabs.harvard.edu/abs/2012ApJ...745..106L} {745, 106}

\bibitem[\protect\citeauthoryear{{Lemaux} et~al.,}{{Lemaux}
  et~al.}{2014}]{lem14}
{Lemaux} B.~C.,  et~al., 2014, \mn@doi [\aap] {10.1051/0004-6361/201323089},
  \href {http://adsabs.harvard.edu/abs/2014A%26A...572A..90L} {572, A90}

\bibitem[\protect\citeauthoryear{{Lemaux} et~al.,}{{Lemaux}
  et~al.}{2017}]{lem17}
{Lemaux} B.~C.,  et~al., 2017, preprint, \href
  {http://adsabs.harvard.edu/abs/2017arXiv170310170L} {} (\mn@eprint {arXiv}
  {1703.10170})

\bibitem[\protect\citeauthoryear{{Lilly} et~al.,}{{Lilly}
  et~al.}{2007}]{lilly07}
{Lilly} S.~J.,  et~al., 2007, \mn@doi [\apjs] {10.1086/516589}, \href
  {http://adsabs.harvard.edu/abs/2007ApJS..172...70L} {172, 70}

\bibitem[\protect\citeauthoryear{{Lin} et~al.,}{{Lin} et~al.}{2016}]{lihwai16}
{Lin} L.,  et~al., 2016, \mn@doi [\apj] {10.3847/0004-637X/817/2/97}, \href
  {http://adsabs.harvard.edu/abs/2016ApJ...817...97L} {817, 97}

\bibitem[\protect\citeauthoryear{{Lubin}, {Gal}, {Lemaux}, {Kocevski}  \&
  {Squires}}{{Lubin} et~al.}{2009}]{lub09}
{Lubin} L.~M.,  {Gal} R.~R.,  {Lemaux} B.~C.,  {Kocevski} D.~D.,   {Squires}
  G.~K.,  2009, \mn@doi [\aj] {10.1088/0004-6256/137/6/4867}, \href
  {http://adsabs.harvard.edu/abs/2009AJ....137.4867L} {137, 4867}

\bibitem[\protect\citeauthoryear{{Ma}, {Ebeling}, {Donovan}  \& {Barrett}}{{Ma}
  et~al.}{2008}]{ma08}
{Ma} C.-J.,  {Ebeling} H.,  {Donovan} D.,   {Barrett} E.,  2008, \mn@doi [\apj]
  {10.1086/589991}, \href {http://adsabs.harvard.edu/abs/2008ApJ...684..160M}
  {684, 160}

\bibitem[\protect\citeauthoryear{{Makovoz} \& {Marleau}}{{Makovoz} \&
  {Marleau}}{2005}]{makovoz06}
{Makovoz} D.,  {Marleau} F.~R.,  2005, \mn@doi [\pasp] {10.1086/432977}, \href
  {http://adsabs.harvard.edu/abs/2005PASP..117.1113M} {117, 1113}

\bibitem[\protect\citeauthoryear{{Maltby} et~al.,}{{Maltby}
  et~al.}{2016}]{maltby16}
{Maltby} D.~T.,  et~al., 2016, \mn@doi [\mnras] {10.1093/mnrasl/slw057}, \href
  {http://adsabs.harvard.edu/abs/2016MNRAS.459L.114M} {459, L114}

\bibitem[\protect\citeauthoryear{{Maraston}}{{Maraston}}{2005}]{maraston05}
{Maraston} C.,  2005, \mn@doi [\mnras] {10.1111/j.1365-2966.2005.09270.x},
  \href {http://adsabs.harvard.edu/abs/2005MNRAS.362..799M} {362, 799}

\bibitem[\protect\citeauthoryear{{Marocco}, {Hache}  \& {Lamareille}}{{Marocco}
  et~al.}{2011}]{Marocco11}
{Marocco} J.,  {Hache} E.,   {Lamareille} F.,  2011, \mn@doi [\aap]
  {10.1051/0004-6361/201016143}, \href
  {http://adsabs.harvard.edu/abs/2011A%26A...531A..71M} {531, A71}

\bibitem[\protect\citeauthoryear{{Martini} et~al.,}{{Martini}
  et~al.}{2013}]{martini13}
{Martini} P.,  et~al., 2013, \mn@doi [\apj] {10.1088/0004-637X/768/1/1}, \href
  {http://adsabs.harvard.edu/abs/2013ApJ...768....1M} {768, 1}

\bibitem[\protect\citeauthoryear{{Masters} et~al.,}{{Masters}
  et~al.}{2010}]{masters10}
{Masters} K.~L.,  et~al., 2010, \mn@doi [\mnras]
  {10.1111/j.1365-2966.2010.16503.x}, \href
  {http://adsabs.harvard.edu/abs/2010MNRAS.405..783M} {405, 783}

\bibitem[\protect\citeauthoryear{{McLean} et~al.,}{{McLean}
  et~al.}{2012}]{mclean12}
{McLean} I.~S.,  et~al., 2012, in Ground-based and Airborne Instrumentation for
  Astronomy IV. p. 84460J, \mn@doi{10.1117/12.924794}

\bibitem[\protect\citeauthoryear{{Melnick} \& {De Propris}}{{Melnick} \& {De
  Propris}}{2013}]{melnick13}
{Melnick} J.,  {De Propris} R.,  2013, \mn@doi [\mnras] {10.1093/mnras/stt199},
  \href {http://adsabs.harvard.edu/abs/2013MNRAS.431.2034M} {431, 2034}

\bibitem[\protect\citeauthoryear{{Merlin} et~al.,}{{Merlin}
  et~al.}{2015}]{merlin15}
{Merlin} E.,  et~al., 2015, \mn@doi [\aap] {10.1051/0004-6361/201526471}, \href
  {http://adsabs.harvard.edu/abs/2015A%26A...582A..15M} {582, A15}

\bibitem[\protect\citeauthoryear{{Miyazaki} et~al.,}{{Miyazaki}
  et~al.}{2002}]{miyazaki02}
{Miyazaki} S.,  et~al., 2002, \mn@doi [\pasj] {10.1093/pasj/54.6.833}, \href
  {http://adsabs.harvard.edu/abs/2002PASJ...54..833M} {54, 833}

\bibitem[\protect\citeauthoryear{{Mok} et~al.,}{{Mok} et~al.}{2013}]{Mok13}
{Mok} A.,  et~al., 2013, \mn@doi [\mnras] {10.1093/mnras/stt251}, \href
  {http://adsabs.harvard.edu/abs/2013MNRAS.431.1090M} {431, 1090}

\bibitem[\protect\citeauthoryear{{Mok} et~al.,}{{Mok} et~al.}{2014}]{mok14}
{Mok} A.,  et~al., 2014, \mn@doi [\mnras] {10.1093/mnras/stt2419}, \href
  {http://adsabs.harvard.edu/abs/2014MNRAS.438.3070M} {438, 3070}

\bibitem[\protect\citeauthoryear{{Moran}, {Ellis}, {Treu}, {Smith}, {Rich}  \&
  {Smail}}{{Moran} et~al.}{2007}]{moran07}
{Moran} S.~M.,  {Ellis} R.~S.,  {Treu} T.,  {Smith} G.~P.,  {Rich} R.~M.,
  {Smail} I.,  2007, \mn@doi [\apj] {10.1086/522303}, \href
  {http://adsabs.harvard.edu/abs/2007ApJ...671.1503M} {671, 1503}

\bibitem[\protect\citeauthoryear{{Moutard} et~al.,}{{Moutard}
  et~al.}{2016}]{thibaud16}
{Moutard} T.,  et~al., 2016, \mn@doi [\aap] {10.1051/0004-6361/201527294},
  \href {http://adsabs.harvard.edu/abs/2016A%26A...590A.103M} {590, A103}

\bibitem[\protect\citeauthoryear{{Muzzin} et~al.,}{{Muzzin}
  et~al.}{2012}]{muz12}
{Muzzin} A.,  et~al., 2012, \mn@doi [\apj] {10.1088/0004-637X/746/2/188}, \href
  {http://adsabs.harvard.edu/abs/2012ApJ...746..188M} {746, 188}

\bibitem[\protect\citeauthoryear{{Muzzin} et~al.,}{{Muzzin}
  et~al.}{2014}]{muz14}
{Muzzin} A.,  et~al., 2014, \mn@doi [\apj] {10.1088/0004-637X/796/1/65}, \href
  {http://adsabs.harvard.edu/abs/2014ApJ...796...65M} {796, 65}

\bibitem[\protect\citeauthoryear{{Newberry}, {Boroson}  \&
  {Kirshner}}{{Newberry} et~al.}{1990}]{newberry90}
{Newberry} M.~V.,  {Boroson} T.~A.,   {Kirshner} R.~P.,  1990, \mn@doi [\apj]
  {10.1086/168413}, \href {http://adsabs.harvard.edu/abs/1990ApJ...350..585N}
  {350, 585}

\bibitem[\protect\citeauthoryear{{Newman} et~al.,}{{Newman}
  et~al.}{2013}]{new13}
{Newman} J.~A.,  et~al., 2013, \mn@doi [\apjs] {10.1088/0067-0049/208/1/5},
  \href {http://adsabs.harvard.edu/abs/2013ApJS..208....5N} {208, 5}

\bibitem[\protect\citeauthoryear{{Noble}, {Webb}, {Muzzin}, {Wilson}, {Yee}  \&
  {van der Burg}}{{Noble} et~al.}{2013}]{Noble13}
{Noble} A.~G.,  {Webb} T.~M.~A.,  {Muzzin} A.,  {Wilson} G.,  {Yee} H.~K.~C.,
  {van der Burg} R.~F.~J.,  2013, \mn@doi [\apj] {10.1088/0004-637X/768/2/118},
  \href {http://adsabs.harvard.edu/abs/2013ApJ...768..118N} {768, 118}

\bibitem[\protect\citeauthoryear{{Noble}, {Webb}, {Yee}, {Muzzin}, {Wilson},
  {van der Burg}, {Balogh}  \& {Shupe}}{{Noble} et~al.}{2016}]{noble16}
{Noble} A.~G.,  {Webb} T.~M.~A.,  {Yee} H.~K.~C.,  {Muzzin} A.,  {Wilson} G.,
  {van der Burg} R.~F.~J.,  {Balogh} M.~L.,   {Shupe} D.~L.,  2016, \mn@doi
  [\apj] {10.3847/0004-637X/816/2/48}, \href
  {http://adsabs.harvard.edu/abs/2016ApJ...816...48N} {816, 48}

\bibitem[\protect\citeauthoryear{{Oemler}}{{Oemler}}{1974}]{oemler74}
{Oemler} Jr. A.,  1974, \mn@doi [\apj] {10.1086/153216}, \href
  {http://adsabs.harvard.edu/abs/1974ApJ...194....1O} {194, 1}

\bibitem[\protect\citeauthoryear{{Oemler}, {Dressler}, {Kelson}, {Rigby},
  {Poggianti}, {Fritz}, {Morrison}  \& {Smail}}{{Oemler} et~al.}{2009}]{oem09}
{Oemler} Jr. A.,  {Dressler} A.,  {Kelson} D.,  {Rigby} J.,  {Poggianti} B.~M.,
   {Fritz} J.,  {Morrison} G.,   {Smail} I.,  2009, \mn@doi [\apj]
  {10.1088/0004-637X/693/1/152}, \href
  {http://adsabs.harvard.edu/abs/2009ApJ...693..152O} {693, 152}

\bibitem[\protect\citeauthoryear{{Oke} \& {Gunn}}{{Oke} \&
  {Gunn}}{1983}]{okengunn83}
{Oke} J.~B.,  {Gunn} J.~E.,  1983, \mn@doi [\apj] {10.1086/160817}, \href
  {http://adsabs.harvard.edu/abs/1983ApJ...266..713O} {266, 713}

\bibitem[\protect\citeauthoryear{{Ouchi} et~al.,}{{Ouchi}
  et~al.}{2004}]{ouchi04}
{Ouchi} M.,  et~al., 2004, \mn@doi [\apj] {10.1086/422207}, \href
  {http://adsabs.harvard.edu/abs/2004ApJ...611..660O} {611, 660}

\bibitem[\protect\citeauthoryear{{Papovich} et~al.,}{{Papovich}
  et~al.}{2010}]{casey10}
{Papovich} C.,  et~al., 2010, \mn@doi [\apj] {10.1088/0004-637X/716/2/1503},
  \href {http://adsabs.harvard.edu/abs/2010ApJ...716.1503P} {716, 1503}

\bibitem[\protect\citeauthoryear{{Pawlik}, {Wild}, {Walcher}, {Johansson},
  {Villforth}, {Rowlands}, {Mendez-Abreu}  \& {Hewlett}}{{Pawlik}
  et~al.}{2016}]{pawlik16}
{Pawlik} M.~M.,  {Wild} V.,  {Walcher} C.~J.,  {Johansson} P.~H.,  {Villforth}
  C.,  {Rowlands} K.,  {Mendez-Abreu} J.,   {Hewlett} T.,  2016, \mn@doi
  [\mnras] {10.1093/mnras/stv2878}, \href
  {http://adsabs.harvard.edu/abs/2016MNRAS.456.3032P} {456, 3032}

\bibitem[\protect\citeauthoryear{{Peng} et~al.,}{{Peng} et~al.}{2010}]{peng10}
{Peng} Y.-j.,  et~al., 2010, \mn@doi [\apj] {10.1088/0004-637X/721/1/193},
  \href {http://adsabs.harvard.edu/abs/2010ApJ...721..193P} {721, 193}

\bibitem[\protect\citeauthoryear{{P{\'e}rez-Montero}
  et~al.,}{{P{\'e}rez-Montero} et~al.}{2009}]{Perez-Montero2009}
{P{\'e}rez-Montero} E.,  et~al., 2009, \mn@doi [\aap]
  {10.1051/0004-6361:200810558}, \href
  {http://adsabs.harvard.edu/abs/2009A%26A...495...73P} {495, 73}

\bibitem[\protect\citeauthoryear{{P{\'e}rez-Montero}
  et~al.,}{{P{\'e}rez-Montero} et~al.}{2013}]{Perez-Montero2013}
{P{\'e}rez-Montero} E.,  et~al., 2013, \mn@doi [\aap]
  {10.1051/0004-6361/201220070}, \href
  {http://adsabs.harvard.edu/abs/2013A%26A...549A..25P} {549, A25}

\bibitem[\protect\citeauthoryear{{Pforr}, {Maraston}  \& {Tonini}}{{Pforr}
  et~al.}{2012}]{janine12}
{Pforr} J.,  {Maraston} C.,   {Tonini} C.,  2012, \mn@doi [\mnras]
  {10.1111/j.1365-2966.2012.20848.x}, \href
  {http://adsabs.harvard.edu/abs/2012MNRAS.422.3285P} {422, 3285}

\bibitem[\protect\citeauthoryear{{Pickles}}{{Pickles}}{1998}]{pickles98}
{Pickles} A.~J.,  1998, \mn@doi [\pasp] {10.1086/316197}, \href
  {http://adsabs.harvard.edu/abs/1998PASP..110..863P} {110, 863}

\bibitem[\protect\citeauthoryear{{Poggianti}, {Smail}, {Dressler}, {Couch},
  {Barger}, {Butcher}, {Ellis}  \& {Oemler}}{{Poggianti} et~al.}{1999}]{pog99}
{Poggianti} B.~M.,  {Smail} I.,  {Dressler} A.,  {Couch} W.~J.,  {Barger}
  A.~J.,  {Butcher} H.,  {Ellis} R.~S.,   {Oemler} Jr. A.,  1999, \mn@doi
  [\apj] {10.1086/307322}, \href
  {http://adsabs.harvard.edu/abs/1999ApJ...518..576P} {518, 576}

\bibitem[\protect\citeauthoryear{{Poggianti} et~al.,}{{Poggianti}
  et~al.}{2009}]{pog09}
{Poggianti} B.~M.,  et~al., 2009, \mn@doi [\apj] {10.1088/0004-637X/693/1/112},
  \href {http://adsabs.harvard.edu/abs/2009ApJ...693..112P} {693, 112}

\bibitem[\protect\citeauthoryear{{Pozzetti} et~al.,}{{Pozzetti}
  et~al.}{2010}]{pozzetti10}
{Pozzetti} L.,  et~al., 2010, \mn@doi [\aap] {10.1051/0004-6361/200913020},
  \href {http://adsabs.harvard.edu/abs/2010A%26A...523A..13P} {523, A13}

\bibitem[\protect\citeauthoryear{{Pr\'{e}vot}, {Lequeux}, {Prevot}, {Maurice}
  \& {Rocca-Volmerange}}{{Pr\'{e}vot} et~al.}{1984}]{prevot84}
{Pr\'{e}vot} M.~L.,  {Lequeux} J.,  {Prevot} L.,  {Maurice} E.,
  {Rocca-Volmerange} B.,  1984, \aap, \href
  {http://adsabs.harvard.edu/abs/1984A%26A...132..389P} {132, 389}

\bibitem[\protect\citeauthoryear{{Puget} et~al.,}{{Puget}
  et~al.}{2004}]{puget04}
{Puget} P.,  et~al., 2004, in {Moorwood} A.~F.~M.,  {Iye} M.,  eds,  \procspie
  Vol. 5492, Ground-based Instrumentation for Astronomy. pp 978--987,
  \mn@doi{10.1117/12.551097}

\bibitem[\protect\citeauthoryear{{Rumbaugh}, {Kocevski}, {Gal}, {Lemaux},
  {Lubin}, {Fassnacht}, {McGrath}  \& {Squires}}{{Rumbaugh}
  et~al.}{2012}]{rum12}
{Rumbaugh} N.,  {Kocevski} D.~D.,  {Gal} R.~R.,  {Lemaux} B.~C.,  {Lubin}
  L.~M.,  {Fassnacht} C.~D.,  {McGrath} E.~J.,   {Squires} G.~K.,  2012, \apj,
  \href {http://adsabs.harvard.edu/abs/2012ApJ...746..155R} {746, 155}

\bibitem[\protect\citeauthoryear{{Rumbaugh}, {Kocevski}, {Gal}, {Lemaux},
  {Lubin}, {Fassnacht}  \& {Squires}}{{Rumbaugh} et~al.}{2013}]{rum13}
{Rumbaugh} N.,  {Kocevski} D.~D.,  {Gal} R.~R.,  {Lemaux} B.~C.,  {Lubin}
  L.~M.,  {Fassnacht} C.~D.,   {Squires} G.~K.,  2013, \mn@doi [\apj]
  {10.1088/0004-637X/763/2/124}, \href
  {http://adsabs.harvard.edu/abs/2013ApJ...763..124R} {763, 124}

\bibitem[\protect\citeauthoryear{{Ryan} Jr. et~al.,}{{Ryan}
  et~al.}{2014}]{Rusl13}
{Ryan} Jr. R.~E.,  et~al., 2014, \mn@doi [\apjl] {10.1088/2041-8205/786/1/L4},
  \href {http://adsabs.harvard.edu/abs/2014ApJ...786L...4R} {786, L4}

\bibitem[\protect\citeauthoryear{{Sanders} et~al.,}{{Sanders}
  et~al.}{2015}]{sanders15}
{Sanders} R.~L.,  et~al., 2015, \mn@doi [\apj] {10.1088/0004-637X/799/2/138},
  \href {http://adsabs.harvard.edu/abs/2015ApJ...799..138S} {799, 138}

\bibitem[\protect\citeauthoryear{{Santos} et~al.,}{{Santos}
  et~al.}{2014}]{santos14}
{Santos} J.~S.,  et~al., 2014, \mn@doi [\mnras] {10.1093/mnras/stt2376}, \href
  {http://adsabs.harvard.edu/abs/2014MNRAS.438.2565S} {438, 2565}

\bibitem[\protect\citeauthoryear{{Santos} et~al.,}{{Santos}
  et~al.}{2015}]{santos15}
{Santos} J.~S.,  et~al., 2015, \mn@doi [\mnras] {10.1093/mnrasl/slu180}, \href
  {http://adsabs.harvard.edu/abs/2015MNRAS.447L..65S} {447, L65}

\bibitem[\protect\citeauthoryear{{Schawinski} et~al.,}{{Schawinski}
  et~al.}{2014}]{kschawin14}
{Schawinski} K.,  et~al., 2014, \mn@doi [\mnras] {10.1093/mnras/stu327}, \href
  {http://adsabs.harvard.edu/abs/2014MNRAS.440..889S} {440, 889}

\bibitem[\protect\citeauthoryear{{Shapley} et~al.,}{{Shapley}
  et~al.}{2015}]{shapley15}
{Shapley} A.~E.,  et~al., 2015, \mn@doi [\apj] {10.1088/0004-637X/801/2/88},
  \href {http://adsabs.harvard.edu/abs/2015ApJ...801...88S} {801, 88}

\bibitem[\protect\citeauthoryear{{Shields}}{{Shields}}{1992}]{shields92}
{Shields} J.~C.,  1992, \mn@doi [\apjl] {10.1086/186598}, \href
  {http://adsabs.harvard.edu/abs/1992ApJ...399L..27S} {399, L27}

\bibitem[\protect\citeauthoryear{{Simcoe}, {Metzger}, {Small}  \&
  {Araya}}{{Simcoe} et~al.}{2000}]{simcoe00}
{Simcoe} R.~A.,  {Metzger} M.~R.,  {Small} T.~A.,   {Araya} G.,  2000, in
  American Astronomical Society Meeting Abstracts \#196. p.~758

\bibitem[\protect\citeauthoryear{{Singh} et~al.,}{{Singh}
  et~al.}{2013}]{singh13}
{Singh} R.,  et~al., 2013, \mn@doi [\aap] {10.1051/0004-6361/201322062}, \href
  {http://adsabs.harvard.edu/abs/2013A%26A...558A..43S} {558, A43}

\bibitem[\protect\citeauthoryear{{Skrutskie} et~al.,}{{Skrutskie}
  et~al.}{2006}]{skrutskie06}
{Skrutskie} M.~F.,  et~al., 2006, \mn@doi [\aj] {10.1086/498708}, \href
  {http://adsabs.harvard.edu/abs/2006AJ....131.1163S} {131, 1163}

\bibitem[\protect\citeauthoryear{{Snyder}, {Cox}, {Hayward}, {Hernquist}  \&
  {Jonsson}}{{Snyder} et~al.}{2011}]{snyder11}
{Snyder} G.~F.,  {Cox} T.~J.,  {Hayward} C.~C.,  {Hernquist} L.,   {Jonsson}
  P.,  2011, \mn@doi [\apj] {10.1088/0004-637X/741/2/77}, \href
  {http://adsabs.harvard.edu/abs/2011ApJ...741...77S} {741, 77}

\bibitem[\protect\citeauthoryear{{Stasi{\'n}ska}, {Cid Fernandes}, {Mateus},
  {Sodr{\'e}}  \& {Asari}}{{Stasi{\'n}ska} et~al.}{2006}]{Stasinska06}
{Stasi{\'n}ska} G.,  {Cid Fernandes} R.,  {Mateus} A.,  {Sodr{\'e}} L.,
  {Asari} N.~V.,  2006, \mn@doi [\mnras] {10.1111/j.1365-2966.2006.10732.x},
  \href {http://adsabs.harvard.edu/abs/2006MNRAS.371..972S} {371, 972}

\bibitem[\protect\citeauthoryear{{Strazzullo} et~al.,}{{Strazzullo}
  et~al.}{2010}]{strazzullo10}
{Strazzullo} V.,  et~al., 2010, \mn@doi [\aap] {10.1051/0004-6361/201015251},
  \href {http://adsabs.harvard.edu/abs/2010A%26A...524A..17S} {524, A17}

\bibitem[\protect\citeauthoryear{{Struck}}{{Struck}}{2006}]{struck06}
{Struck} C.,  2006, {Galaxy Collisions -- Dawn of a New Era}.
p.~115, \mn@doi{10.1007/3-540-30313-8_4}

\bibitem[\protect\citeauthoryear{{Swinbank}, {Balogh}, {Bower}, {Zabludoff},
  {Lucey}, {McGee}, {Miller}  \& {Nichol}}{{Swinbank}
  et~al.}{2012}]{swinbank12}
{Swinbank} A.~M.,  {Balogh} M.~L.,  {Bower} R.~G.,  {Zabludoff} A.~I.,  {Lucey}
  J.~R.,  {McGee} S.~L.,  {Miller} C.~J.,   {Nichol} R.~C.,  2012, \mn@doi
  [\mnras] {10.1111/j.1365-2966.2011.20082.x}, \href
  {http://adsabs.harvard.edu/abs/2012MNRAS.420..672S} {420, 672}

\bibitem[\protect\citeauthoryear{{Taniguchi}, {Shioya}  \&
  {Murayama}}{{Taniguchi} et~al.}{2000}]{taniguchi00}
{Taniguchi} Y.,  {Shioya} Y.,   {Murayama} T.,  2000, \mn@doi [\aj]
  {10.1086/301520}, \href {http://adsabs.harvard.edu/abs/2000AJ....120.1265T}
  {120, 1265}

\bibitem[\protect\citeauthoryear{{Taranu}, {Hudson}, {Balogh}, {Smith},
  {Power}, {Oman}  \& {Krane}}{{Taranu} et~al.}{2014}]{taranu14}
{Taranu} D.~S.,  {Hudson} M.~J.,  {Balogh} M.~L.,  {Smith} R.~J.,  {Power} C.,
  {Oman} K.~A.,   {Krane} B.,  2014, \mn@doi [\mnras] {10.1093/mnras/stu389},
  \href {http://adsabs.harvard.edu/abs/2014MNRAS.440.1934T} {440, 1934}

\bibitem[\protect\citeauthoryear{{Terlevich} \& {Melnick}}{{Terlevich} \&
  {Melnick}}{1985}]{terlevich85}
{Terlevich} R.,  {Melnick} J.,  1985, \mn@doi [\mnras]
  {10.1093/mnras/213.4.841}, \href
  {http://adsabs.harvard.edu/abs/1985MNRAS.213..841T} {213, 841}

\bibitem[\protect\citeauthoryear{{Thomas} et~al.,}{{Thomas}
  et~al.}{2016}]{romain16}
{Thomas} R.,  et~al., 2016, preprint, \href
  {http://adsabs.harvard.edu/abs/2016arXiv160201841T} {} (\mn@eprint {arXiv}
  {1602.01841})

\bibitem[\protect\citeauthoryear{{Tody}}{{Tody}}{1993}]{tody93}
{Tody} D.,  1993, in {Hanisch} R.~J.,  {Brissenden} R.~J.~V.,   {Barnes} J.,
  eds,  Astronomical Society of the Pacific Conference Series Vol. 52,
  Astronomical Data Analysis Software and Systems II. p.~173

\bibitem[\protect\citeauthoryear{{Tomczak} et~al.,}{{Tomczak}
  et~al.}{2016}]{AA16}
{Tomczak} A.~R.,  et~al., 2016, in prep.

\bibitem[\protect\citeauthoryear{{Tran}, {Franx}, {Illingworth}, {Kelson}  \&
  {van Dokkum}}{{Tran} et~al.}{2003}]{tran03}
{Tran} K.-V.~H.,  {Franx} M.,  {Illingworth} G.,  {Kelson} D.~D.,   {van
  Dokkum} P.,  2003, \mn@doi [\apj] {10.1086/379804}, \href
  {http://adsabs.harvard.edu/abs/2003ApJ...599..865T} {599, 865}

\bibitem[\protect\citeauthoryear{{Tran}, {Franx}, {Illingworth}, {van Dokkum},
  {Kelson}  \& {Magee}}{{Tran} et~al.}{2004}]{tran04}
{Tran} K.-V.~H.,  {Franx} M.,  {Illingworth} G.~D.,  {van Dokkum} P.,  {Kelson}
  D.~D.,   {Magee} D.,  2004, \mn@doi [\apj] {10.1086/421237}, \href
  {http://adsabs.harvard.edu/abs/2004ApJ...609..683T} {609, 683}

\bibitem[\protect\citeauthoryear{{Tran}, {Franx}, {Illingworth}, {van Dokkum},
  {Kelson}, {Blakeslee}  \& {Postman}}{{Tran} et~al.}{2007}]{tran07}
{Tran} K.-V.~H.,  {Franx} M.,  {Illingworth} G.~D.,  {van Dokkum} P.,  {Kelson}
  D.~D.,  {Blakeslee} J.~P.,   {Postman} M.,  2007, \mn@doi [\apj]
  {10.1086/513738}, \href {http://adsabs.harvard.edu/abs/2007ApJ...661..750T}
  {661, 750}

\bibitem[\protect\citeauthoryear{{Tran} et~al.,}{{Tran} et~al.}{2010}]{tran10}
{Tran} K.-V.~H.,  et~al., 2010, \mn@doi [\apjl] {10.1088/2041-8205/719/2/L126},
  \href {http://adsabs.harvard.edu/abs/2010ApJ...719L.126T} {719, L126}

\bibitem[\protect\citeauthoryear{{Treu}, {Ellis}, {Kneib}, {Dressler}, {Smail},
  {Czoske}, {Oemler}  \& {Natarajan}}{{Treu} et~al.}{2003}]{treu03}
{Treu} T.,  {Ellis} R.~S.,  {Kneib} J.-P.,  {Dressler} A.,  {Smail} I.,
  {Czoske} O.,  {Oemler} A.,   {Natarajan} P.,  2003, \mn@doi [\apj]
  {10.1086/375314}, \href {http://adsabs.harvard.edu/abs/2003ApJ...591...53T}
  {591, 53}

\bibitem[\protect\citeauthoryear{{Treu} et~al.,}{{Treu} et~al.}{2005}]{treu05}
{Treu} T.,  et~al., 2005, \mn@doi [\apj] {10.1086/444585}, \href
  {http://adsabs.harvard.edu/abs/2005ApJ...633..174T} {633, 174}

\bibitem[\protect\citeauthoryear{{Trouille}, {Barger}  \&
  {Tremonti}}{{Trouille} et~al.}{2011}]{trouille11}
{Trouille} L.,  {Barger} A.~J.,   {Tremonti} C.,  2011, \mn@doi [\apj]
  {10.1088/0004-637X/742/1/46}, \href
  {http://adsabs.harvard.edu/abs/2011ApJ...742...46T} {742, 46}

\bibitem[\protect\citeauthoryear{{Veilleux}, {Kim}, {Sanders}, {Mazzarella}  \&
  {Soifer}}{{Veilleux} et~al.}{1995}]{veilleux95}
{Veilleux} S.,  {Kim} D.-C.,  {Sanders} D.~B.,  {Mazzarella} J.~M.,   {Soifer}
  B.~T.,  1995, \mn@doi [\apjs] {10.1086/192158}, \href
  {http://adsabs.harvard.edu/abs/1995ApJS...98..171V} {98, 171}

\bibitem[\protect\citeauthoryear{{Vergani} et~al.,}{{Vergani}
  et~al.}{2010}]{vergani10}
{Vergani} D.,  et~al., 2010, \mn@doi [\aap] {10.1051/0004-6361/200912802},
  \href {http://adsabs.harvard.edu/abs/2010A%26A...509A..42V} {509, A42}

\bibitem[\protect\citeauthoryear{{Vikhlinin}, {van Speybroeck}, {Markevitch},
  {Forman}  \& {Grego}}{{Vikhlinin} et~al.}{2002}]{vikhlinin02}
{Vikhlinin} A.,  {van Speybroeck} L.,  {Markevitch} M.,  {Forman} W.~R.,
  {Grego} L.,  2002, \mn@doi [\apjl] {10.1086/344591}, \href
  {http://adsabs.harvard.edu/abs/2002ApJ...578L.107V} {578, L107}

\bibitem[\protect\citeauthoryear{{Vulcani} et~al.,}{{Vulcani}
  et~al.}{2013}]{vulcani13}
{Vulcani} B.,  et~al., 2013, \mn@doi [\aap] {10.1051/0004-6361/201118388},
  \href {http://adsabs.harvard.edu/abs/2013A%26A...550A..58V} {550, A58}

\bibitem[\protect\citeauthoryear{{Wang} et~al.,}{{Wang}
  et~al.}{2016}]{taowang16}
{Wang} T.,  et~al., 2016, preprint, \href
  {http://adsabs.harvard.edu/abs/2016arXiv160407404W} {} (\mn@eprint {arXiv}
  {1604.07404})

\bibitem[\protect\citeauthoryear{{Webb} et~al.,}{{Webb} et~al.}{2013}]{webb13}
{Webb} T.~M.~A.,  et~al., 2013, \mn@doi [\aj] {10.1088/0004-6256/146/4/84},
  \href {http://adsabs.harvard.edu/abs/2013AJ....146...84W} {146, 84}

\bibitem[\protect\citeauthoryear{{Werner} et~al.,}{{Werner}
  et~al.}{2004}]{wer04}
{Werner} M.~W.,  et~al., 2004, \mn@doi [\apjs] {10.1086/422992}, \href
  {http://adsabs.harvard.edu/abs/2004ApJS..154....1W} {154, 1}

\bibitem[\protect\citeauthoryear{{Wetzel}, {Tinker}, {Conroy}  \& {van den
  Bosch}}{{Wetzel} et~al.}{2013}]{wetzel13}
{Wetzel} A.~R.,  {Tinker} J.~L.,  {Conroy} C.,   {van den Bosch} F.~C.,  2013,
  \mn@doi [\mnras] {10.1093/mnras/stt469}, \href
  {http://adsabs.harvard.edu/abs/2013MNRAS.432..336W} {432, 336}

\bibitem[\protect\citeauthoryear{{Wild}, {Walcher}, {Johansson}, {Tresse},
  {Charlot}, {Pollo}, {Le F{\`e}vre}  \& {de Ravel}}{{Wild}
  et~al.}{2009}]{wild09}
{Wild} V.,  {Walcher} C.~J.,  {Johansson} P.~H.,  {Tresse} L.,  {Charlot} S.,
  {Pollo} A.,  {Le F{\`e}vre} O.,   {de Ravel} L.,  2009, \mn@doi [\mnras]
  {10.1111/j.1365-2966.2009.14537.x}, \href
  {http://adsabs.harvard.edu/abs/2009MNRAS.395..144W} {395, 144}

\bibitem[\protect\citeauthoryear{{Wild} et~al.,}{{Wild} et~al.}{2014}]{wild14}
{Wild} V.,  et~al., 2014, \mn@doi [\mnras] {10.1093/mnras/stu212}, \href
  {http://adsabs.harvard.edu/abs/2014MNRAS.440.1880W} {440, 1880}

\bibitem[\protect\citeauthoryear{{Wootten} \& {Thompson}}{{Wootten} \&
  {Thompson}}{2009}]{wootten09}
{Wootten} A.,  {Thompson} A.~R.,  2009, \mn@doi [IEEE Proceedings]
  {10.1109/JPROC.2009.2020572}, \href
  {http://adsabs.harvard.edu/abs/2009IEEEP..97.1463W} {97, 1463}

\bibitem[\protect\citeauthoryear{{Wu}, {Gal}, {Lemaux}, {Kocevski}, {Lubin},
  {Rumbaugh}  \& {Squires}}{{Wu} et~al.}{2014}]{pwu14}
{Wu} P.-F.,  {Gal} R.~R.,  {Lemaux} B.~C.,  {Kocevski} D.~D.,  {Lubin} L.~M.,
  {Rumbaugh} N.,   {Squires} G.~K.,  2014, \mn@doi [\apj]
  {10.1088/0004-637X/792/1/16}, \href
  {http://adsabs.harvard.edu/abs/2014ApJ...792...16W} {792, 16}

\bibitem[\protect\citeauthoryear{{Yan}, {Newman}, {Faber}, {Konidaris}, {Koo}
  \& {Davis}}{{Yan} et~al.}{2006}]{yan06}
{Yan} R.,  {Newman} J.~A.,  {Faber} S.~M.,  {Konidaris} N.,  {Koo} D.,
  {Davis} M.,  2006, \mn@doi [\apj] {10.1086/505629}, \href
  {http://adsabs.harvard.edu/abs/2006ApJ...648..281Y} {648, 281}

\bibitem[\protect\citeauthoryear{{Yan} et~al.,}{{Yan} et~al.}{2009}]{yan09}
{Yan} R.,  et~al., 2009, \mn@doi [\mnras] {10.1111/j.1365-2966.2009.15192.x},
  \href {http://adsabs.harvard.edu/abs/2009MNRAS.398..735Y} {398, 735}

\bibitem[\protect\citeauthoryear{{Yan} et~al.,}{{Yan} et~al.}{2011}]{renbin11}
{Yan} R.,  et~al., 2011, \mn@doi [\apj] {10.1088/0004-637X/728/1/38}, \href
  {http://adsabs.harvard.edu/abs/2011ApJ...728...38Y} {728, 38}

\bibitem[\protect\citeauthoryear{{Yesuf}, {Faber}, {Trump}, {Koo}, {Fang},
  {Liu}, {Wild}  \& {Hayward}}{{Yesuf} et~al.}{2014}]{yesuf14}
{Yesuf} H.~M.,  {Faber} S.~M.,  {Trump} J.~R.,  {Koo} D.~C.,  {Fang} J.~J.,
  {Liu} F.~S.,  {Wild} V.,   {Hayward} C.~C.,  2014, \mn@doi [\apj]
  {10.1088/0004-637X/792/2/84}, \href
  {http://adsabs.harvard.edu/abs/2014ApJ...792...84Y} {792, 84}

\bibitem[\protect\citeauthoryear{{York} et~al.,}{{York} et~al.}{2000}]{york00}
{York} D.~G.,  et~al., 2000, \mn@doi [\aj] {10.1086/301513}, \href
  {http://adsabs.harvard.edu/abs/2000AJ....120.1579Y} {120, 1579}

\bibitem[\protect\citeauthoryear{{Yuan}, {Kewley}  \& {Richard}}{{Yuan}
  et~al.}{2013}]{Yuan13}
{Yuan} T.-T.,  {Kewley} L.~J.,   {Richard} J.,  2013, \mn@doi [\apj]
  {10.1088/0004-637X/763/1/9}, \href
  {http://adsabs.harvard.edu/abs/2013ApJ...763....9Y} {763, 9}

\bibitem[\protect\citeauthoryear{{Zabludoff}, {Zaritsky}, {Lin}, {Tucker},
  {Hashimoto}, {Shectman}, {Oemler}  \& {Kirshner}}{{Zabludoff}
  et~al.}{1996}]{zabludoff96}
{Zabludoff} A.~I.,  {Zaritsky} D.,  {Lin} H.,  {Tucker} D.,  {Hashimoto} Y.,
  {Shectman} S.~A.,  {Oemler} A.,   {Kirshner} R.~P.,  1996, \mn@doi [\apj]
  {10.1086/177495}, \href {http://adsabs.harvard.edu/abs/1996ApJ...466..104Z}
  {466, 104}

\bibitem[\protect\citeauthoryear{{Zahid}, {Kewley}  \& {Bresolin}}{{Zahid}
  et~al.}{2011}]{Zahid2011}
{Zahid} H.~J.,  {Kewley} L.~J.,   {Bresolin} F.,  2011, \mn@doi [\apj]
  {10.1088/0004-637X/730/2/137}, \href
  {http://adsabs.harvard.edu/abs/2011ApJ...730..137Z} {730, 137}

\bibitem[\protect\citeauthoryear{{Zeimann} et~al.,}{{Zeimann}
  et~al.}{2013}]{greg13}
{Zeimann} G.~R.,  et~al., 2013, \mn@doi [\apj] {10.1088/0004-637X/779/2/137},
  \href {http://adsabs.harvard.edu/abs/2013ApJ...779..137Z} {779, 137}

\bibitem[\protect\citeauthoryear{{Zeimann} et~al.,}{{Zeimann}
  et~al.}{2015}]{greg15}
{Zeimann} G.~R.,  et~al., 2015, \mn@doi [\apj] {10.1088/0004-637X/798/1/29},
  \href {http://adsabs.harvard.edu/abs/2015ApJ...798...29Z} {798, 29}

\bibitem[\protect\citeauthoryear{{Ziparo} et~al.,}{{Ziparo}
  et~al.}{2014}]{ziparo14}
{Ziparo} F.,  et~al., 2014, \mn@doi [\mnras] {10.1093/mnras/stt1901}, \href
  {http://adsabs.harvard.edu/abs/2014MNRAS.437..458Z} {437, 458}

\bibitem[\protect\citeauthoryear{{Zwaan}, {Kuntschner}, {Pracy}  \&
  {Couch}}{{Zwaan} et~al.}{2013}]{zwaan13}
{Zwaan} M.~A.,  {Kuntschner} H.,  {Pracy} M.~B.,   {Couch} W.~J.,  2013,
  \mn@doi [\mnras] {10.1093/mnras/stt496}, \href
  {http://adsabs.harvard.edu/abs/2013MNRAS.432..492Z} {432, 492}

\bibitem[\protect\citeauthoryear{{van der Burg} et~al.,}{{van der Burg}
  et~al.}{2013}]{vanderburg13}
{van der Burg} R.~F.~J.,  et~al., 2013, \mn@doi [\aap]
  {10.1051/0004-6361/201321237}, \href
  {http://adsabs.harvard.edu/abs/2013A%26A...557A..15V} {557, A15}

\makeatother
\end{thebibliography}




\appendix

\section{\normalsize{Simultanous Spectral Energy Distribution Fitting of Spectra and Photometry}}

Here we describe the process of coadding the broadband magnitudes and the simultaneous fitting of the coadded spectra and photometry used to generate
the results presented in \S\ref{KAevolution}. In order to coadd the broadband photometry, a scaling factor was applied to the photometry of each galaxy 
such that the average flux density of the combined $I^{+}$ and $Z^{+}$ bands was unity. This scaling factor was applied to all bands prior to coaddition. 
This choice was motivated by our desire to match the normalization of the spectral coadding process described in \S\ref{KAevolution}, as the central 
observed-frame wavelength coverage of our DEIMOS spectroscopy roughly lies at the border between the $I^{+}$ and $Z^{+}$ bands. The (normalized)
flux density for each band was then calculated by an inverse variance-weighted mean after removing, for each galaxy, bands in which that galaxy went
undetected ($<3\sigma$). An small (3\%) additional systematic uncertainty was included in the formal random uncertainty of the coadded flux density in each
band to account for uncertainty in the photometric zero-points (see, e.g., \citealt{dreadOlivier06, brammer08}).

The resultant coadded spectra and photometry for the traditional K+A and KAIROS/K+A-H$\alpha$ samples were then compared to a suite of synthetic spectral models 
(BC03, \citealt{maraston05, CB07}) after linearly interpolating over the [OII] feature. For each
prescription a variety of SFHs ($\psi(t)$), stellar-phase metallicities, and stellar continuum extinctions were employed, with each model generated for ages in the
range $\log(t_{SB}$)=7-9.6 in 24 steps roughly equally spaced in $log(t_{SB})$. For the spectral comparison, coadded spectra were degraded
through spline interpolation to match the plate scale of the models. For the photometric comparison, synthetic model magnitudes were created by convolving
models with the appropriate filter curves shifted to the average redshift of the sample. Only those wavelength ranges for which 100\% of the galaxies
contributed to the coadded spectra and photometry were considered in the models. The probability of a certain parameter/model combination was calculated
at each age step as $e^{- (0.5\chi^{2}_{\nu, spec}+0.5\chi^{2}_{\nu, phot})/2}$, where $\chi^{2}_{\nu, spec}$ and $\chi^{2}_{\nu, phot}$ are the reduced $\chi^2$
calculated from the comparison of coadded spectra and photometry, respectively, to the model at this step (see \citealt{romain16} for details).

While this machinery can be, in principle,
used to generate combined probabilities for all combinations of parameters and models, such a large range of allowable values quickly becomes
computationally untenable and massively underconstrained astrophysically. Instead, we chose for this analysis to place two strong constraints on
the models used. The first is that the SFH of the model(s) used for the fit be exponentially declining with a short e-folding time. Such a SFH does
not preclude the possibility of star formation in a continuous or bursty mode prior to this decline, but this prior requires that
the cessation of star formation, once begun, proceed rapidly (here we choose $\tau=10$ Myr, where $\psi(t)\propto \tau^{-1}\exp{-t/\tau}$). Such a SFH
essentially defines a K+A galaxy (e.g., \citealt{pog99}) and thus represents a reasonable assumption.

The second constraint
is placed on the stellar-phase metallicity, a parameter which, if left unconstrained, can wreak havoc on age measurements (e.g., \citealt{fumagalli16}).
In the absence of the ability to measure the gas- or stellar-phase metallicity directly, the mean stellar masses of the traditional K+A and
KAIROS/K+A-H$\alpha$ samples can be used to place strong constraints on the gas-phase metallicity from
the mass-metallicity relation. Such constraints are, in turn, linked to the average stellar-phase metallicity by virtue of the rapid formation of the
luminosity-dominant stellar population recently formed in our samples. Results from analysing the largest samples of star-forming galaxies at $z\sim0.8$
currently available place the average gas-phase metallicity of galaxies at stellar masses equal to the average of both of our samples at approximately solar
\citep{Lamareille2009,Perez-Montero2009,Zahid2011, Yuan13, Perez-Montero2013}. Thus, we explicitly impose that the models used be generated with 
what is referred to as $Z=Z_{\odot}$ in the stellar synthesis models.
In addition, we adopt the prescription of \citet{CB07} for this fitting as these models, generally, provided a slightly better fit to our data than
either the BC03 or Maraston models for a given combination of parameters. It is crucial to note that the relative comparisons made in this paper
are \emph{completely insensitive} to variations in $\psi(t)$, stellar-phase metallicity, and prescription as long as the same choice is made for both
samples, and only result in absolute offsets in $t_{SB}$. These relative comparisons also hold if we instead choose a SFH parameterised by a galaxy
undergoing a rejuvination event of variable strength (5-20\% by stellar mass) at a variety of different times (2-3 Gyr) after the onset of its initial
star formation event or a delayed $\tau$ model.



\bsp	
\label{lastpage}
\end{document}